\documentclass[11pt]{article}

\usepackage{amsmath}
\usepackage{amsfonts}
\usepackage{amssymb}
\usepackage{pdflscape}
\usepackage{hhline}
\usepackage{wrapfig}

\usepackage{hyperref}

\def\be{\begin{eqnarray}}
\def\ee{\end{eqnarray}}
\def\nn{\nonumber}

\def\Bm{\mathcal{B}}

\def\Dm{\mathcal{D}}
\def\Fm{\mathcal{F}}
\def\Gm{\mathcal{G}}
\def\Hm{\mathcal{H}}
\def\Jm{\mathcal{J}}
\def\Km{\mathcal{K}}

\def\Pm{\mathcal{P}}
\def\Qm{\mathcal{Q}}
\def\Rm{\mathcal{R}}
\def\Vm{\mathcal{V}}
\def\Xm{\mathcal{X}}
\def\Zm{\mathcal{Z}}

\def\tRm{\tilde{\mathcal{R}}}

\def\Mfr{\mathfrak{M}}

\def\Pfr{\mathfrak{P}}

\def\Qfr{\mathfrak{Q}}

\def\Rfr{\mathfrak{R}}

\def\Zfr{\mathfrak{Z}}

\def\hPfr{\hat{\mathfrak{P}}}
\def\hRfr{\hat{\mathfrak{R}}}

\def\brr{\right>}
\def\ckl{\left<}

\def\rank{\mathrm{rank\,}}
\def\dim{\mathrm{dim\,}}
\def\Tr{\mathrm{Tr\,}}
\def\ker{\mathrm{ker\,}}
\def\Im{\mathrm{Im\,}}
\def\coIm{\mathrm{coIm\,}}

\def\Id{\mathrm{Id}}
\def\IdId{\mathrm{Id}\otimes\mathrm{Id}}
\def\PV{{\Large\mbox{$\Pi$}}^{\wedge}}

\newcommand{\one}[2]{\hspace{#1}\begin{array}{c}\footnotesize\mbox{I}\\[-2mm]\end{array}\hspace{#2}}
\newcommand{\two}[2]{\hspace{#1}\begin{array}{c}\footnotesize\mbox{II}\\[-2mm]\end{array}\hspace{#2}}

\def\fp{{\scriptstyle+}}
\def\fm{{\scriptstyle-}}

\def\fpp{{\scriptstyle++}}
\def\fmm{{\scriptstyle--}}
\def\fpm{{\scriptstyle+-}}
\def\fmp{{\scriptstyle-+}}

\def\Rs{\hspace{0em}\rule{0pt}{0.9em}^{*}\hspace{-0.2em}\Rm}
\def\Rp{\hspace{0em}\rule{0pt}{0.9em}^{+}\hspace{-0.2em}\Rm}
\def\Rn{\hspace{0em}\rule{0pt}{0.9em}^{-}\hspace{-0.2em}\Rm}
\def\Rw{\hspace{0em}\rule{0pt}{0.9em}^{\circ}\hspace{-0.2em}\Rm}
\def\Rb{\hspace{0em}\rule{0pt}{0.9em}^{\bullet}\hspace{-0.2em}\Rm}

\def\Qlcup{{\rule{0em}{0.7em}^{*}\hspace{-0.2em}{\Qm\rule{-0.75em}{1.15em}}^{\lcup}}
\rule{-0.3em}{0em}}%^i_j}
\def\Qrcup{{\rule{0em}{0.7em}^{*}\hspace{-0.2em}{\Qm\rule{-0.75em}{1.15em}}^{\lcup}}
\rule{-0.3em}{0em}} %^i_j}
\def\Qlcap{{\rule{0em}{0.7em}^{*}\hspace{-0.2em}{\Qm\rule{-0.65em}{1.15em}}^{\lcap}}
\rule{-0.3em}{0em}} %^i_j}
\def\Qrcap{{\rule{0em}{0.7em}^{*}\hspace{-0.2em}{\Qm\rule{-0.65em}{1.15em}}^{\lcap}}
\rule{-0.3em}{0em}} %^i_j}

\def\Zs{\hspace{-0.2em}\rule{0pt}{0.8em}^{*}\hspace{-0.2em}\Zm}
\def\Zbs{\hspace{-0.2em}\rule{0pt}{0.8em}^{*}\hspace{-0.2em}\bar\Zm}
\def\Zbps{\hspace{-0.2em}\rule{0pt}{0.8em}^{*^{\hspace{-0.1em}\prime}}\hspace{-0.35em}\bar\Zm}

\def\Rfrp{\hspace{-0.2em}\rule{0pt}{0.9em}^{+}\hspace{-0.2em}\Rfr}
\def\Rfrn{\hspace{-0.2em}\rule{0pt}{0.9em}^{-}\hspace{-0.2em}\Rfr}

\def\c{\times}
\def\cc{\times\!\!\times}

\def\Lcup{{\unitlength=0.2em
\begin{picture}(7,3)(4,0)
\qbezier(5,1.5)(7.5,-1)(10,1.5)
\put(5,1.5){\vector(-1,1){0.6}}
\end{picture}}}

\def\Rcup{{\unitlength=0.2em
  \begin{picture}(7,3)(5,0)
\qbezier(5,1.5)(7.5,-1)(10,1.5)
\put(10,1.2){\vector(1,1){0.6}}
\end{picture}}}

\def\Lcap{{\unitlength=0.2em
  \begin{picture}(7,3)(4,1)
\qbezier(5,1.5)(7.5,4)(10,1.5)
\put(5,1.2){\vector(-1,-1){0.6}}
\end{picture}}}

\def\Rcap{{\unitlength=0.2em
  \begin{picture}(7,3)(5,1)
\qbezier(5,1.5)(7.5,4)(10,1.5)
\put(10,1.5){\vector(1,-1){0.6}}
\end{picture}}}

\def\lcup{{\unitlength=0.15em
\begin{picture}(7,3)(5,0)
\qbezier(5,1.5)(7.5,-1)(10,1.5)
\put(5,1.5){\vector(-1,1){0.6}}
\end{picture}}}

\def\lcap{{\unitlength=0.15em
  \begin{picture}(7,3)(5,1)
\qbezier(5,1.5)(7.5,4)(10,1.5)
\put(5,1.2){\vector(-1,-1){0.6}}
\end{picture}}}

\newcommand{\pcr}{
\begin{picture}(24,20)(-12,-12)
\qbezier(-6,-6)(0,0)(6,6)
\qbezier(-6,6)(-4,4)(-3,3)\qbezier(6,-6)(4,-4)(3,-3)
\qbezier(6,6)(4,6)(2,6)\qbezier(6,6)(6,4)(6,2)
\qbezier(-6,6)(-4,6)(-2,6)\qbezier(-6,6)(-6,4)(-6,2)
\put(-11,-9){$i$}\put(-11,2){$l$}\put(8,-9){$j$}\put(8,2){$k$}
\end{picture}
}

\newcommand{\Qncr}{
\begin{picture}(24,20)(-12,-12)
\qbezier(-6,-6)(0,0)(6,6)
\qbezier(-6,6)(-4,4)(-3,3)\qbezier(6,-6)(4,-4)(3,-3)
\qbezier(6,6)(4,6)(2,6)\qbezier(6,6)(6,4)(6,2)
\qbezier(6,-6)(4,-6)(2,-6)\qbezier(6,-6)(6,-4)(6,-2)
\put(-12,-9){$i$}\put(-12,2){$j$}\put(8,-9){$l$}\put(8,2){$k$}
\end{picture}
}

\newcommand{\ncrQ}{
\begin{picture}(24,20)(-12,-12)
\qbezier(-6,-6)(0,0)(6,6)
\qbezier(-6,6)(-4,4)(-3,3)\qbezier(6,-6)(4,-4)(3,-3)
\qbezier(-6,-6)(-4,-6)(-2,-6)\qbezier(-6,-6)(-6,-4)(-6,-2)
\qbezier(-6,6)(-4,6)(-2,6)\qbezier(-6,6)(-6,4)(-6,2)
\put(-13,-9){$k$}\put(-11,2){$l$}\put(8,-9){$j$}\put(8,2){$i$}
\end{picture}
}

\newcommand{\QpcrQ}{
\begin{picture}(24,20)(-12,-12)
\qbezier(-6,-6)(0,0)(6,6)
\qbezier(-6,6)(-4,4)(-3,3)\qbezier(6,-6)(4,-4)(3,-3)
\qbezier(-6,-6)(-4,-6)(-2,-6)\qbezier(-6,-6)(-6,-4)(-6,-2)
\qbezier(6,-6)(4,-6)(2,-6)\qbezier(6,-6)(6,-4)(6,-2)
\put(-13,-9){$k$}\put(-12,2){$j$}\put(8,-9){$l$}\put(8,2){$i$}
\end{picture}
}

\newcommand{\ncr}{
\begin{picture}(24,20)(-12,-12)
\qbezier(-6,-6)(-4,-4)(-3,-3)\qbezier(6,6)(4,4)(3,3)
\qbezier(-6,6)(0,0)(6,-6)
\qbezier(6,6)(4,6)(2,6)\qbezier(6,6)(6,4)(6,2)
\qbezier(-6,6)(-4,6)(-2,6)\qbezier(-6,6)(-6,4)(-6,2)
\put(-12,-9){$j$}\put(-13,2){$k$}\put(8,-9){$i$}\put(8,2){$l$}
\end{picture}
}

\newcommand{\Qpcr}{
\begin{picture}(24,20)(-12,-12)
\qbezier(-6,-6)(-4,-4)(-3,-3)\qbezier(6,6)(4,4)(3,3)
\qbezier(-6,6)(0,0)(6,-6)
\qbezier(6,6)(4,6)(2,6)\qbezier(6,6)(6,4)(6,2)
\qbezier(6,-6)(4,-6)(2,-6)\qbezier(6,-6)(6,-4)(6,-2)
\put(-11,-9){$j$}\put(-10,2){$i$}\put(8,-9){$k$}\put(8,2){$l$}
\end{picture}
}

\newcommand{\pcrQ}{
\begin{picture}(24,20)(-12,-12)
\qbezier(-6,-6)(-4,-4)(-3,-3)\qbezier(6,6)(4,4)(3,3)
\qbezier(-6,6)(0,0)(6,-6)
\qbezier(-6,-6)(-4,-6)(-2,-6)\qbezier(-6,-6)(-6,-4)(-6,-2)
\qbezier(-6,6)(-4,6)(-2,6)\qbezier(-6,6)(-6,4)(-6,2)
\put(-11,-9){$l$}\put(-13,2){$k$}\put(8,-9){$i$}\put(8,2){$j$}
\end{picture}
}

\newcommand{\QncrQ}{
\begin{picture}(24,20)(-12,-12)
\qbezier(-6,-6)(-4,-4)(-3,-3)\qbezier(6,6)(4,4)(3,3)
\qbezier(-6,6)(0,0)(6,-6)
\qbezier(-6,-6)(-4,-6)(-2,-6)\qbezier(-6,-6)(-6,-4)(-6,-2)
\qbezier(6,-6)(4,-6)(2,-6)\qbezier(6,-6)(6,-4)(6,-2)
\put(-11,-9){$l$}\put(-11,2){$i$}\put(8,-9){$k$}\put(8,2){$j$}
\end{picture}
}

\newcommand{\wcr}{
\begin{picture}(24,20)(-12,-12)
\qbezier(-6,-6)(-4,-4)(-3,-3)\qbezier(6,6)(4,4)(3,3)
\qbezier(-6,6)(-4,4)(-3,3)\qbezier(6,-6)(4,-4)(3,-3)
\qbezier(6,6)(4,6)(2,6)\qbezier(6,6)(6,4)(6,2)
\qbezier(-6,6)(-4,6)(-2,6)\qbezier(-6,6)(-6,4)(-6,2)
\put(0,0){\circle{7.2}}
\put(-12,-9){$i$}\put(-12,2){$l$}\put(10,-9){$k$}\put(10,2){$j$}
\end{picture}
}

\newcommand{\bcr}{
\begin{picture}(24,20)(-12,-12)
\qbezier(-6,-6)(-4,-4)(-3,-3)\qbezier(6,6)(4,4)(3,3)
\qbezier(-6,6)(-4,4)(-3,3)\qbezier(6,-6)(4,-4)(3,-3)
\qbezier(6,6)(4,6)(2,6)\qbezier(6,6)(6,4)(6,2)
\qbezier(-6,6)(-4,6)(-2,6)\qbezier(-6,6)(-6,4)(-6,2)
\put(0,0){\circle*{7.2}}
\put(-12,-9){$i$}\put(-12,2){$l$}\put(10,-9){$k$}\put(10,2){$j$}
\end{picture}
}

\newcommand{\lmin}{
\begin{picture}(18,15)(-9,-9)
\qbezier(-6,6)(-6,0)(0,0)\qbezier(6,6)(6,0)(0,0)
\put(1,-1){\circle*{3}}
\qbezier(-6,6)(-2,3)(2,4)\qbezier(-6,6)(-6,4)(-7,-2)
\put(-6,-9){$j$}\put(3,-9){$i$}
\end{picture}
}

\newcommand{\rmin}{
\begin{picture}(18,15)(-9,-9)
\qbezier(-6,6)(-6,0)(0,0)\qbezier(6,6)(6,0)(0,0)
\put(1,-1){\circle*{3}}
\qbezier(6,6)(2,3)(-2,4)\qbezier(6,6)(6,4)(7,-2)
\put(-6,-9){$i$}\put(1,-9){$j$}
\end{picture}
}

\newcommand{\lmax}{
\begin{picture}(18,18)(-9,-9)
\qbezier(-6,-6)(-6,0)(0,0)\qbezier(6,-6)(6,0)(0,0)
\put(1,-1){\circle*{3}}
\qbezier(-6,-6)(-2,-3)(2,-4)\qbezier(-6,-6)(-6,-4)(-7,2)
\put(-5,2){$j$}\put(3,2){$i$}
\end{picture}
}

\newcommand{\rmax}{
\begin{picture}(18,18)(-9,-9)
\qbezier(-6,-6)(-6,0)(0,0)\qbezier(6,-6)(6,0)(0,0)
\put(1,0){\circle*{3}}
\qbezier(6,-6)(2,-3)(-2,-4)\qbezier(6,-6)(6,-4)(7,2)
\put(-5,2){$i$}\put(3,3){$j$}
\end{picture}
}

\newcommand{\Pcr}{
{\unitlength=0.3mm
\begin{picture}(24,24)(-12,-3)
\qbezier(-6,-6)(0,0)(6,6)
\qbezier(-6,6)(-4,4)(-3,3)\qbezier(6,-6)(4,-4)(3,-3)
\qbezier(6,6)(4,6)(2,6)\qbezier(6,6)(6,4)(6,2)
\qbezier(-6,6)(-4,6)(-2,6)\qbezier(-6,6)(-6,4)(-6,2)
\end{picture}
}
}

\newcommand{\Ncr}{
{\unitlength=0.3mm
\begin{picture}(24,24)(-12,-3)
\qbezier(-6,-6)(-4,-4)(-3,-3)\qbezier(6,6)(4,4)(3,3)
\qbezier(-6,6)(0,0)(6,-6)
\qbezier(6,6)(4,6)(2,6)\qbezier(6,6)(6,4)(6,2)
\qbezier(-6,6)(-4,6)(-2,6)\qbezier(-6,6)(-6,4)(-6,2)
\end{picture}
}
}

\newcommand{\Wcr}{
{\unitlength=0.3mm
\begin{picture}(24,0)(-12,-3)
\qbezier(-6,-6)(-4,-4)(-3,-3)\qbezier(6,6)(4,4)(3,3)
\qbezier(-6,6)(-4,4)(-3,3)\qbezier(6,-6)(4,-4)(3,-3)
\qbezier(6,6)(4,6)(2,6)\qbezier(6,6)(6,4)(6,2)
\qbezier(-6,6)(-4,6)(-2,6)\qbezier(-6,6)(-6,4)(-6,2)
\put(0,0){\circle{7.2}}
\end{picture}
}
}

\newcommand{\Bcr}{
{\unitlength=0.3mm
\begin{picture}(24,0)(-12,-3)
\qbezier(-6,-6)(-4,-4)(-3,-3)\qbezier(6,6)(4,4)(3,3)
\qbezier(-6,6)(-4,4)(-3,3)\qbezier(6,-6)(4,-4)(3,-3)
\qbezier(6,6)(4,6)(2,6)\qbezier(6,6)(6,4)(6,2)
\qbezier(-6,6)(-4,6)(-2,6)\qbezier(-6,6)(-6,4)(-6,2)
\put(0,0){\circle*{7.2}}
\end{picture}
}
}

\newcommand{\Lmin}{
{\unitlength=0.3mm
\begin{picture}(18,0)(-9,0)
\qbezier(-6,6)(-6,0)(0,0)\qbezier(6,6)(6,0)(0,0)
\put(1,-1){\circle*{3}}
\qbezier(-6,6)(-2,3)(2,4)\qbezier(-6,6)(-6,4)(-7,-2)
\end{picture}
}
}

\newcommand{\Rmin}{
{\unitlength=0.3mm
\begin{picture}(18,0)(-9,0)
\qbezier(-6,6)(-6,0)(0,0)\qbezier(6,6)(6,0)(0,0)
\put(1,-1){\circle*{3}}
\qbezier(6,6)(2,3)(-2,4)\qbezier(6,6)(6,4)(7,-2)
\end{picture}
}
}

\newcommand{\Lmax}{
{\unitlength=0.3mm
\begin{picture}(18,0)(-9,-9)
\qbezier(-6,-6)(-6,0)(0,0)\qbezier(6,-6)(6,0)(0,0)
\put(1,-1){\circle*{3}}
\qbezier(-6,-6)(-2,-3)(2,-4)\qbezier(-6,-6)(-6,-4)(-7,2)
\end{picture}
}
}

\newcommand{\Rmax}{
{\unitlength=0.3mm
\begin{picture}(18,0)(-9,-9)
\qbezier(-6,-6)(-6,0)(0,0)\qbezier(6,-6)(6,0)(0,0)
\put(1,0){\circle*{3}}
\qbezier(6,-6)(2,-3)(-2,-4)\qbezier(6,-6)(6,-4)(7,2)
\end{picture}
}
}

\newcommand{\figpcr}[2]{
\put(#1,#2){
\qbezier(-6,-6)(0,0)(6,6)
\qbezier(-6,6)(-4,4)(-3,3)\qbezier(6,-6)(4,-4)(3,-3)
\qbezier(6,6)(4,6)(2,6)\qbezier(6,6)(6,4)(6,2)
\qbezier(-6,6)(-4,6)(-2,6)\qbezier(-6,6)(-6,4)(-6,2)
}
}

\newcommand{\figvarpcr}[2]{
\put(#1,#2){
\qbezier(-6,-6)(0,0)(6,6)
\qbezier(-6,6)(-4,4)(-3,3)\qbezier(6,-6)(4,-4)(3,-3)
\put(-1,-1){\qbezier(6,6)(4,6)(2,6)\qbezier(6,6)(6,4)(6,2)}
\put(1,-1){\qbezier(-6,6)(-4,6)(-2,6)\qbezier(-6,6)(-6,4)(-6,2)}
}
}

\newcommand{\figQpcr}[2]{
\put(#1,#2){
\qbezier(-6,-6)(-4,-4)(-3,-3)\qbezier(6,6)(4,4)(3,3)
\qbezier(-6,6)(0,0)(6,-6)
\qbezier(6,6)(4,6)(2,6)\qbezier(6,6)(6,4)(6,2)
\qbezier(6,-6)(4,-6)(2,-6)\qbezier(6,-6)(6,-4)(6,-2)
}
}

\newcommand{\figpcrQ}[2]{
\put(#1,#2){
\qbezier(-6,-6)(-4,-4)(-3,-3)\qbezier(6,6)(4,4)(3,3)
\qbezier(-6,6)(0,0)(6,-6)
\qbezier(-6,-6)(-4,-6)(-2,-6)\qbezier(-6,-6)(-6,-4)(-6,-2)
\qbezier(-6,6)(-4,6)(-2,6)\qbezier(-6,6)(-6,4)(-6,2)
}
}

\newcommand{\figQpcrQ}[2]{
\put(#1,#2){
\qbezier(-6,-6)(0,0)(6,6)
\qbezier(-6,6)(-4,4)(-3,3)\qbezier(6,-6)(4,-4)(3,-3)
\qbezier(-6,-6)(-4,-6)(-2,-6)\qbezier(-6,-6)(-6,-4)(-6,-2)
\qbezier(6,-6)(4,-6)(2,-6)\qbezier(6,-6)(6,-4)(6,-2)
}
}

\newcommand{\figncr}[2]{
\put(#1,#2){
\qbezier(-6,-6)(-4,-4)(-3,-3)\qbezier(6,6)(4,4)(3,3)
\qbezier(-6,6)(0,0)(6,-6)
\qbezier(6,6)(4,6)(2,6)\qbezier(6,6)(6,4)(6,2)
\qbezier(-6,6)(-4,6)(-2,6)\qbezier(-6,6)(-6,4)(-6,2)
}
}

\newcommand{\figQncrQ}[2]{
\put(#1,#2){
\qbezier(-6,-6)(-4,-4)(-3,-3)\qbezier(6,6)(4,4)(3,3)
\qbezier(-6,6)(0,0)(6,-6)
\qbezier(-6,-6)(-4,-6)(-2,-6)\qbezier(-6,-6)(-6,-4)(-6,-2)
\qbezier(6,-6)(4,-6)(2,-6)\qbezier(6,-6)(6,-4)(6,-2)
}
}

\newcommand{\figwcr}[2]{
\put(#1,#2){
\qbezier(-6,-6)(-4,-4)(-3,-3)\qbezier(6,6)(4,4)(3,3)
\qbezier(-6,6)(-4,4)(-3,3)\qbezier(6,-6)(4,-4)(3,-3)
\put(0,0){\circle{7.2}}
}
}

\newcommand{\figbcr}[2]{
\put(#1,#2){
\qbezier(-6,-6)(-4,-4)(-3,-3)\qbezier(6,6)(4,4)(3,3)
\qbezier(-6,6)(-4,4)(-3,3)\qbezier(6,-6)(4,-4)(3,-3)
\put(0,0){\circle*{7.2}}
}
}

\newcommand{\figei}[6]
{\put(#1,#2)
{#3{40}{40}#4{60}{40}#5{40}{10}#6{60}{10}
\put(50,0){
\qbezier(-5,5)(0,0)(5,5)
}
\put(50,10){
\qbezier(-5,5)(0,10)(5,5)
}
\put(50,30){
\qbezier(-5,5)(0,0)(5,5)
}
\put(50,40){
\qbezier(-5,5)(0,10)(5,5)
}
\qbezier(35,16)(35,25)(50,25)
\qbezier(65,35)(75,25)(50,25)
\qbezier(35,4)(25,0)(25,15)\qbezier(25,15)(25,25)(35,35)
\qbezier(65,4)(75,0)(80,15)\qbezier(35,60)(28,50)(35,45)\qbezier(80,15)(90,85)(35,60)
\qbezier(65,45)(75,55)(75,40)\qbezier(75,40)(75,28)(65,15)
}
}

\newcommand{\figsncr}[2]{
\figpcr{#1}{#2}
\put(#1,#2){
\put(0,-4){\qbezier(-4,0)(0,0)(4,0)}
}}

\newcommand{\figdbcr}[2]{
\figpcr{#1}{#2}
\put(#1,#2){
\put(0,-4){\qbezier(-4,0)(0,0)(4,0)}
\put(0,-5){\qbezier(-5,0)(0,0)(5,0)}
\put(0.5,-4.5){\circle*{3}}
}}

\newcommand{\figsncup}[2]{
\put(#1,#2){
\qbezier(-5,0)(0,-5)(5,0)
\put(0,0){\qbezier(-5,0)(0,0)(5,0)}
}
}

\newcommand{\figsncap}[2]{
\put(#1,#2){
\qbezier(-5,0)(0,5)(5,0)
\put(0,0){\qbezier(-5,0)(0,0)(5,0)}
}
}

\newcommand{\figlmin}[2]{
\put(#1,#2){
\qbezier(-6,6)(-6,0)(0,0)\qbezier(6,6)(6,0)(0,0)
\put(1,-1){\circle*{3}}
\qbezier(-6,6)(-2,3)(2,4)\qbezier(-6,6)(-6,4)(-7,-2)
}
}

\newcommand{\figrmin}[2]{
\put(#1,#2){
\qbezier(-6,6)(-6,0)(0,0)\qbezier(6,6)(6,0)(0,0)
\put(1,-1){\circle*{3}}
\qbezier(6,6)(2,3)(-2,4)\qbezier(6,6)(6,4)(7,-2)
}
}

\newcommand{\figlmax}[2]{
\put(#1,#2){
\qbezier(-6,-6)(-6,0)(0,0)\qbezier(6,-6)(6,0)(0,0)
\put(1,-1){\circle*{3}}
\qbezier(-6,-6)(-2,-3)(2,-4)\qbezier(-6,-6)(-6,-4)(-7,2)
}
}

\newcommand{\figrmax}[2]{
\put(#1,#2){
\qbezier(-6,-6)(-6,0)(0,0)\qbezier(6,-6)(6,0)(0,0)
\put(1,0){\circle*{3}}
\qbezier(6,-6)(2,-3)(-2,-4)\qbezier(6,-6)(6,-4)(7,2)
}
}

\newcommand{\Arrow}[5]{
\begin{picture}(0,0)
\put(#4,#5){\vector(#1,#2){#3}}
\end{picture}
}

\def\c{\times}
\def\cc{\times\!\!\times}

\def\en{{
\footnotesize{
\begin{array}{c}|\end{array}}}
}
\def\sn{{
\footnotesize{
\begin{array}{c}\uparrow\end{array}}}
}
\def\db{{
\footnotesize{
\begin{array}{c}\uparrow\hspace{-0.8mm}\uparrow\\[-2.5mm]\bullet\\[-0.5mm]\end{array}}}
}

\title{Towards formalization of the soliton counting technique for the
  Khovanov--Rozansky invariants in the deformed $\Rm$-matrix
  approach}
\author{\textbf{A.Anokhina}\footnote{ {\small \textit{ITEP, Moscow, Russia}};
anokhina@itep.ru}\date{ }}

\begin{document}

\numberwithin{equation}{section}

\maketitle

\phantom{k}
\vspace{-7cm}

\begin{center}
\hfill ITEP/TH-29/17\\
\end{center}

\vspace{4cm}

\begin{abstract}
  We consider recently developed Cohomological Field Theory (CohFT)
  soliton counting diagram technique for Khovanov (Kh) and
  Khovanov--Rozansky (KhR) invariants~\cite{MG,
    Gal}. %in a more general framework of different approaches to the same invariants, putting assent on the deformed $\Rm$-matrix approach earlier developed by ourselves~\cite{AM}.
  Although the expectation  to obtain a new way for computing the invariants
  has not yet come true, we demonstrate that soliton counting
  technique can be totally formalized at an intermediate stage, at
  least in particular cases. We present the corresponding algorithm,
  based on the approach involving deformed $\Rm$-matrix and minimal
  positive division, developed previously in~\cite{AM}. We start from
  a detailed review of the minimal positive division approach,
  comparing it with other methods, including the rigorous mathematical
  treatment~\cite{KhR}. Pieces of data obtained within our approach
  are presented in the Appendices.
\end{abstract}

\tableofcontents

\section{Introduction\label{sec:int}}
\begin{table}
\caption{\label{tab:goodinv}
Constructively defined knot invariants}
{\arraycolsep=0mm
\begin{tabular}{ccccc}
\boxed{
\begin{tabular}{p{5cm}}Knot diagram\\
+\\
representation of a Lie algebra
\end{tabular}
}
&$\rightarrow$&
\boxed{
\begin{tabular}{p{3cm}}
Algebraic manipulations
\end{tabular}}
&$\rightarrow$&
\boxed{
\begin{tabular}{p{5cm}}
Knot invariant: Alexander, Jones, HOMFLY, Kauffman, Vassiliev, coloured versions, \ldots
\end{tabular}}
\end{tabular}}
\end{table}

\begin{table}
\caption{\label{tab:badinv}
Homological knot invariants}
{\arraycolsep=1mm
$
\begin{array}{ccccccc}
\boxed{
\begin{tabular}{p{1.6cm}}
The knot diagram
\end{tabular}
}
&&&&&&
\boxed{
\begin{tabular}{p{1.6cm}}
Knot invarinat
\end{tabular}
}\\[4mm]
\Downarrow
&&&&&&
\Uparrow
\\
\boxed{
\begin{tabular}{p{2cm}}
Resolution hypercube
\end{tabular}}
&\Rightarrow&
\boxed{
\begin{tabular}{p{3cm}}
Sequence of linear spaces
\end{tabular}}
&\Rightarrow&
\boxed{
\begin{tabular}{p{2.2cm}}
Complex
\end{tabular}}
&\rightarrow&
\boxed{
\begin{tabular}{p{2.5cm}}
Computing the homology
\end{tabular}}
\\[2mm]
\&
&&&&{\scriptstyle ?}\!\uparrow
&&\&\phantom{\Bigg|}\\[-3mm]
\boxed{
\begin{tabular}{p{2.2cm}}
Representation structure on vertices
\end{tabular}}
&\rightarrow&
\boxed{
\begin{tabular}{p{2.5cm}}
Spaces as representation spaces
\end{tabular}}
&\stackrel{?}{\rightarrow}&
\boxed{
\begin{tabular}{p{2.5cm}}
``Representation-friendly'' nilpotent linear maps
\end{tabular}}
&\stackrel{?}{\rightarrow}&
\boxed{
\begin{tabular}{p{2.5cm}}
Decomposing the homology
\end{tabular}}
\\[-2mm]
\uparrow&&
&&&&{\scriptstyle ?}\!\downarrow
\\
\boxed{
\begin{tabular}{p{3cm}}
Representation of Lie algebra
\end{tabular}}
&
\multicolumn{5}{c}{
\begin{tabular}{cp{5cm}}
$\Rightarrow$&Standard construction\\
$\rightarrow$&Proposed conjecture \\
$\stackrel{?}{\rightarrow}$&Particular cases analysed
\end{tabular}
}
&
\boxed{
\begin{tabular}{p{2.5cm}}
``Differential expansion'' for the knot invariant
\end{tabular}}
\end{array}
$}
\end{table}

Cohomological knot invariants are rather young by knot theory
standards. Khovanov invariants (denoted Kh below)~\cite{Khov} were
proposed less two decades ago and their Khovanov--Rozanski (KhR)
version~\cite{KhR} was introduced less than a decade ago.  Both
approaches associate a function (precisely, a polynomial) to a knot
(or a link), so that this function is invariant under arbitrary
continuous space transformations. They belong to the wide and
well-studied class of polynomial knot invariants~\cite{KauffTB, katlas, knbook}, although drastically differing from the other ones
in their properties. Differences between various
knot %(and more general topological~\cite{???} %????Sham, Dan,PtDbar
% )
invariants are illustrated in
Tables~\ref{tab:goodinv}~and~\ref{tab:badinv}.

The first and key special point of Khovanov, KhR and other invariants
of the same kind~\cite{KnHomLec} %Dan, PtDbar
is a separate structure associated with a knot, called \textit{knot
  homology.} The knot polynomial is a generating function for the
basis vectors in homologies of a certain complex (and, even worse, in
the KhR case the complex of other complexes). This sort of definition,
which is multi-level and at many points implicit, causes severe
obstacles both for general analysis and for explicit
computations. Searches for an alternative approach are naturally quite
popular (see, e.g., the references in Table~\ref{tab:KhRsumm}).

A closely related quantity is the \textit{superpolynomial} of the
knot~\cite{GSchV}, which remains among the most obscure issues of the
knot theory.  The superpolynomials are often studied together with the
KhR invariants, the connection being two-fold. On the one hand, the
superpolynomial is by its only strict (yet not very practical)
definition an analytic continuation of the KhR polynomial (similarly
to the HOMFLY invariant, which is an analytic continuation of the
discrete set of polynomials generalising Jones
polynomial\footnote{However, the naively performed analytical
  continuation, working well in the HOMFLY case, runs into certain
  problems in the KhR case~\cite{RasmKhR, DGN, AM, GorLew}, see
  sec.~\ref{sec:finN},~\ref{sec:summ}). %???Mor
}.  On the other hand, alternative approaches developed for
superpolynomials, although being neither general nor mathematically
strict in the most cases, proved themselves highly useful both for
computations and investigation of general properties of the
theory. Hence, the interference of these two subjects may shed light
on both super- and KhR polynomials.

A new inspiration in the subject comes from the recently proposed
\textit{cohomological field theory}~\cite{Witt}, (``CohFT''
henceforth) associated with Khovanonov and KhR invarinats in the same
way as Chern-Simons theory is associated with Jones and HOMFLY
invarinats~\cite{Schw, WittJ, WittH}.

In the context of the recent progress of various approaches to the KhR
invariants we wish to recall our own research~\cite{AM} and compare it
with rigorous mathematical treatment, as well as with alternative
methods, including the newly proposed CohFT approach.

\

Various issues concerned here, are summarized and supplied with
bibliographic references in Table~\ref{tab:KhRsumm}. The rightmost
column of the table reflects the structure of the present text.

\subsection{Different appoaches to KhR invariants\label{sec:KhRsumm}}

\begin{table}
\caption{Summary of the discussed approaches, with bibliographic and internal references\label{tab:KhRsumm}}
\begin{tabular}{|p{5cm}|p{3cm}|p{5cm}|p{2.5cm}|}
  \hline\multicolumn{3}{|c|}{}&\\[-4mm]
  \multicolumn{3}{|c|}{Khovanov polynomial}&$N=2$\\
  \hline&&&\\[-4mm]
  Introduced in&\cite{Khov}&&\\
  \cline{1-3}&&&\\[-4mm]
  Computational technique \par developed in&\cite{BarNat}&&\\
  \cline{1-3}&&&\\[-4mm]
  Table of results, together with computer code&\cite{katlas}&Presented for all prime knots (up to 11 crossings) and links (up to 11 crossings); in principle, computed for any knots&\\
  \hline&&&\\[-4mm]
  Reviewed, e.g., in&\cite{DM1, DM2}, \cite{SatNaw, KnHomLec}&&\\
  \hline
  \hline\multicolumn{3}{|c|}{}&\\[-4mm]
  \multicolumn{3}{|c|}{Khovanov-Rozansky polynomial}&$N\in\mathbb{Z}_+$\\
  \hline&&&\\[-4mm]
  The definition introduced in&\cite{KhR}&&\\
  \cline{1-3}&&&\\[-4mm]
  Applied to explicit \par computations in&\cite{RasmKhR}&``Thin'' knots up to 9 crossings&\\
  \cline{2-3}&&&\\[-4mm]
  &\cite{CarMuf}&Knots and links up to 6 crossings, mostly for particular vales of $N$&\\
  \hline&&\multicolumn{2}{c|}{}\\[-4mm]
  The appoach reviewed, e.g., in&\cite{SatNaw, KnHomLec}&\multicolumn{2}{c|}{}\\
  \hline\multicolumn{4}{|c|}{}\\[-4mm]
  \multicolumn{4}{|p{5cm}|}{Attemts of modification}\\
  \hline&&&\\[-4mm]
  Tensor-like formalism&\cite{DM3}&Simplest examples, 2-strand torus knots, twist knots&\\
  \hline&&&\\[-4mm]
  $\Rm$-matrix bases formalism&\cite{AM}&	2 and 3-strand torus knots, 3- and 4-strand knots and links up to 6 crossings, two-component links from two antiparallel strands&ssec.~\ref{sec:gen}, \ref{sec:repth}\\
  \hline&&&\\[-4mm]
  Positive division technique&\cite{AM}&&ssec.~\ref{sec:dec}, \ref{sec:div}, \ref{sec:Gal} \\
  \hline&&&\\[-4mm]
  CoHFT approach&\cite{Witt, MG, Gal}&&sec.~\ref{sec:Gal}\\
  \hline
  \hline\multicolumn{3}{|c|}{}&\\[-4.5mm]
  \multicolumn{3}{|c|}{Superpolynomial}&$0<N_0\le N\in\mathbb{C}$\\
  \hline&&&\\[-4mm]
  Introduced in&\cite{GSchV}&&\\
  \hline&&&\\[-4mm]
  Elvolution method&\cite{DMMSS}, \cite{MMM3}, \cite{AntM2}&&sec.~\ref{sec:dexp}\\
  \hline&&&\\[-4mm]
  Differential expansion&\cite{DGN}, \cite{MMM3}, \cite{supsup}, \cite{Art}, \cite{MorKon2}&&sec.~\ref{sec:dexp}\\
  \hline
  \hline&&&\\[-4mm]
  Finite $N$ problem&\cite{RasmKhR}, \cite{DGN}, \cite{AM}, \cite{GorLew}&&ssec.~\ref{sec:finN}, \ref{sec:summ} %??? Mor
  \\
  \hline
  \hline&&&\\[-4mm]
  coloured generalizations&\cite{KhovCol}, \cite{DMMSS}, \cite{MMM3}, \cite{supsup}, \cite{Art}, \cite{IndSup}, \cite{KnHomLec},\cite{Dan}&&sec.~\ref{sec:summ}\\
  \hline
\end{tabular}
\end{table}

\subsection{Conjectures and expectation s\label{sec:Khconj}}
All the viewpoints on the KhR invariants (or superpolynomials)
differing from the original one generally aim, speaking most strictly
and naively, to obtain the desired invariants just by performing some
algebraic manipulations, i.e., to pass by the construction of the
complex and the computation of the homologies. In other words, one
tries to invent a treatment more in the spirit of the standard knot
invariant computations~\cite{KauffTB}.%  where many other knot
% invariants, including the ones mentioned in Tab.~\ref{tab:goodinv} are
% obtained.

Because the KhR formalism applies to a complex of complexes, the expectation 
for its simplification lies on two levels,

\begin{enumerate}
\item{\textbf{The weak expectation }: \underline{to find an explicit
      representation for the spaces and maps in the KhR complex.}
\par %\indent
In sec.~\ref{sec:gen} we suggest a way to do this relying on the
$\Rm$-matrix formalism (sec.~\ref{sec:repth} contains some further
details). The same method was in fact implicitly used in~\cite{AM}.}

\item{\textbf{The strong expectation }: \underline{to skip computing the
      homologies.}
\par
This is the main idea of the minimal positive division technique,
developed in~\cite{AM} which we present in sec.~\ref{sec:div} in its
abstract form. In~\ref{sec:Gal} we attempt to combine it with the
CohFT formalism from~\cite{Gal}. }
\label{item:1}

\end{enumerate}

\section{A sketch of the general construction\label{sec:gen}}

\begin{table}
  \caption{Sketch of the hypercube calculus for knot invariants\label{tab:sketch}}
\begin{tabular}{|p{4cm}|p{4.5cm}|p{3.5cm}|p{2.5cm}|}
\hline&\multicolumn{2}{|c|}{}&\\[-4.2mm]
Resolution hypercube $\Sigma$&
\multicolumn{2}{|c|}{Complex
$\Vm_0\stackrel{\hat d_1}{\longrightarrow} \Vm_1\stackrel{\hat d_2}{\longrightarrow}
\ldots\stackrel{\hat d_n}{\longrightarrow}\Vm_n$}&sec.~\ref{sec:kndef}\\
\hline&&&\\[-4mm]
Vertex $*$
&Colouring (resolution) $*$ of the diagram $\Dm$ of the knot $\Km$,&
Operator\par $\Zbs\equiv q^{\hat\Delta}$ (\ref{Qm}, \ref{Zcol}, \ref{Zbcol})&sec.~\ref{sec:hypsp}\\
\cline{3-4}&&&\\[-4mm]
& %\small{
\vspace{-5mm}
\begin{itemize}
\item{$n=\nu_*+\bar\nu_*$ \mbox{four-valent} vertices,\linebreak
    $\nu_*$ of $\Wcr$ type,\par $\bar\nu$ of $\Bcr$ type;}
\item{$2m$ turning points,\par
%\linebreak 
$m$ of type %either 
$\Lcup$ or $\Lcap$,% $\Lmin$ or $\Rmax$,
\par $m$ of type %either 
$\Rcup$ or $\Rcap$.%$\Lmax$ or $\Rmin$.
}
\end{itemize}
%}
&Representation space $$\Vm^{*}\subset V^{\circ m}=\Im \Zbs$$&ssec.~\ref{sec:hypsp}, \ref{sec:grad}, \ref{sec:repth} \\
\cline{2-4}&&&\\[-4mm]
&Edge $i$&
Representation space $V_i$&sec.~\ref{sec:kndef}\\
\cline{2-4}&&&\\[-4mm]
&A crossing/ \par four-valent vertex $\ckl k,l|i,j\brr$
&Operator $V_i\otimes V_j\mapsto V_k\otimes V_l$&sec.~\ref{sec:gen}\\
\cline{2-4}&&&\\[-4mm]
&Turning point/ \par two-valent vertex $\ckl \bar{i}|i\brr$&Operator $V_{\bar i}\mapsto V_i$&sec.~\ref{sec:hypsp}\\
\cline{2-4}&&&\\[-4mm]
&$\Wcr$ vertex&Identity operator $\IdId$&sec.~\ref{sec:hypsp}\\
\cline{2-4}&&&\\[-4mm]
&$\Bcr$ vertex&``Double'' projector \par $[2]_q\PV$ (\ref{Rdec})&sec.~\ref{sec:hypsp}\\
\cline{2-4}&&&\\[-4mm]
&Ignored turning point,\par of type $\Lcup$ or $\Lcap$
&Identity operator $\Id$&sec.~\ref{sec:hypsp}\\
\cline{2-4}&&&\\[-4mm]
&\multicolumn{1}{|p{4.8cm}|}{Acknowledged turning point, of type $\Rcup$}&Operator $\Qrcup$ (\ref{Qm})&ssec.~\ref{sec:RQpol}, \ref{sec:hypsp}\\
\cline{2-4}&&&\\[-4mm]
&\multicolumn{1}{|p{4.8cm}|}{Acknowledged turning point, of type $\Rcap$}
&Operator $\Qrcap$ (\ref{Qm})&ssec.~\ref{sec:RQpol}, \ref{sec:hypsp}\\
\hline
Directed edge
$*\longrightarrow*^{\prime}$,\par
$\nu_*=\nu_{*^{\prime}}+1$, $\bar\nu_*=\bar\nu_{*^{\prime}}-1$
&\multicolumn{2}{p{10cm}|}
{Morphism
$\Mfr^{*}_{*^{\prime}}:\ \Vm^{*}\to \Vm^{*^{\prime}}$,\par
%&\multicolumn{2}{c|}{
commuting with the grading operator,\hspace{3mm}
$\Mfr^{*}_{*^{\prime}}\hat\Delta=\hat\Delta \Mfr^{*}_{*^{\prime}}$
}&sec.~\ref{sec:cmp}\par sec.~\ref{sec:grad}\\
\hline&\multicolumn{2}{c|}{}&\\[-4mm]
Hyperplane $\Xi_k=\bigcup\limits_{\bar\nu_*=k}\hspace{-2mm}*$&
\multicolumn{2}{c|}{
Representation space
$\bigoplus\limits_{\bar\nu_*=k}\hspace{-2mm}\Vm^*$
}&ssec.~\ref{sec:cmp}\\
\hline&\multicolumn{2}{c|}{}&\\[-4mm]
Two subsequent hyperplanes, $\left(\Xi_k,\Xi_{k+1}\right)$&
\multicolumn{2}{c|}{
Differential
$\hat d_k= %??? Dan
\bigoplus\limits^{\nu_*=k}_{\nu_{*^{\prime}}=k+1}\hspace{-3mm}{\Mfr}_{*^{\prime}}^*$,}&sec.~\ref{sec:cmp}\\
&\multicolumn{2}{c|}{satisfying the nilpotency condition $\hat d_{k+1}\hat d_k=0$}&\\
\hline
\end{tabular}
\end{table}

\subsection{The necessary notions of knot and representation theory\label{sec:kndef}}
Here we briefly review the necessary notions of the knot
theory. Details can be found in any knot theory textbook, e.g.,
in~\cite{KauffTB}.
\paragraph{Oriented knot in $\mathbb{R}_3$.}
A knot $\mathcal K$ is by definition an embedding of the oriented
circle (e.g., a counterclockwise direction is selected) in the
three-dimensional flat space
\be
\Km:\
S_1\hookrightarrow\mathbb{R}_3,
\label{knot}
\ee
considered up to continuous transformations of the space
$\mathbb{R}_3$.

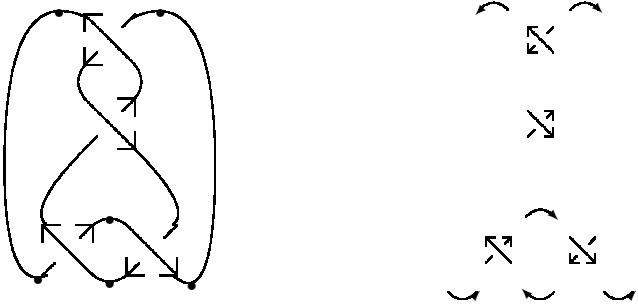
\begin{figure}
\unitlength=0.56mm
\begin{picture}(100,75)
\figpcrQ{50}{60}\figQpcr{50}{40}
\figncr{40}{10}\figQncrQ{60}{10}
% % % % % % % % % % % % % % %
\put(50,0){
\qbezier(-5,5)(0,0)(5,5)
}
\put(50,10){
\qbezier(-5,5)(0,10)(5,5)
}
% % % % % % % % % % % % % % %
\put(40,50){
\qbezier(5,-5)(0,0)(5,5)
}
\put(50,50){
\qbezier(5,-5)(10,0)(5,5)
}
% % % % % % % % % % % % % % % % % % % % %
%\qbezier(35,16)(35,25)(50,25)
%\qbezier(55,35)(65,25)(50,25)
% % % % % % % % % % % % % % % % % % %
\qbezier(35,16)(30,20)(45,35)
\qbezier(65,16)(70,20)(55,35)
\qbezier(35,4)(25,0)(25,30) \qbezier(25,30)(25,75)(45,65)
\qbezier(65,4)(75,-5)(75,30) \qbezier(75,30)(75,75)(55,65)
% % % % % % % % % % % % % % % % % % %
\put(38,66.4){\circle*{2}}\put(62,66.4){\circle*{2}}
\put(50,17.2){\circle*{2}}\put(50,2){\circle*{2}}
\put(33,3){\circle*{2}}\put(69.5,1.5){\circle*{2}}
\end{picture}
\begin{picture}(100,60)
{\unitlength=0.28mm
\figpcrQ{100}{120}\figQpcr{100}{80}
\figncr{80}{20}\figQncrQ{120}{20}
}
% % % % % % % % % % % % % % %
%\put(50,0){
%\qbezier(-5,5)(0,0)(5,5)
%}
%\put(50,10){
%\qbezier(-5,5)(0,10)(5,5)
%}
%% % % % % % % % % % % % % % %
%\put(40,50){
%\qbezier(5,-5)(0,0)(5,5)
%}
%\put(50,50){
%\qbezier(5,-5)(10,0)(5,5)
%}
%% % % % % % % % % % % % % % % % % % % % %
%%\qbezier(35,16)(35,25)(50,25)
%%\qbezier(55,35)(65,25)(50,25)
%% % % % % % % % % % % % % % % % % % %
%\qbezier(35,16)(30,20)(45,35)
%\qbezier(65,16)(70,20)(55,35)
%\qbezier(35,4)(25,0)(25,30) \qbezier(25,30)(25,75)(45,65)
%\qbezier(65,4)(75,-5)(75,30) \qbezier(75,30)(75,75)(55,65)
% % % % % % % % % % % % % % % % % % %
\put(34,66.4){$\Lcap$}\put(57,66.4){$\Rcap$}
\put(46.5,17.2){$\Rcap$}\put(45,-2){$\Lcup$}
\put(28,-2){$\Rcup$}\put(65,-2){$\Rcup$}
\end{picture}
\caption{
Diagram of the figure-eight knot and the types of the crossings and turning points.
\label{fig:ei}}
\end{figure}

\paragraph{Diagram of the oriented knot.}
The knot can be represented by a knot diagram $\Dm$, which is a planar projection
\be
\Dm: \ \Km\ \to\ \mathbb{R}_2\label{kndiag}
\ee
that distinguishes over- and undercrossings of the segments %???
(see example in Fig.\ref{fig:ei}). The selected direction on the knot is preserved on the knot diagram.

As a result, $\Dm$ is a planar oriented graph with four-valent
vertices (self-crossings), each one having type $\Pcr$ or $\Ncr$ up to continuous
planar continuous transformations.

\paragraph{Extrema on the knot diagram.}
In addition, select a direction in the projection plane. Then the
special points of the knot projection include, apart from the
crossings, the turning points with respect to the direction. Each of the
turning-points is treated as a two-valent vertex on the knot diagram,
of type $\Lcup$,~$\Lcap$, $\Rcup$~or~$\Rcap$. %$\Lmin$, $\Rmin$, $\Lmax$, or $\Rmax$.

\begin{figure}
\unitlength=0.56mm

\begin{picture}(100,100)
\figpcr{60}{40}\figQpcrQ{40}{40}
\figncr{40}{10}\figQncrQ{60}{10}
% % % % % % % % % % % % % % %
\put(50,0){
\qbezier(-5,5)(0,0)(5,5)
}
\put(50,10){
\qbezier(-5,5)(0,10)(5,5)
}
%% % % % % % % % % % % % % % %
\put(50,30){
\qbezier(-5,5)(0,0)(5,5)
}
\put(50,40){
\qbezier(-5,5)(0,10)(5,5)
}
% % % % % % % % % % % % % % % % % % % %
%\put(40,50){
%\qbezier(5,-5)(0,0)(5,5)
%}
%\put(50,50){
%\qbezier(5,-5)(10,0)(5,5)
%}
%% % % % % % % % % % % % % % % % % % % % %
\qbezier(35,16)(35,25)(50,25)
\qbezier(65,35)(75,25)(50,25)
%% % % % % % % % % % % % % % % % % % %
%\qbezier(35,16)(30,20)(45,35)
%\qbezier(65,16)(70,20)(55,35)
\qbezier(35,4)(25,0)(25,15)\qbezier(25,15)(25,25)(35,35)
\qbezier(65,4)(75,0)(80,15)\qbezier(35,60)(28,50)(35,45)\qbezier(80,15)(90,85)(35,60)
\qbezier(65,45)(75,55)(75,40)\qbezier(75,40)(75,28)(65,15)
% % % % % % % % % % % % % % % % % % % % %
\put(38,61){\qbezier(0,0)(0,2)(0,4)\qbezier(0,0)(2,0)(4,0)}
% % % % % % % % % % % % % % % % % % % % % %
\put(60.5,66){\circle*{2}}
\put(72,49){\circle*{2}}\put(50,47){\circle*{2}}
\put(50.5,32){\circle*{2}}
\put(50,17.5){\circle*{2}}
\put(32,2.5){\circle*{2}}\put(50,2.4){\circle*{2}}\put(70,2.5){\circle*{2}}
% % % % % % % % % % % % % % % % % % % % % % % % % %
\put(27,-2.5){$i$}\put(35,-2){$i^{\prime}$}
\put(41,-2.5){$j^{\prime}$}\put(52,-2.5){$j$}
\put(62,-2.5){$m$}\put(73,-2.5){$m^{\prime}$}
% % % % % % % % % % % % % % % % % % % % % % % % % % %
\put(30,16.5){$k^{\prime}$}\put(70,16.5){$p^{\prime}$}
\put(45,16.5){$l$}\put(54,16.5){$l^{\prime}$}
% % % % % % % % % % % % % % % % % % % % % % % % % % %
\put(35,28){$i$}\put(60,28){$k^{\prime}$}
\put(45,28){$u$}\put(52,28){$u^{\prime}$}
% % % % % % % % % % % % % % % % % % % % % % % % % % % %
\put(33.5,48){$m^{\prime\prime}$}
\put(45,48){$v^{\prime}$}\put(52,48){$v$}
\put(64,48){$p$}\put(69,38){$p^{\prime}$}
% % % % % % % % % % % % % % % % % % % % % % % % % % % % %
\put(49,59){$m^{\prime\prime}$}\put(61,59){$m^{\prime}$}
\end{picture}
\begin{picture}(100,75)
{\unitlength=0.3mm
\figpcr{120}{80}\figQpcrQ{80}{80}
\figncr{80}{20}\figQncrQ{120}{20}
}
% % % % % % % % % % % % % % %
{\unitlength=0.6mm
\put(50,0){
\qbezier(-5,5)(0,0)(5,5)
}
\put(50,10){
\qbezier(-5,5)(0,10)(5,5)
}
%% % % % % % % % % % % % % % %
\put(50,30){
\qbezier(-5,5)(0,0)(5,5)
}
\put(50,40){
\qbezier(-5,5)(0,10)(5,5)
}
% % % % % % % % % % % % % % % % % % % %
%\put(40,50){
%\qbezier(5,-5)(0,0)(5,5)
%}
%\put(50,50){
%\qbezier(5,-5)(10,0)(5,5)
%}
%% % % % % % % % % % % % % % % % % % % % %
\qbezier(35,16)(35,25)(50,25)
\qbezier(65,35)(75,25)(50,25)
%% % % % % % % % % % % % % % % % % % %
%\qbezier(35,16)(30,20)(45,35)
%\qbezier(65,16)(70,20)(55,35)
\qbezier(35,4)(25,0)(25,15)\qbezier(25,15)(25,25)(35,35)
\qbezier(65,4)(75,0)(80,15)\qbezier(35,60)(28,50)(35,45)\qbezier(80,15)(90,85)(35,60)
\qbezier(65,45)(75,55)(75,40)\qbezier(75,40)(75,28)(65,15)
% % % % % % % % % % % % % % % % % % % % %
\put(50,68){$\Lcap$}
\put(62,49){$\Lcap$}\put(47,47){$\Rcap$}
\put(48,28){$\Rcup$}
\put(42,17.5){$\Lcap$}
\put(27,-2){$\Rcup$}\put(42,-2){$\Lcup$}\put(66,-2){$\Rcup$}
}
\end{picture}
\begin{picture}(100,60)
%{\unitlength=0.3mm
\put(60,38){+}\put(40,38){+}
\put(40,7){---}\put(60,7){---}
%}
% % % % % % % % % % % % % % %
%{\unitlength=0.6mm
%\put(50,0){
%\qbezier(-5,5)(0,0)(5,5)
%}
%\put(50,10){
%\qbezier(-5,5)(0,10)(5,5)
%}
%%% % % % % % % % % % % % % % %
%\put(50,30){
%\qbezier(-5,5)(0,0)(5,5)
%}
%\put(50,40){
%\qbezier(-5,5)(0,10)(5,5)
%}
%% % % % % % % % % % % % % % % % % % % %
%%\put(40,50){
%%\qbezier(5,-5)(0,0)(5,5)
%%}
%%\put(50,50){
%%\qbezier(5,-5)(10,0)(5,5)
%%}
%%% % % % % % % % % % % % % % % % % % % % %
%\qbezier(35,16)(35,25)(50,25)
%\qbezier(65,35)(75,25)(50,25)
%%% % % % % % % % % % % % % % % % % % %
%%\qbezier(35,16)(30,20)(45,35)
%%\qbezier(65,16)(70,20)(55,35)
%\qbezier(35,4)(25,0)(25,15)\qbezier(25,15)(25,25)(35,35)
%\qbezier(65,4)(75,0)(80,15)\qbezier(35,60)(28,50)(35,45)\qbezier(80,15)(90,85)(35,60)
%\qbezier(65,45)(75,55)(75,40)\qbezier(75,40)(75,28)(65,15)
% % % % % % % % % % % % % % % % % % % % %
\put(50,68){$\Lcap$}
\put(47,47){$\Rcap$}\put(68,47){$\Lcap$}
\put(48,30){$\Rcup$}
\put(45,16){$\Lcap$}
\put(27,-2){$\Rcup$}\put(47,-2){$\Lcup$}\put(68,-2){$\Rcup$}
%}
\end{picture}
\caption{Diagram that represents the same knot $4_1$ as Fig.\ref{fig:ei} and contains only two kinds of crossings.\label{fig:varei}}
\end{figure}
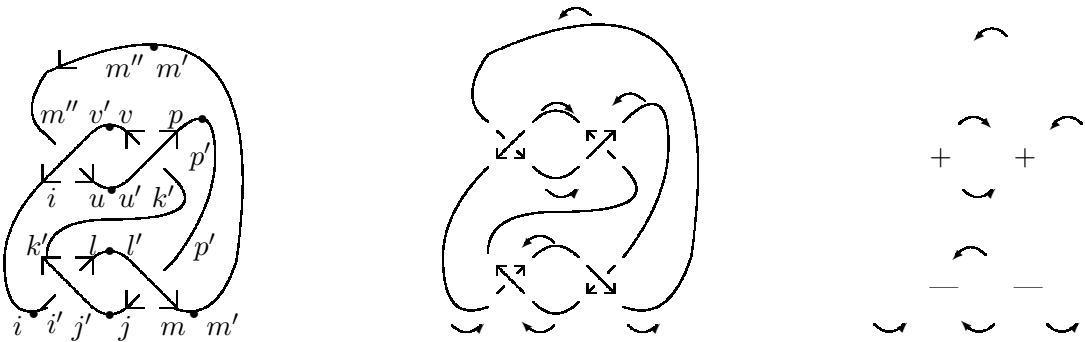

\paragraph{Representation space on the knot.}
The knot is associated with a representation space $V$ of a Lie algebra $\mathfrak{g}$,
\be
\Km\to Y\in\mathrm{rep}\, \mathfrak{g}\label{repknot}
\ee
i.e., one can imagine the linear space $V$ suspended over each point of the knot. Then one can associate each edge on the knot diagram with the space $V$.

\paragraph{Tensor representation for knot invariants.}
As a result, a two-valent vertex (a turning point) of the knot diagram can be related to a linear operator $\Qm:\ V\to V$, and a four-valent vertex (a crossing) can be related to a linear operator $\Rm:\ V\otimes V\to V\otimes V$.

\paragraph{The knot diagram as a diagram of the tensor contraction.}
The entire diagram $\Dm$ represents then the tensor contraction of these operators. To calculate this contraction explicitly, one should
\begin{enumerate}
\item{
Associate to each edge of $D$ an integer valued label (the number of the basis vector in the space $V$).}
\item{Multiply the components of $\Rm$ and $\Qm$ corresponding to all the vertices.}
\item{Take the sum over all values of the labels (the summation sign is usually omitted in the formulas).}
\end{enumerate}

For some specially chosen operators $\Rm$ and $\Qm$, the tensor
contraction related to the knot diagram turns out to be a topological
invariant. Discussing the corresponding constraints (briefly
summarised in App.~\ref{app:RQ}) on the operators and their general
solutions is beyond the scope of our survey. We just write down
and use below all the needed explicit expressions.

\subsection{Explicit expression for knot invariants\label{sec:RQpol}}
Here we present the key points of the $\Rm$-matrix approach to knot
invariants (proposed in~\cite{Tur, ReshTur}; see, e.g.\
textbook~\cite{KauffTB}, or either of the papers~\cite{MorSm, MMM2,
  Ano} for a detailed review).
\paragraph{Tensor contraction from knot diagram.}
The knot diagram must be drawn so that
\begin{itemize}
\item{both arrows in each crossing have the projections on the
    preferred direction of the same sign.}
\end{itemize}
In particular, the crossings never coincide with the turning
points. One can bring any knot diagram to the required form with the
help of continuous planar transformations~\cite{KauffTB}, e.g.\ the
knot represented by the diagram in Fig.~\ref{fig:ei} is represented by
the diagram in fig.~\ref{fig:varei} as well.

Then, the desired knot invariant is given by the
expression\footnote{We assume the summation over repeated indices.}
\be
\Hm=
\prod_{\lmin}\ \Qlcup^i_j\prod_{\lmax}\ \Qlcap^i_j\prod_{\rmin}\ \Qrcup^i_j\prod_{\rmax}\ \Qrcap^i_j
%\hspace{-1mm}
\prod_{\pcr}
\hspace{-2mm}
\Rp
{}^{ij}_{kl}
\prod_{\ncr}
\hspace{-2.5mm}
\Rn
{}^{ij}_{kl}
%\equiv
%\Zm^I_J\Qm^J_I\equiv\bar\Zm^J_I\,,
%\hspace{2cm}
%*\in
%%{\scriptstyle*}=
%\left\{{\scriptstyle+},{\scriptstyle-}\right\}
%%\left\{+,-\right\}
.\label{HRm}
\ee
E.g., knot diagram in Fig.~\ref{fig:varei} corresponds to the contraction
\be
H^{(\mathrm{fig.\,\ref{fig:varei}})}=
\Rn^{j^{\prime}i^{\prime}}_{k^{\prime}l}\Rn^{l^{\prime}p^{\prime}}_{mj}
\Rp^{v^{\prime}m^{\prime\prime}}_{iu}\Rp^{u^{\prime}k^{\prime}}_{pv}
\ \ \Qrcup^i_{i^{\prime}}\ \ \Qlcup^j_{j^{\prime}}\ \ \Qrcup^m_{m^{\prime}}
\ \ \Qrcap^l_{l^{\prime}}
\ \ \Qrcup^{u^{\prime}}_u
\ \ \Qlcap_v^{v^{\prime}}\ \ \Qrcap^p_{p^{\prime}}
\ \ \Qrcap^{m^{\prime}}_{{m^{\prime\prime}}}.
\label{Hei}\ee

\paragraph{Particular solution for crossing and turning point operators.}
Here we consider particular series of solutions for the crossing and
turning point operators. Namely, the solution labelled by a positive
integer $N$ corresponds to the Lie group $\mathfrak{g} = su_N$, $V$
being the space of the fundamental representation\footnote{The
  solution for any representation of a general Lie group is explicitly
  written down and discussed, e.g.\ in~\cite{KauffTB, MorSm,
    KlimSch}.}, and (in some specially chosen basis) all the non-zero
components of the operators have the explicit form\footnote{We have
  already fixed an ambiguity in the definition of the $\Qm$ operators
  (see App.~\ref{app:RQ}). Note that many sources,
  including~\cite{KauffTB, MorSm} fix it differently.}
\be
\begin{array}{cc}
\Pcr&\Ncr\\[2mm]
\left\{\begin{array}{rlcc}
\Rp^{ii}_{ii}&=&q,&1\le i\le 1,\\
\Rp^{ij}_{ji}&=&-1,&1\le i\ne j\le N,\\
\Rp^{ij}_{ij}&=&q-q^{-1},&1\le i<j\le N,
\end{array}\right.
&
\left\{\begin{array}{rlcc}
\Rn^{ii}_{ii}&=&q^{-1},&1\le i\le 1,\\
\Rn^{ij}_{ji}&=&-1,&1\le i\ne j\le N,\\
\Rn^{ij}_{ij}&=&q^{-1}-q,&1\le i<j\le N,
\end{array}\right.
\end{array}
\label{Rm}
\ee
and
\be
\arraycolsep=8mm
\begin{array}{cccc}
\Rcap&\Rcup&\Lcap&\Lcup\\
\Qrcap^i_i=q^{N-2i+1},&\Qrcup^i_i=q^{2i-1-N},&\Qlcap^i_i=1,&\Qlcup^i_i=1,\\[2mm]
\multicolumn{4}{c}{
%&&&
1\le i\le N.}
\end{array}
\label{Qm}
\ee Note that $\Qlcap^i_j=\Qrcup^i_j=\delta^i_j$, hence all the
corresponding turning points can just be ignored in (\ref{HRm}), and
we do so henceforth.

\subsection{Resolution hypercube\label{sec:hyp}}
Now we turn to the first notion which is used both in
Khovanov~\cite{Khov} and KhR~\cite{KhR} constructions, as well as in
the alternative approaches under development~\cite{DM3, AM}.
\begin{figure}
\unitlength=0.4mm
\arraycolsep=0mm
$
\begin{array}{lccccccl}
\Vm_0= %\bigoplus
&&&
\multicolumn{2}{c}{
\unitlength=0.2mm
\begin{picture}(80,80)
\figei{0}{0}{\figwcr}{\figwcr}{\figwcr}{\figwcr}
\end{picture}
}&&&
\hspace{5mm}\boxed{1}\\
&&&
\multicolumn{2}{c}{
\hspace{4.5mm}
\begin{array}{c}
\Vm^{\circ\circ\circ\circ}\\
1\!\cdot\! 1\!\cdot\! 1\!\cdot\! 1 \\
=1
\\[-8mm]
\end{array}
}
\\\\
\begin{array}{c}
\\[2mm]
\hat d_0\\[-2mm]
\end{array}
\Arrow{0}{-1}{100}{2}{40}&&
\hspace{1.5cm}
\begin{array}{c}
\\[-1cm]
\mathfrak{M}^{\circ\circ\circ\circ}_{\circ\circ\circ\bullet}
\end{array}
\hspace{-1.5cm}
&
\multicolumn{2}{c}{
\Arrow{-2}{-1}{55}{-12}{15}
\Arrow{-1}{-1}{20}{0}{10}
\Arrow{1}{-1}{20}{15}{10}
\Arrow{2}{-1}{60}{24}{16}
}
\\
% % % % % % % % % % % % % % % % % % % % % % % % % %
&&
\unitlength=0.2mm
\begin{picture}(80,80)
\figei{0}{0}{\figwcr}{\figwcr}{\figwcr}{\figbcr}
\end{picture}
&
\unitlength=0.2mm
\begin{picture}(80,80)
\figei{0}{0}{\figwcr}{\figwcr}{\figbcr}{\figwcr}
\end{picture}
&
\unitlength=0.2mm
\begin{picture}(80,80)
\figei{0}{0}{\figwcr}{\figbcr}{\figwcr}{\figwcr}
\end{picture}
&
\unitlength=0.2mm
\begin{picture}(80,80)
\figei{0}{0}{\figbcr}{\figwcr}{\figwcr}{\figwcr}
\end{picture}\\
\Vm_1=\bigoplus&
&
\hspace{4.5mm}
\begin{array}{c}
\Vm^{\circ\circ\circ\bullet}\\
1\!\cdot\! 1\!\cdot\! 1\!\cdot\! q^{-1} \\
=q^{-1}
\end{array}
&
\hspace{4.5mm}
\begin{array}{c}
\Vm^{\circ\circ\bullet\circ}\\
1\!\cdot\! 1\!\cdot\! q^{-1}\!\cdot\! 1 \\
=q^{-1}
\end{array}
&
\hspace{4.5mm}
\begin{array}{c}
\Vm^{\circ\bullet\circ\circ}\\
1\!\cdot\! q\!\cdot\! 1\!\cdot\! 1\\
=q
\end{array}
&
\hspace{4.5mm}
\begin{array}{c}
\Vm^{\bullet\circ\circ\circ}\\
q\!\cdot\! 1\!\cdot\! 1\!\cdot\! 1 \\
=q
\end{array}
&&
\hspace{5mm}\boxed{T}
\\
\begin{array}{c}
\\[5mm]
\hat d_1\\[-5mm]
\end{array}
\Arrow{0}{-1}{75}{2}{20}
&&&
\Arrow{-3}{-1}{90}{-15}{20}
\Arrow{2}{-1}{50}{8}{12}
\Arrow{3}{-1}{90}{21}{18}
\\
% % % % % % % % % % % % % % % % % % % % % % % % % % %
&
\unitlength=0.2mm
\begin{picture}(80,80)
\figei{0}{0}{\figwcr}{\figwcr}{\figbcr}{\figbcr}
\end{picture}
&
\unitlength=0.2mm
\begin{picture}(80,80)
\figei{0}{0}{\figwcr}{\figbcr}{\figwcr}{\figbcr}
\end{picture}
&
\unitlength=0.2mm
\begin{picture}(80,80)
\figei{0}{0}{\figbcr}{\figwcr}{\figwcr}{\figbcr}
\end{picture}
&
\unitlength=0.2mm
\begin{picture}(80,80)
\figei{0}{0}{\figwcr}{\figbcr}{\figbcr}{\figwcr}
\end{picture}
&
\unitlength=0.2mm
\begin{picture}(80,80)
\figei{0}{0}{\figbcr}{\figwcr}{\figbcr}{\figwcr}
\end{picture}
&
\unitlength=0.2mm
\begin{picture}(80,80)
\figei{0}{0}{\figbcr}{\figbcr}{\figwcr}{\figwcr}
\end{picture}\\
\Vm_2=\bigoplus
&
\hspace{4.5mm}
\begin{array}{c}
\Vm^{\circ\circ\bullet\bullet}\\
1\!\cdot\! 1\!\cdot\! q^{-1}\!\cdot\! q^{-1} \\
=q^{-2}
\end{array}
&
\hspace{4.5mm}
\begin{array}{c}
\Vm^{\circ\bullet\circ\bullet}\\
1\!\cdot\! q\!\cdot\! 1\!\cdot\! q^{-1} \\
=1
\end{array}
&
\hspace{4.5mm}
\begin{array}{c}
\Vm^{\bullet\circ\circ\bullet}\\
q\!\cdot\! 1\!\cdot\! 1\!\cdot\! q^{-1} \\
=1
\end{array}
&
\hspace{4.5mm}
\begin{array}{c}
\Vm^{\circ\bullet\bullet\circ}\\
1\!\cdot\! q\!\cdot\! q^{-1}\!\cdot\! 1 \\
=1
\end{array}
&
\hspace{4.5mm}
\begin{array}{c}
\Vm^{\bullet\circ\bullet\circ}\\
q\!\cdot\! 1\!\cdot\! q^{-1}\!\cdot\! 1 \\
=1
\end{array}
&
\hspace{4.5mm}
\begin{array}{c}
\Vm^{\bullet\bullet\circ\circ}\\
q\!\cdot\! q\!\cdot\! 1\!\cdot\! 1 \\
=q^2
\end{array}
&\hspace{5mm}\boxed{T^2}\\
\begin{array}{c}
\\[5mm]
\hat d_2\\[-2mm]
\end{array}
\Arrow{0}{-1}{80}{2}{25}
&
\Arrow{3}{-1}{105}{15}{20}
&&
\Arrow{0}{-1}{27}{8}{15}
&&
\Arrow{-3}{-1}{100}{10}{18}
\\
% % % % % % % % % % % % % % % % % % % % % % % % % % % %
&&
\unitlength=0.2mm
\begin{picture}(80,80)
\figei{0}{0}{\figwcr}{\figbcr}{\figbcr}{\figbcr}
\end{picture}
&
\unitlength=0.2mm
\begin{picture}(80,80)
\figei{0}{0}{\figbcr}{\figwcr}{\figbcr}{\figbcr}
\end{picture}
&
\unitlength=0.2mm
\begin{picture}(80,80)
\figei{0}{0}{\figbcr}{\figbcr}{\figwcr}{\figbcr}
\end{picture}
&
\unitlength=0.2mm
\begin{picture}(80,80)
\figei{0}{0}{\figbcr}{\figbcr}{\figbcr}{\figwcr}
\end{picture}\\
\Vm_3=\bigoplus&&
\hspace{4.5mm}
\begin{array}{c}
\Vm^{\circ\bullet\bullet\bullet}\\
1\!\cdot\! q\!\cdot\! q^{-1}\!\cdot\! q^{-1} \\
=q^{-1}
\end{array}
&
\hspace{4.5mm}
\begin{array}{c}
\Vm^{\bullet\circ\bullet\bullet}\\
q\!\cdot\! 1\!\cdot\! q^{-1}\!\cdot\! q^{-1} \\
=q^{-1}
\end{array}
&
\hspace{4.5mm}
\begin{array}{c}
\Vm^{\bullet\bullet\circ\bullet}\\
q\!\cdot\! q\!\cdot\! 1\!\cdot\! q^{-1} \\
=q
\end{array}
&
\hspace{4.5mm}
\begin{array}{c}
\Vm^{\bullet\bullet\bullet\circ}\\
q\!\cdot\! q\!\cdot\! q^{-1}\!\cdot\! 1 \\
=q
\end{array}
&&\hspace{5mm}\boxed{T^3}\\
\begin{array}{c}
\\[5mm]
\hat d_3\\[-5mm]
\end{array}
\Arrow{0}{-1}{80}{2}{20}
&&
\Arrow{3}{-1}{75}{5}{10}
&
\Arrow{1}{-1}{20}{5}{10}
&
\Arrow{-1}{-1}{20}{5}{10}
&
\Arrow{-2}{-1}{64}{5}{15}\\
% % % % % % % % % % % % % % % % % % % % % % % % % %
&&&
\multicolumn{2}{c}{
\unitlength=0.2mm
\begin{picture}(80,80)
\figei{0}{0}{\figbcr}{\figbcr}{\figbcr}{\figbcr}
\end{picture}
}\\
\Vm_4= %\bigoplus
&&&
\multicolumn{2}{c}{
\hspace{4.5mm}
\begin{array}{c}
\Vm^{\bullet\bullet\bullet\bullet}\\
q\!\cdot\! q\!\cdot\! q^{-1}\!\cdot\! q^{-1} \\
=1
\end{array}
}&&&\hspace{5mm}\boxed{T^4}
\end{array}
$
\caption{The resolution hypercube for the figure-eight knot
  (Fig.~\ref{fig:varei}). \label{fig:eihyp}}
\end{figure}
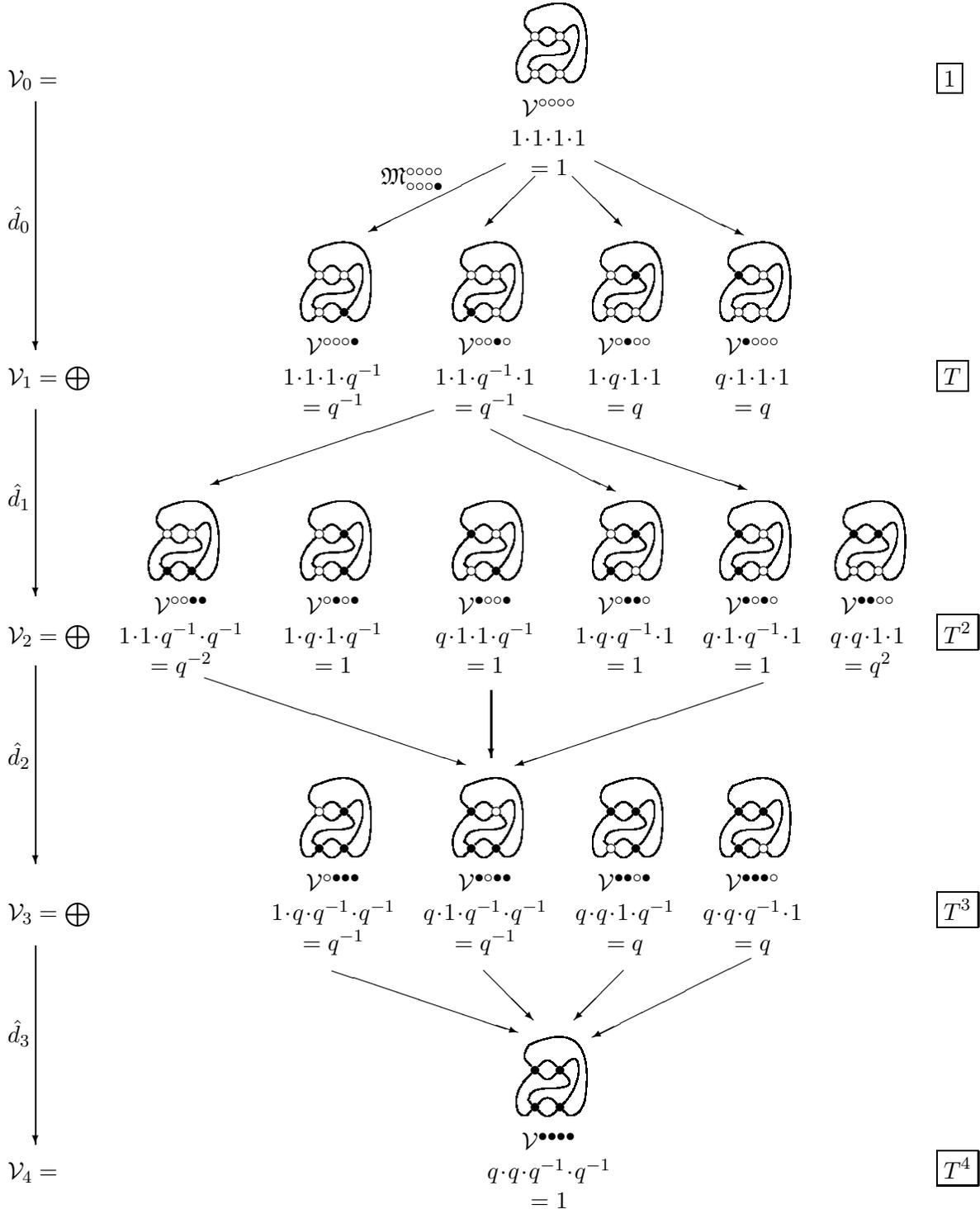

\paragraph{Decomposition for crossing operators.}
The operators (\ref{Rm}) can be identically rewritten as linear
combinations of the identity operator and a projector (an operator
$\PV$ such that $\PV\PV=\PV$), namely,
\be
\arraycolsep=1mm
\begin{array}{rcrclcrcl}
\Rp&=&q\Rw&-&\Rb&=&q\IdId&-&[2]_q\PV,\\
\Rn&=&q^{-1}\Rw&-&\Rb&=&q^{-1}\IdId&-&[2]_q\PV.
\end{array}\label{Rdec}
\ee
%for the crossings of types $\Pcr$ and $\Ncr$, respectively.
Formulas~(\ref{Rdec}) do not follow from the general group theoretical
and topological constraints. They are the special properties of the
simplest solution to the constraints, which satisfies characteristic
equations~(\ref{Rev}) (see App.~\ref{app:skein}).

%\be
%\Rp\Rn=\Rn\Rp=\IdId,\hspace{1cm} \PV\PV=\PV.\nn\\
%\ee

%In the following construction, the summands in (\ref{Rdec}) are considered as operators,

\paragraph{Hypercube representation for the knot invariant.}
If one substitutes the decomposition~(\ref{Rdec}) of the crossing
operators in~(\ref{HRm}) and expands the product, the terms of the
expansion are enumerated by various \textit{colourings}
$*\in\left\{\circ, \bullet\right\}^n$ of the knot diagram, i.e., by
the partitions of the $\circ$ and $\bullet$ signs (associated with two
summands in~(\ref{Rdec})) over the $n$ crossings.

One can treat the types of crossings as the coordinates, $\circ=0$, $\bullet=1$. The expansion term associated to the colouring $*$ thus corresponds to the point of $n$-dimensional space with coordinates $*(\circ=0,\bullet=1)$, so that all the $2^n$ points %can then be placed in a vertex of 
form an $n$-dimensional hypercube. %??? clarify $2^n$ colourings ???. 
In turn, an edge (assumed to be directed) of the hypercube connects
a pair of colourings %is connected by an arrow corresponding 
related by %to
a flip $\circ\to\bullet$. A vertex
$*\in\mathrm{perm}\{\underbrace{\circ \ldots \circ}_{\nu_*} ,
\underbrace{\bullet \ldots \bullet}_{\bar \nu_*} \}$ is then connected by
$\nu_*$ incoming and $\bar \nu_*=n-\nu_{*}$ outgoing arrows to the
total of $n$ vertices from the sets
$\left\{*^{\prime}\right\}=\mathrm{perm}\{\underbrace{\circ \ldots
  \circ}_{\nu_*+1} , \underbrace{\bullet \ldots \bullet}_{\bar
  \nu_*-1}\}$ and
$\left\{*^{\prime\prime}\right\}=\mathrm{perm}\{\underbrace{\circ
  \ldots \circ}_{\nu_*-1} , \underbrace{\bullet \ldots \bullet}_{\bar
  \nu_*+1}\}$, respectively.
  %??? 

As a result,
\begin{itemize}
\item{ Expression (\ref{HRm}) for the knot invariant is expanded as a
    sum over the vertices of the directed graph, called a
    \textit{resolution hypercube}\footnote{In the graphical
      representation for the tensor contractions, the expansion terms
      are associated with the various \textit{resolutions} of the knot
      diagram.}.}
\end{itemize}
For example, Fig.~\ref{fig:eihyp} illustrates the hypercube for the
diagram in Fig.~\ref{fig:varei}.

\paragraph{The generating function for the colourings}  is defined as a
formal series in the new variable $T$ as follows
\be
\Zfr(q,T)\equiv\Zm\big(\Rm(q)\rightarrow\Rfr(q,T)\big)=
\sum_{*\in\big\{\circ,\bullet\big\}^n}\hspace{-3mm}T^{\nu_*}\Zs,\hspace{1cm}
\bar\nu_*={\scriptstyle\#}\left\{\bullet\, {\scriptstyle\in}\, *\right\},
\label{Zfr}
\ee
where $\Zs$ stands for the expansion term related to the colouring $*$.

Expansion~(\ref{Zfr}) is formally obtained from expression~(\ref{HRm})
for the knot invariant by substituting the crossing operators with \be
\arraycolsep=1mm
\begin{array}{rcrclcrcl}
\Rfrp(q,T)&\equiv&q\Rw(q)&+&T\Rb(q)& %=&q\IdId&+T&[2]_q\PV,
\\
\Rfrn(q,T)&\equiv&q^{-1}\Rw(q)&+&T\Rb(q)&
%=&q^{-1}\IdId&+T&[2]_q\PV
,
\end{array}\label{Rfrdec}
\ee
or, equivalently, their matrix elements with
\be
\begin{array}{cc}
\left\{\begin{array}{rlcc}
\Rm^{ii}_{ii}&\mapsto&q,&1\le i\le 1,\\
\Rm^{ij}_{ji}&\mapsto&T,&1\le i\ne j\le N,\\
\Rm^{ij}_{ij}&\mapsto&q+q^{-1}T,&1\le i<j\le N,\\
\Rm^{ji}_{ji}&\mapsto&q(1+T),&1\le i<j\le N,
\end{array}\right.
&
\left\{\begin{array}{rlcc}
\tRm^{ii}_{ii}&\mapsto&q^{-1},&1\le i\le 1,\\
\tRm^{ij}_{ji}&\mapsto&T,&1\le i\ne j\le N,\\
\tRm^{ij}_{ij}&\mapsto&q^{-1}+qT,&1\le i<j\le N,\\
\tRm^{ji}_{ji}&\mapsto&q^{-1}(1+T),&1\le i<j\le N.
\end{array}\right.
\end{array}
\label{TRm}
\ee

\subsection{Vector spaces at the hypercube vertices\label{sec:hypsp}}
The next step introduces representation theory data into the
construction (see, e.g.,~\cite{KlimSch} for the necessary
background). The following presentation is equivalent to the standard
presentation of the Khovanov construction~\cite{Khov} for the
particular case of the $\mathfrak{su}_2$ algebra (details can be found
in~\cite{KauffTB}). It is very plausible that the KhR
construction~\cite{KhR} in the more general case of $\mathfrak{su}_N$
is also reproduced, although in somewhat different notions%???
\footnote{Two resolutions of a crossing in~\cite{KhR} are
  identified with two terms in the decomposition~(\ref{Rdec}) for the
  $\mathfrak{su}_N$ $\Rm$-matrix in the fundamental
  representation~\cite{KlimSch}. Then the spaces in the vertices of
  the resolution hypercube (and hence the spaces in the resulting
  complex), are described in~\cite{KhR} as the complexes of certain
  commutative polynomial rings. At the same time, these complexes
  describe $\mathfrak{su}_N$ representation spaces in terms of the BGG
  resolvents~\cite{PrSig}. The author is indebted to I.~Danilenko for
  pointing out this observation and providing the reference.
%???Dan
\label{foot:Dan}}.

\paragraph{The operators related to a colouring.}
The $\Zbs$ contribution to~(\ref{HRm}), related to the colouring $*$
(with $\nu_*$ resolutions $\circ$, $\bar\nu_*$ resolutions $\bullet$,
$n=\nu_*+\bar\nu_-$ of them in total), identically equals to the
contraction of two operators (we omit the operators
$\Qlcap^i_j=\Qlcup^i_j=\delta^i_j$).
The first of these operators is given by
\be
\Qm^{\circ
  m}\big|^I_J=\prod_{\rmin}\,\Qrcup^i_j\prod_{\rmax}\,\Qrcap^i_j\label{mQm}
\ee
where $m$ is the number of the ``left-to-right'' turning points on the
knot diagram\footnote{Then the knot diagram has $2m$ turning points in total. Indeed,
  the numbers of turning points of various kinds on a closed curve
  satisfy $ \left\{\begin{array}{rcl}
      \#\left\{{\scriptstyle\Lcap}\right\}+\#\left\{\Rcap\right\}&=&\#\left\{\Lcup\right\}+\#\left\{\Rcup\right\}\\
      \#\left\{\Lcap\right\}+\#\left\{\Lcup\right\}&=&\#\left\{\Rcap\right\}+\#\left\{\Rcup\right\}
\end{array}\right.
$,
and the sum (resp.\ difference) %??? 
of these equations yields
$\#\Lcap=\#\Rcup$ (resp.\ $\#\Lcap = \#\Rcup$).\label{foot:half}} (those are
the only turning points one takes into account). The second operator is
\be
\Zs^I_J\equiv
q^{\nu_+-\nu_-}
\prod_{\wcr}
\hspace{-2mm}
\Rw
{}^{ij}_{kl}
\prod_{\bcr}
\hspace{-2.5mm}
\Rb
{}^{ij}_{kl}
\ , \
\hspace{1cm}
*\in\big\{\circ,\bullet\big\}^n,\label{Zcol}
\ee
where we introduce the number $\nu_+$ (resp.\ $\nu_-$) of $\Pcr$
(resp.\ $\Ncr$) crossings resolved into $\Rw$
\be
%\hspace{5mm}
\nu_+={\scriptstyle\#}\Big\{
\Pcr\rightarrow\Wcr
\, {\scriptstyle\in}\, \Dm^*\Big\},\ \
\nu_-={\scriptstyle\#}\Big\{
{\unitlength=0.3mm\Ncr\rightarrow\Bcr}
\, {\scriptstyle\in}\, \Dm^*\Big\}.
\ee
The power of $q$ factor appears as a price for expressing both $\Rp$ and $\Rn$ through the same pair $\Rw$ and $\Rb$.
The multi-indices $I$ (resp.\ $J$) are defined as the sets of the labels
attached to the edges coming in (resp.\ going out of) the turning points, i.e.,
\be
I=\Big\{
%\!\bigcup_{\lmin}\!i\!\bigcup_{\lmax}\!i
\!\bigcup_{\rmin}\!\bigcup_{\rmax}\!i
\Big\},\ \
J=\Big\{
%\!\bigcup_{\lmin}\!j\!\bigcup_{\lmax}\!j
\bigcup_{\rmin}\!\bigcup_{\rmax}\!j
\Big\}.
\label{Zind}\ee
Hence, both operators $\Qm^{\circ m}$ and $\Zs$ are defined on the
tensor power $V^{\circ m}$ of the original space $V$, $m$ being the
number of the ``left-to-right'' turning points in the knot diagram. In
the following we also consider the composition of these two operators
\be
\Zbs\equiv\Qm\Zm.\label{Zbcol}
\ee

Contraction~(\ref{HRm}) can then be  equivalently presented as a trace
\be
\Hm=\Tr_{V^{\circ m}} \Qm\Zm=\Tr_{V^{\circ m}}Q^{\circ m}\Zm=\Tr_{V^{\circ m}}\bar{\Zm}.
\label{HTr}
\ee

\subsection{Spaces at the vertices as graded spaces\label{sec:grad}}
The Khovanov~\cite{Khov} and the KhR~\cite{KhR} constructions both
essentially use one more notion related to representation theory of
quantum groups~\cite{KlimSch}. Namely, a special \textit{graded}
basis is considered in each vertex of the hypercube. Below we give the
$\Rm$-matrix description of these spaces implicitly used for explicit
computations in~\cite{AM}.
\paragraph{Spaces at the vertices as image spaces.}
One can show that\footnote{\label{foot:imsp}
The operators $\Rs$ 
  ($*\in\big\{\circ,\bullet\big\}$) defined in (\ref{Rdec}), satisfy
  $\Rs^2\sim\Rs$, so that, if $y\in\Im\Rs$, then $y=\Rs^2x=\Rs x\in
  \Rs(V)$ and hence the same is true for $\Zs$, which by definition
  (\ref{Zcol}) is a tensor product of the operators $\Rs$. The same is
  also true for the operator $\Zbs=\Qm \Zs$, because $\Qm\Zs=\Zs\Qm$
  for any $\Zs$ (due to the ``built in'' representation theory
  properties of these operators~\cite{KlimSch}) so that
  $$
  x=\Qm\Zs
  \left(\Qm\Zs y\right)=\Qm^2\Zs^2 y\Rightarrow x=\Qm^2\Zs
  z=\Qm\Zs\left(\Qm y\right),
  $$
  and
  $$\Zbs\left(\Vm\right)\equiv
  \Qm\Zs\left(\Vm\right)\subset
  \Zbs^{-1}\Vm\subset\Vm\Rightarrow\Vm\subset\Zbs\Vm
  $$
  The relations above imply that the operator $\Zbs$ is
  invertible. Hence, $\Zbs\left(\Vm\right)=\Vm$, as claimed.}  \be
%\boxed{
\Zbs\big(\Zbs\left(\Vm\right)\big)= \Zbs\left(\Vm\right)\subset V^{\Dm}, %.
\ee
i.e., one can restrict the operator $\Zbs$, originally defined on the
space $V^{\circ m}$, to its image where it acts as an isomorphism:
\be
\boxed{\Zbs:\ \Vm^*\equiv \left.\Im\Zbs\right|_{V^{\circ m}} \to \Vm^*}\label{Zop}
\ee
Henceforth, we
\begin{itemize}
\item{
relate a colouring $*$ to the space $\Vm^*$ defined by~(\ref{Zop}), with $\Zbs$ defined in~(\ref{Zcol},~\ref{Zbcol}) and put this space in the corresponding hypercube vertex.
}
\end{itemize}

\paragraph{Basis of eigenvectors as a graded basis.}
The operator $\Zbs$ has a basis of eigenvectors in the subspace $\Vm^*$, the eigenvalues being $q$ powers,\footnote{%???Dan
Similarly to the previous property, these facts follow from the
definitions~(\ref{Zcol}),~(\ref{Zbcol}) and the commutation relation $\Qm\Zs=\Zs \Qm$.}
\be
\left\{\Xm^*_{\alpha}:\ \Zbs\Xm^*_{\alpha}=q^{\Delta_{\alpha}}\Xm^*_{\alpha}\right\}_{I=1}^{\dim{\Vm^*}}.\label{Grbas}
\ee
Hence, one can consider the linear space
\be
v^*\equiv\mathrm{span}\left\{q^{\Delta_{\alpha}}\right\}.
\ee
The contraction
\be
\Pfr^*(q)\equiv\Tr_{\Vm^*}\Zbs  \label{Pfrcol}
\ee
is then the generating function for the basis of monomials in $q$.
The  two-variable generating function,
\be
\Pfr(q,T)\equiv \Hm\big(\Rm\to\Rfr\big)=\sum_*T^{\nu_*}\Pfr^*(q).\label{Pfrgen}
\ee
then coincides with the \textit{primary polynomial} introduced in~\cite{AM}.
\begin{itemize}
\item{ Relations (\ref{Pfrcol},~\ref{Pfrgen}) associate each term in
    the primary polynomial with a vector of the linear space placed at
    one of the resolution hypercube vertices.  }
\end{itemize}

\subsection{From the resolution hypercube to the complex\label{sec:cmp}}
The last step of the Khovanov~\cite{Khov}/KhR~\cite{KhR} construction
consists of relating the resolution hypercube arrows to certain maps
(the \textit{morphisms}). These maps must respect the graded bases
associated with the hypercube vertices, and this severely constraints
the form of the morphisms~\footnote{The morphisms are described
  in~\cite{KhR} as ring homomorphisms.%???Dan
}.

The resolution hypercube is then transformed into the Khovanov/KhR
complex in a canonical way~\cite{Khov, KhR}.

\paragraph{Maps associated with edges as maps of the graded spaces.}

An edge of the hypercube is by definition oriented from a colouring $*$
to a colouring $*^{\prime}$ obtained by the change $\circ\to\bullet$ at
only one vertex of the knot diagram. The edge corresponds to a \textit{morphism}, i.e., to a linear map
\be
\mathfrak{M}^*_{*^{\prime}}:\Vm^{*}\to\Vm^{*^{\prime}}.\label{Mfr}
\ee
Here we do not determine this map explicitly\footnote{See \cite{KhR}~for the original definition, \cite{CarMuf, RasmKhR}~for adaptations to explicit computations, and~\cite{DM3} for an attempt
%???s, Mor
to give an alternative definition.}, but formulate an essential
constraint on which we heavily rely in what follows\footnote{Some further properties of the maps are discussed in sec.~\ref{sec:repth}}.
%???Dan
Namely,

\begin{itemize}
\item{The maps acting along the edges must commute with the maps $\Zbs$ of the spaces in the hypercube vertices,}
\end{itemize}
\be
\boxed{\mathfrak{M}^*_{*^{\prime}}\Zbs=\Zbps\mathfrak{M}^*_{*^{\prime}}}.\label{MfrZbs}
\ee
%}
%\end{itemze}
Relation~(\ref{MfrZbs}) implies that an eigenvector $\Xm\in\Vm^*$ of $\Zbs$, is mapped to an eigenvector $\Xm^{\prime}\in\Vm^{*^{\prime}}$ of $\Zbps$ with the same eigenvalue,
\be
\Zbs\Xm=q^{\Delta}\Xm\ \Rightarrow\
\Zbps\Xm^{\prime}=
\Zbps\mathfrak{M}^*_{*^{\prime}}\Xm=\mathfrak{M}^*_{*^{\prime}}\Zbs\Xm=
q^{\Delta}\mathfrak{M}^*_{*^{\prime}}\Xm
=q^{\Delta}\Xm^{\prime}.
\label{grad}
\ee
The same statement in terms of the gradings introduced above reads:
\begin{itemize}
\item{Morphism~(\ref{Mfr}) maps a graded vector $\Xm\in\Vm^*$ to the
    graded vector $\Xm^{\prime}\in\Vm^{*^{\prime}}$ with the same
    grading\footnote{In fact, only that a graded vector is mapped to a
      graded vector is essential, while the equality of the gradings
      ($\Delta=\Delta^{\prime}$) is related to the freedom in the
      definition of the morphisms (i.e., there are different
      possibilities that ultimately yield the same knot invariant). A
      more conventional choice (in particular, used in~\cite{Khov,
        BarNat, KhR, CarMuf, DM1, DM2, DM3, Dan}%%%??? Dan
      ) is such that the morphisms lower the grading by one:
      $\Delta^{\prime}=\Delta-1$.}.}
\end{itemize}

\paragraph{The differentials compatible with the grading .}
To obtain the complex, one should first construct a sequence of linear spaces that are direct sums of the spaces at separate hypercube vertices. Namely, the sum for the space $\Vm_k$ includes all colourings with $\nu_*=k$ instances of $\bullet$, 
\be
\Vm_k=\bigoplus_{\nu_*=k}\equiv\Vm^*,\hspace{1cm} k=\overline{0,n},\label{cmpsp}
\ee
so that the space can be associated to the hypercube section spanned
by the vertices with exactly $\nu_*=k$ of the coordinates being equal
to $0$.
The differential $\hat d_k$ is then associated to the pair of the successive sections with $k$ differing by $1$. This operator acts from the space $\Vm_{k-1}$ to the
space $\Vm_k$ as a sum of the corresponding morphisms (with proper %???
signs),
\be
%???Dan
\oplus^{\nu_*=k}_{\nu_{*^{\prime}}=k+1}
(-1)^{ %???
\mathrm{par}*+\mathrm{par} *^{\prime}}\Mfr^*_{*^{\prime}}
\equiv\hat d_k:
\ \Vm_k\ \to\ \Vm_{k+1},\ \ \ k=\overline{0,n-1},\label{diff}
\ee

where $\mathrm{par}*\in\{0,1\}$ is the number of steps from
the initial vertex $\{\circ^n\}$ to the vertex $*$ \textbf{modulo} $2$.
%
%stands for the parity of the corresponding colouring %???
(two colourings thus have distinct parities, whenever differing by the only permutation $\Bcr\leftrightarrow\Wcr$).
%; see, either of the cited papers
%(e.g.~\cite{Khov}) for the precise definition ???number of steps from
%a given hypercube vertex???).

The differentials defined by~(\ref{diff}) are by construction nilpotent,
\be
\hat d_k\hat d_{k-1}=0,\ \ k=\overline{1,n},\label{nild}
\ee
and commute with the colouring operators $\Zbs$ (because all the morphisms do),
\be
\Zm_{k+1}\hat d_k=\hat d_{k+1}\Zm_{k}\hspace{1cm}\mbox{with}\hspace{1cm}\Zm_k\equiv\bigoplus_{\nu_*=k}\Vm^*
\label{diffcomm}.
\ee
In other words,
\begin{itemize}
\item{
Starting from the resolution hypercube, the linear spaces at the
vertices and the morphisms along the directed edges, one constructs a
sequence of spaces associated with entire hyperplanes %(??? write more about geometric picture ???) 
and sequence of nilpotent
grading-preserving\footnote{See the above footnote in the definition
  of the morphisms.} linear maps, the \textit{differentials}. The
resulting sequence provides the \textit{complex}~\cite{GelMan} associated with the
knot diagram.
}
\end{itemize}

\subsection{Graded basis in the homologies\label{sec:grbas}}

In Eq.~\eqref{Grbas} we introduced graded bases in the hypercube
vertices. There is, however, a freedom in the definition of such a
basis: one can make a linear transformation mixing the vectors of
the same grading. Below we describe a special basis, which is the union of the graded bases in the image, co-image and the homology subspaces of the differentials in the complex. A more detailed derivation is given in~\ref{app:spbas}.
\

Nilpotency condition~(\ref{nilp}) implies that
\be
x\in\Im\hat d_{k}\subset \Vm_k\ \stackrel{\mathrm{def}}{\Leftrightarrow}\ x=\hat d_ky
\hspace{5mm}\Rightarrow\hspace{5mm}
x\in\ker\hat d_{k+1}\subset \Vm_{k+1}\ \stackrel{\mathrm{def}}{\Leftrightarrow}\ \hat d_{k+1}x=0
\ee
The converse is generally wrong. 
The homology space is by definition~\cite{GelMan} the
\textit{factor space}
\be
H_k\stackrel{\mathrm{def}}{=\!=}\ker\hat d_{k+1}/\hat d_k\subset \Vm_{k+1},
\hspace{5mm}\mbox{i.e.,}\hspace{5mm}
x\in H_k\ \Leftrightarrow\ \hat d_k x=0\ \&\  \hat d_ky\ne x,\ \forall y\in \Vm_k,
\label{Hdef}
\ee
i.e., %\footnote{ Rigorously speaking, 
each vector $x$ in the  homology %subspace 
  $H_k$ is a kernel vector defined up to an arbitrary vector
  from the image subspace, $x \sim x+\hat d_ky$, $\forall y\in V_k$. %} 
  Hence, one can identify the homology with any subspace  $H_k\in\ker \hat d_k$ of the kernel such that $\dim H_k=\dim\ker \hat d_k-\dim\Im \hat d_{k-1}$.  

%???two definitions of the factor space???
Just using the definitions we have the decomposition
\be
\Vm_k=\Im\hat d_k\oplus\coIm \hat d_{k+1}\oplus H_k,\label{Vkdec}
\ee
where $x\in\coIm\hat d\stackrel{\mathrm{def}}{\Leftrightarrow}\hat d x\ne 0$.

A non-trivial consequence of the property~(\ref{diffcomm}) of the
differentials is that
\begin{itemize}
\item{Each of the three spaces in the decomposition~(\ref{Vkdec}) contains a graded basis.}
\end{itemize}
Precisely, a basis in the space $\Vm_k$ can be composed of three
groups of vectors (see App.~\ref{app:spbas}),
\be
%\left\{x_{k,i}\right\}_{i=1}^{\dim\Vm_k}=
\begin{array}{lclcl}
\left\{y_{k,i}:\ \hat d_{k+1}y_{k,i}\ne 0\right\}_{i=1}^{\dim\Im\hat d_{k+1}}
&\cup&
\left\{z_{k,i}=\hat d_ky_{k-1,i}\right\}_{i=1}^{\dim\Im \hat d_k}
&\cup&
\Big\{h_{k,i}\Big\}_{i=1}^{\dim\Vm_k-\dim\Im \hat d_k-\dim\Im\hat d_{k+1}},\\[4mm]
\hat\Delta y_{k,i}=\Delta_{k,i}y_{k,i}
&&
\hat\Delta z_{k,i}=\Delta_{k-1,i}z_{k,i}
&&
\hat\Delta h_{i,k}=\Delta^{\prime}_{i,k}h_{i,k}.
\end{array}
\label{basdec}\ee

\subsection{The geometric meaning of the positive integer decomposition\label{sec:dec}}
Here we address the final issue of this section, namely the derivation
of the positive integer decomposition for the primary polynomial
from~\cite{AM}.

\

Calculating the trace~(\ref{HTr}) in the basis~(\ref{basdec}) yields the following decomposition of the generating function
\be
\Pfr(q,T)
=
\sum_{k=0}^nT^k\Tr_{\Vm_k} q^{\hat\Delta}
\equiv
\sum_{k=0}^nT^k\dim_q\Vm
=
\sum_{k=0}^nT^k\left(
\Tr_{\coIm \hat d_{k+1}} q^{\hat\Delta}+
\Tr_{\Im\hat d_k} q^{\hat\Delta}+
\Tr_{H_k} q^{\hat\Delta}
\right)\stackrel{(\ref{basdec})}{=\!=}\nn\\=
\sum_{k=0}^nT^k\left(
\sum_{i=1}^{\dim\Im\hat d_{k+1}} q^{\Delta_{k,i}}+
\sum_{i=1}^{\dim\Im\hat d_k} q^{\Delta_{k-1,i}}+
\sum_{i=1}^{\dim H_k} q^{\Delta^{\prime}_{k,i}}
\right)=\nn\\=
\sum_{k=0}^{n-1}\left(T^k+T^{k+1}\right)
\sum_{i=1}^{\dim\Im\hat d_{k+1}} q^{\Delta_{k,i}}+
\sum_{k=0}^nT^k\sum_{i=1}^{\dim H_k} q^{\Delta^{\prime}_{k,i}}=
\left(1+T\right)\Jm(q,T)+\Pm(q,T).
\label{Prdec}\ee
Note that the coefficient of each power $q^{\Delta}$, both in $\Jm(q,T)$ and $\Pm(q,T)$, as well as in $\Pfr(q,T)$, is nothing but the multiplicity of the basis vector from the corresponding subspace
($\oplus_{k=1}^n\Im\hat d_k$, $\oplus_{k=0}^nH_k$ or $\oplus_{k=0}^n\Vm_k$, respectively) with the grading $\Delta$. This leads to the essential property of the obtained decomposition,
\begin{itemize}
\item{All three quantities in decomposition~(\ref{Prdec}), the dividend $\Pfr(q,T)$, the quotient $\Jm(q,T)$ and the remainder $\Pm(q,T)$ are sums  of the $q$ and $T$ powers with \textit{positive integer} coefficients.}
\end{itemize}

\section{Minimal positive division instead of computing the homology\label{sec:div}}
In this section, we wish to discuss the positive integer division
approach as a shortcut in general homology computations, without
reference to the knot theory applications.

Through the section, the spaces $V$, the maps $\hat d$ and all the
related quantities are not associated anyhow with the particular
complexes arising from the knot theory. Here we do not consider graded linear spaces, since a graded complex (see sec.~\ref{sec:grad}) is always a direct sum of ordinary complexes. Also we do not consider any %no???Mir, Zen
additional structures on the linear spaces, such as the structure of
Lie algebra representation (this is the case of the next section, sec.~\ref{sec:repth}).

\subsection{General definitions\label{sec:cmpdef}}
\subsubsection{Complex, homologies and decomposition of the generating functions.~\label{sec:hom}}
Despite the fact that the most relevant definitions and theorems have
already been given above, in sec.~\ref{sec:gen}, we present them below
in a compressed form for convenience.

Generally, a complex~\cite{GelMan} is defined as a sequence of linear spaces and linear maps,
\be
0\stackrel{\hat d_0}{\longrightarrow}
V_0\stackrel{\hat d_1}{\longrightarrow}
V_1\stackrel{\hat d_2}{\longrightarrow}
V_2\stackrel{\hat d_3}{\longrightarrow}
\ldots
\stackrel{\hat d_n}{\longrightarrow}
V_n\stackrel{\hat d_{n+1}}{\longrightarrow}
0,
\label{gencmp}
\ee
such that any pair of the subsequent maps satisfies the \textit{nilpotency} condition:
\be
\hat d_2\hat d_1=\hat d_3\hat d_2=\ldots=\hat d_n\hat d_{n-1}=0.
\label{nilp}
\ee
Condition~(\ref{nilp}) in particular implies that $\Im\hat d_k\subset\ker\hat d_{k+1}$ ($k=\overline{0,n}$). 

The homology in term $k$ is the \textit{factor space}
\be
H_k\equiv \ker\hat d_{k+1}/\Im\hat d_k,
\label{Homdef}
\ee which can be understood as a
subspace %(???not subspace, but factor space???)
of kernel vectors, in which any two vectors differing by an image
vector %is by definition
are treated as the same homology element, 
\be
x\in H_k&\Rightarrow& \hat d_{k+1}x=0,\label{HomExpl}\\
x\sim y\in H_k&\Leftarrow&\hat x-y=\hat d_kz \
\left(\stackrel{(\ref{nilp})}{\Rightarrow} \hat d_{k+1} x=\hat d_{k+1}
  y=0\right).  \nn\ee The generating functions \be
\Fm\equiv\sum_{k=0}^n T^k\dim V_k,\ \ \Jm(T)\equiv\sum_{k=0}^n T^k\dim
\Im \hat d_k,\ \ \Pm\equiv\sum_{k=0}^n T^k\dim H_k
\label{genfunc}
\ee
in formal variable $T$ for the dimensions of the linear spaces, the image subspaces and the homologies
satisfy the relation (a particular case of~(\ref{Prdec}), when $q=1$ so that $\dim_q\to\dim$)
\be
\boxed{\Fm(T)=\Pm(T)+\Jm(T)(1+T).}
\label{gendec}
\ee
The function $\Pm(T)$ is referred to as the \textit{Poincare
  polynomial} of the complex; in particular, $\chi\equiv\Pm(T=-1)$ is
the \textit{Euler character} of the complex. In the following, we call~(\ref{gendec}) a \textit{positive integer decomposition} for the generating function, because
\begin{itemize}
\item{
All the coefficients of the $T$ powers in $\Fm(T)$, $\Pm(T)$ and $\Jm(T)$ are by definition positive integer numbers (the dimensions of the corresponding subspaces).
}
\end{itemize}
We study the following question:
\begin{itemize}
\item{
To what extent can one recover the remainder $\Pm(T)$ from the dividend $\Fm(T)$ \textbf{avoiding} the straightforward computing the homology?
}
\end{itemize}
The realization of the Strong expectation  formulated in sec.~\ref{sec:Khconj}
is associated with the progress along this way.

\

Treated as an isolated algebraic problem, the above question can
possess no satisfactory answer, not being even formulated
rigorously. Below we attempt to formulate the problem more precisely,
employing various additional considerations.

\subsubsection{Integer and polynomial division.\label{sec:poldiv}}
The polynomial division is a straightforward extension of the integer
division\footnote{It is in fact the \emph{multivariable} division,
  which is applied to the KhR calculus, but we do not discuss this
  generalization here.}. Both problems are formulated in the table
below (the given quantities are underlined, the quantities to be
determined are marked by the question sign).
\be
\arraycolsep=1mm
\begin{array}{|ccccccc|ccccccc|}
\hline
\multicolumn{7}{|p{6cm}|}{}&\multicolumn{7}{|p{6cm}|}{}\\[-4mm]
\multicolumn{7}{|p{6cm}|}{\hspace{2.2cm}Integer division}&
\multicolumn{7}{|p{7cm}|}{\hspace{1cm}Polynomial division (variable $T$)}\\
\hline
\multicolumn{7}{|p{6cm}|}{}&\multicolumn{7}{|p{6cm}|}{}\\[-4mm]
\multicolumn{7}{|c|}{m,k,q,p\in \mathbb{Z}}&
\multicolumn{7}{|c|}{\Fm(T),\, \Gm(T),\, \Jm(T),\, \Pm(T)\in \mathrm{Pol}(T)}\\
\hline&&&&&&&&&&&&&\\[-2mm]
m&=&k&\cdot &q&+&p&
\Fm(T)&=&\Gm(T)&\cdot &\Jm(T)&+&\Pm(T)\\
\mbox{\underline{dividend}}&&\mbox{\underline{divisor}}&&\mbox{quotient}&&\mbox{remainder}&
\mbox{\underline{dividend}}&&\mbox{\underline{divisor}}&&\mbox{quotient}&&\mbox{remainder}
\\[-1mm]
&&&&\mbox{\small{?}}&&\mbox{\small{?}}&&&&&\mbox{\small{?}}&&\mbox{\small{?}}\\
\hline
\multicolumn{5}{|c}{}&\multicolumn{2}{|c|}{}&\multicolumn{7}{c|}{}\\[-4.5mm]
\multicolumn{5}{|c}{\mbox{ambiguity}}
&
\multicolumn{2}{|c|}{\mbox{constraint}}&
\multicolumn{7}{c|}{\mbox{ambiguity}} %&
%\multicolumn{2}{|c|}{\mbox{constraint}}
\\
\hline
\multicolumn{5}{|c}{}&\multicolumn{2}{|c|}{}&\multicolumn{7}{c|}{}\\[-4mm]
\multicolumn{5}{|c}{
%\hspace{-12mm}
k\rightarrow q+u,\hspace{2mm} p\rightarrow p-ku,\hspace{2mm} u\in\mathbb{Z}
%\hspace{-12mm}
}
&
\multicolumn{2}{|c|}{0\le p<q}&
\multicolumn{7}{c|}{\Jm\rightarrow \Jm+f(T),\ \Pm\rightarrow \Pm-f(T)Q,\ f(T)\in\mathrm{Pol}(\mathbb{Z}|T)} %&
\\
\hline
\end{array}
\label{tab:div}
\ee
In both cases the equality in the third row does not define the
unknown quantity uniquely. Namely, the equality still holds if both the
quotient and the remainder are subjected to the transformation in the
last row. In the integer case, the ambiguity is conventionally fixed
by the constraint given in the same line. The polynomial case is
essentially distinct at this point. Namely, none of the formal ways to
select the unique remainder (now defined up to a polynomial instead of
just an integer) \textit{a priory} produces a quantity adequate for
our purposes.

%\subsubsection{Euclid algorithm in terms of complexes\label{sec:Eucl}}
%As a first illustration to the interplay between the division procedure and the homology computing, we present in %details the following textbook example \cite{???}.
%??? Mor

\subsubsection{Division as homology computation\label{sec:cmpdiv}}
Now we give an elementary illustration of the main idea we use, namely
of the correspondence between the homology calculus and the polynomial
division, with a freedom in definition of the remainder (see the last
line in Tab.~\ref{tab:div} and the comment in the end of
sec.~\ref{sec:poldiv}). Below we take a simple particular case of
division problem, and construct two different complexes with the
homologies producing two different remainders.

In the context of knot polynomials (see sec.~\ref{sec:Gal}), we in fact deal with the very special case of the right column of Tab.~\ref{tab:div}, when all the polynomials have only positive integer coefficients, and the divisor is the binomial $(1+T)$.
The general problem we consider is to present a given polynomial $\Fm(T)$ in the form
\be
&\boxed{\Fm(T)=(1+T)\Jm(T)+\Pm(T)}\label{divprob}\\&
\Fm(T)=\sum\limits_{k=0}^n a_kT^k,\ \
\Jm(T)=\sum\limits_{k=0}^m b_kT^k,\ \
\Pm(T)=\sum\limits_{k=0}^l c_kT^k,
\hspace{1cm}\{a_k,b_k,c_k\}\subset\mathbb{Z}_+.\nn
\ee
Large freedom in the definition of the remainder yet remains even in this particular case, as we will see in the examples below.

In the particular case described below the division problem is
straightforwardly related to the problem of computing the
homologies. Namely, extracting the $\sim (1+T)$ part is equivalent to
matching the monomial terms in the pairs $(x,Tx)$, each monomial
entering only one pair. Some monomials remain unpaired and represent
the homologies. E.g., $\Fm(T)=3+4T+T^2$ admits two matchings,
each with one unpaired monomial,
%??? Dan
\be
\begin{array}{c|c}
\multicolumn{2}{c}{\underline{3+4T+2T^2}=}\\[2mm]
\arraycolsep=0.5mm
\begin{array}{ccccccc}
&&3&+&3T\\
&&&+&T&+&T^2\\
&&\bullet&\Arrow{1}{0}{20}{-10}{3}&\bullet&+&\boxed{T^2}\\
&&\bullet&\Arrow{1}{0}{20}{-10}{3}&\bullet\\
&&\bullet&\Arrow{1}{0}{20}{-10}{3}&\bullet&\hat d_1\\
&&&\hat d_0&\bullet&\Arrow{1}{0}{20}{-10}{3}&\bullet\\
&&&&&&\bullet\\
\hline\\[-4mm]
(1+T)&\cdot&(3&+&T)&+&\boxed{T^2}
\end{array}&
%\hspace{3cm}
\arraycolsep=0.5mm
\begin{array}{ccccccc}
1&+\\
2&+&2T\\
\bullet&+&2T&+&2T^2\\
\bullet&\Arrow{1}{0}{20}{-10}{3}&\bullet\\
\bullet&\Arrow{1}{0}{20}{-10}{3}&\bullet&\hat d_1\\
&\hat d_0&\bullet&\Arrow{1}{0}{20}{-10}{3}&\bullet\\
&&\bullet&\Arrow{1}{0}{20}{-10}{3}&\bullet\\
\hline\\[-4mm]
\boxed{1}&+&(2&+&2T)&\cdot&(1+T)
\end{array}
\end{array}
\label{divcmp}
\ee In either diagram in~(\ref{divcmp}), a bullet is placed in the
column $(k+1)$ for each $T^k$ term in $\Fm$; the arrows label the
matching pairs. The positions of the arrows matching column $k$ with
column $(k+1)$ can then be represented by the matrix $\hat d_{k-1}$
such that $d_{k-1}^{ij}=1$ if bullet $i$ in column $k$ matches bullet
$j$ in column $(k+1)$, and $d_{k-1}^{ij}=0$ otherwise. Then, the two
particular diagrams in~(\ref{divcmp}) give rise to the matrices
\be
\hat d_1=\left(
\begin{array}{cccc}
0&0&0&1\\0&0&0&0
\end{array}
\right),\
\hat d_0=\left(
\begin{array}{ccc}
1&0&0\\0&1&0\\0&0&1\\0&0&0
\end{array}
\right),\
\mbox{and}
\
\hat d_1=\left(
\begin{array}{cccc}
0&0&1&0\\0&0&0&1
\end{array}
\right),\
\hat d_0=\left(
\begin{array}{ccc}
0&1&0\\0&0&1\\0&0&0\\0&0&0
\end{array}
\right).
\label{divdiff}
\ee In both cases $\hat d_1\hat d_0 = 0$, since each bullet enters no
more than one pair. Hence, if a column is associated with a linear
space (and bullet with a basis vector), the arrows define the
differentials with the matrices~(\ref{divdiff}). One thus obtains the
complex, with the homology spanned by the basis vectors corresponding
to the unpaired bullet (related to the remainder in the division
problem).

\subsection{``Ambigiuos'' and ``unambigiuos'' minimal remainders\label{sec:amb}}
As already discussed in sec.~\ref{sec:poldiv}, decomposition (\ref{divprob}) is not unique. The first and the most naive constraint we impose on the remainder is
\begin{itemize}
\item{The remainder $\Pm(T)$ contains the minimal possible number of different $T$ powers.}
\end{itemize}
Although this requirement defines the unique remainder only in
particular cases, it turns out to be very useful in application to
knot polynomials (see~\cite{AM} and sec.~\ref{sec:Gal}). Moreover, all
further more involved constraints will be developments of this elementary one.

Below we consider in details the simplest examples, when $\Fm(T)$
contains up to three subsequent $T$ powers\footnote{If the $T$ powers
  are not subsequent, i.e., $a_k=0$ for some $k$, then problem
  (\ref{divprob}) should be solved separately for the polynomials
  $\Fm^{\prime}(T)=\sum\limits_{k^{\prime}=0}^{k-1} a_{k^{\prime}}$
  and $\Fm^{\prime\prime}(T)=\sum\limits_{k+1}^n a_{k^{\prime}}$ (such
  that $\Fm(T)=\Fm^{\prime}(T)+\Fm^{\prime\prime}(T)$); this is a
  consequence of the positive coefficients requirement.}. The
ambiguity in the remainder depends on the coefficients in the
dividend. Explicit relations are provided in Tabs.\footnote{In the
  latter two cases we omit the inessential $T$ power factor in $\Fm$
  by setting the smallest $T$ power to zero.}
(\ref{Fm12}--\ref{Fm3})~(\ref{Fm12}),~(\ref{Fm3}). As one can see from
Tab.~(\ref{Fm3}), already the case of the three-term dividend appears
to be rather non-trivial.
\be
\begin{array}{|c|c|}
\hline \multicolumn{2}{|c|}{}\\[-4mm]
\multicolumn{2}{|c|}{\Fm(T)=aT^k}\\[0.2mm]
\hline\multicolumn{2}{|c|}{}\\[-4.2mm]
%\mbox{Case}&\
%\cline{2-2}&\\[-4.2mm]
%&aT^k\\
%\hline
%\multicolumn{2}{c|}{}\\[-4mm]
\mbox{Type}&\mbox{Unambigious}\\
\hline&\\[-4.2mm]
\Jm(T)&0\\
\hline&\\[-4.2mm]
\Pm(T)&a_0T^k\\
\hline
\end{array}\hspace{1cm}
%\ee\be
\begin{array}{|c|c|c|c|}
\hline \multicolumn{4}{|c|}{}\\[-4mm]
\multicolumn{4}{|c|}{\Fm(T)=a_0+a_1T, \hspace{5mm} \chi\equiv a_0-a_1}\\[0.2mm]
\hline
&\multicolumn{3}{c|}{}\\[-4mm]
\mbox{Type}&\multicolumn{3}{c|}{\mbox{Unambigious}}\\
\hline &&&\\[-4.2mm]
\mbox{Case}&\chi>0&\chi=0&\chi<0\\
\cline{2-4}&&&\\[-4.2mm]
&\chi+a_1(1+T)&a_0(1+T)&(-\chi)T+a_0(1+T)\\
\hline&&&\\[-4.2mm]
\Jm(T)&a_1&a_0=a_1&a_0\\
\hline&&&\\[-4.2mm]
\Pm(T)&\chi&0&(-\chi)T\\
\hline
\end{array}
\label{Fm12}\ee

\be
\arraycolsep=0.5mm
\begin{array}{|c|c|c|c||c|c|}
\hline \multicolumn{6}{|c|}{}\\[-4mm]
\multicolumn{6}{|c|}{\Fm(T)=a_0+a_1T+a_2T^2, \hspace{5mm} \chi\equiv a_0-a_1+a_2}\\[0.2mm]
\hline
&\multicolumn{3}{c||}{}&\multicolumn{2}{c|}{}\\[-4mm]
\mbox{Type}&
\multicolumn{3}{c||}{\mbox{Unambigious}}
&\multicolumn{2}{c|}{\mbox{Ambigious}}\\
\hline&&\multicolumn{4}{|c|}{}\\[-4.2mm]
\mbox{Case}&
\chi\le0&
%a_0+a_2=a_1&
\multicolumn{4}{|c|}{\chi>0}\\
\cline{3-6}&&&\\[-4.2mm]
&&a_0\le a_1<a_2&a_2\le a_1<a_0&a_1<a_0, a_1<a_2&a_0\le a_1, a_2\le a_1\\
%&(a_1-a_0-a_2)T+(1+T)(a_0+cT)&(1+T)(a_0+cT)&\left(p+(a_2-a_1+a_0-p)T^2\right)+(1+T)\big((a_0-p)+(a_1-a_0+p)T\big)\\
\hline&&&&&\\[-4.2mm]
\Jm(T)&a_0+a_2T&a_0+(a_1-a_0)T&(a_1-a_2)+a_2T&
(a_0-p)+(p-a_0+a_1)T&
\begin{array}{c}(a_1-a_2)+a_2T;\\ a_0+(a_1-a_0)T\end{array}\\
\hline&&&&&\\[-4.2mm]
\Pm(T)&(-\chi) T&\chi T^2&\chi
&p+(\chi-p)T^2&\chi;\\[0.5mm]
\cline{5-5}&&&&&\\[-4mm]
&&&&0<a_0-a_1\le p \le a_0 <\chi&\chi  T^2\\[0.5mm]
\hline&&&&&\\[-4.5mm]
\hline&&&&&\\[-4mm]
\mbox{E.g.}
&1+3T+T^2&1+T+2T^2&2+T+T^2&
3+2T+4T^2=&
3+4T+2T^2=\\
\cline{2-6}
&&&&&\\[-4mm]
&=T+&=T^2+&=1+&
\left(1+4T^2\right)+2(1+T)&
T^2+(1+T)(1+2T)\\
&(1+T)^2&(1+T)^2&(1+T)^2&
=\left(2+3T^2\right)+(1+T)^2&
=1+2(1+T)^2\\
\hline
\end{array}
\label{Fm3}\ee

\subsection{Properties of maps as ``selection rules''  for the ``minimal remainder''\label{sec:delems}}

In this section we recall the homology computation problem (formulated
in Sec.~\ref{sec:hom}) underlying the minimal positive division
problem we consider (formulated in Sec.~\ref{sec:poldiv}). Below we
explore, in the simplest cases, how the particular known properties of
the differentials may help to determine the ``minimal remainder'',
when it is ambiguous.

\subsubsection{Ranks of the differentials\label{sec:dranks}}
The maps $\hat d$ in complex~(\ref{gencmp}) can not be isomorphisms
for generic dimensions of the spaces $V$. E.g., in the case of a two-term complex,
\be
0\stackrel{\hat d_0}{\longrightarrow}
V_0\stackrel{\hat d_1}{\longrightarrow}
V_1\stackrel{\hat d_2}{\longrightarrow}
0,
\ee
the rank of the only non-zero map $\hat d_1$ must satisfy
%presented by a rectangular matrix $M_1$ of the dimension $\dim V_0$ by$\dim V_1$, so that
%\be
%\rank\hat d_1\equiv\rank M_1\le\min \left(\dim V_0,\dim V_1 \right),
%\label{ranks2}
%\ee
%or, in the invariant terms,
\be
\rank \hat d_1\equiv \dim \Im \hat d_1 \subseteq  V_1 \le \min(\dim V_0,\dim V_1).
\ee
Starting from the next case of a three-term complex,
\be
0\stackrel{\hat d_0}{\longrightarrow}
V_0\stackrel{\hat d_1}{\longrightarrow}
V_1\stackrel{\hat d_2}{\longrightarrow}
V_2\stackrel{\hat d_3}{\longrightarrow}
0,
\ee
the nilpotency condition also becomes non-trivial, producing additional constraints. In particular,
$\hat d_2\hat d_1=0$ implies that
\be
\Im\hat d_1\subset \ker \hat d_2
\Rightarrow%\nn\\
\rank \hat d_1\equiv\dim\Im\hat d_1\le
\dim \ker \hat d_2=
\dim V_1-\dim\Im\hat d_2\equiv \dim V_1-\rank \hat d_2,
%\ \ \Rightarrow\ \  \nn\\
%\boxed{\rank\hat d_1+\rank\hat d_2\le\dim V_1}
\ee
i.e.,
\be
\rank \hat d_1+\rank \hat d_2\le\dim V_1,
\label{rank3}
\ee
and also
\be
\rank\hat d_1\le\min \left(\dim V_0,\dim V_1 \right),\ \
\rank\hat d_2\le\min \left(\dim V_1,\dim V_2 \right),
\ee
similarly to the previous case.
%As a result,
%\be
%\ee
%For generic complex (\ref{gencmp}), one can derive the condition
%\footnote{}
%\cite{???}.
%\be
%???
%\hat d_3.
%\label{ranks}
%\ee
%Hence,

%The homology of the complex is defined just as a ``measure of the degeneracy''

The ranks of the differentials for various choices of minimal
remainders in the division problem~(\ref{Fm12}--\ref{Fm3}) are given
in Tables~(\ref{ranks2}),~(\ref{ranks3}) (we omit the trivial one-term case).
\be
\begin{array}{|c|c|c|c|}
\hline
\multicolumn{4}{|c|}{}\\[-4.2mm]
\multicolumn{4}{|c|}
{\dim V_0=v_0,\ \dim V_1=v_1,\hspace{8mm} \chi=v_0-v_1}\\
\hline
\mbox{Case}&\chi>0&\chi=0&\chi<0\\&&&\\[-4.2mm]
\hline&&&\\[-4mm]
\max\rank \hat d_1&v_0&v_0=v_1&v_1\\
\hline&&&\\[-4.2mm]
\min\dim H_0\subset V_0&0&0&-\chi\\[0.2mm]
\hline&&&\\[-4.2mm]
\min\dim H_1\subset V_1&\chi&0&0\\[0.2mm]
\hline
\end{array}
\label{ranks2}\ee

%\be
%\begin{array}{|c|c|c|c|}
%\end{array}
%\ee

\be
\arraycolsep=0.5mm
\begin{array}{|c|c|c|c||c|c|c|}
\hline
\multicolumn{7}{|c|}{}\\[-4.2mm]
\multicolumn{7}{|c|}
{\dim V_0=v_0,\ \dim V_1=v_1,\ \dim V_2=v_2, \hspace{5mm} \chi=v_0-v_1+v_2}\\
\hline&&\multicolumn{5}{c|}{}\\[-4mm]
\mbox{Case}&\chi\le 0&\multicolumn{5}{c|}{\chi>0}\\
&&\multicolumn{5}{c|}{}\\[-4.2mm]
\cline{3-7}&&&&&\multicolumn{2}{c|}{}\\[-4mm]
&&v_0\!\le\! v_1\!<\!v_2&v_2\!\le\! v_1\!<\!v_0
&v_1\!<\! v_0,\ v_1\!<\!v_2&
\multicolumn{2}{c|}{v_0\!\le\!v_1,\, v_2\!\le\!v_1}\\
\hline&&&&&&\\[-4mm]
%\max
\rank \hat d_1&v_0&v_0&v_1-v_2&v_0-p&v_1-v_2&v_0\\
\hline&&&\\[-4.2mm]
%\max
\rank \hat d_2&v_2&v_1-v_0&v_2&p-v_0+v_1&v_2&v_1-v_0\\
\hline&&&&&&\\[-4.8mm]
\hline&&&&&&\\[-4.2mm]
%\min
\dim H_0&0&0&\chi&0\!<\!v_0-v_1\!\le\!p\le v_0&\chi&0\\[0.2mm]
\hline&&&&&&\\[-4.2mm]
%\min
\dim H_1&-\chi&0&0&0&0&0\\[0.2mm]
\hline&&&&&&\\[-4.2mm]
%\min
\dim H_2&0&\chi&0&0\!<\!v_2-v_1\!\le\!\chi-p&\hspace{5mm}0\hspace{5mm}&\chi\\[0.2mm]
\hline
\end{array}
\label{ranks3}\ee
%
%\be
%\begin{array}{|c|c|c|c|c|c|}
%
%\end{array}
%\ee

\subsubsection{Particular values of matrix elements\label{sec:ranks}}
More detailed data include the values of the particular matrix
elements. To determine explicitly just several of them it is sufficient to determine the ranks of the differentials, and hence the ``minimal remainder''.
For example,
\be
%\forall k\hspace{5mm}
\rank\!\!
\left(
\begin{array}{cc}
k&k
\end{array}
\right)=
\left\{\begin{array}{rl}
0,&k=0\\
1,&k\ne 0
\end{array}
\right.;
\hspace{5mm}
%\forall k,p\hspace{2mm}
\rank\!\!
\left(
\begin{array}{ccc}
1&-1&k\\
-1&1&p
\end{array}
\right)=
\left\{\begin{array}{rl}
0,&k+p=0\\
1,&k+p\ne 0
\end{array}
\right.
\ee
However, the general case is different, e.g.,
\be
\begin{array}{ccc}
\forall k\hspace{2mm}
\rank\!\!
\left(
\begin{array}{cc}
1&k
\end{array}
\right)=1;
&%\hspace{5mm}
\forall k\hspace{2mm}
\rank\!\!
\left(
\begin{array}{ccc}
1&-1&k\\
-1&1&1-k
\end{array}
\right)=2;
& %\hspace{5mm}
\forall k\hspace{2mm}
\rank\!\!
\left(
\begin{array}{ccc}
1&-1&k\\
-1&1&-k
\end{array}
\right)=1;\\\\
&\forall k,p\hspace{2mm}
\rank\!\!
\left(
\begin{array}{ccc}
1&0&k\\
0&1&p
\end{array}
\right)=2.
\end{array}
\ee

\subsection{``Multilevel'' division as a way to fix ambiguities\label{sec:levdiv}}
A further ``improvement'' consists in artificially splitting the
initial division problem into two (or more) levels. Namely, suppose
the original polynomial is given in the form of the decomposition (the
summation index $Y$ running over a given set of values)
\be
\Fm(T)=\sum_Y\Fm_Y(T) C_Y(T).\label{Fdec}
\ee
Then, the division can be first performed for each $\Fm_Y(T)$ separately, and then for the linear combinations of the resulting remainders, namely
\be
\mbox{I.}&&\Fm_Y(T)=\Pm_Y(T)+(1+T)\Jm_Y(T).
\label{divlev1}\\
\mbox{II.}&&\Pm_{\one{-3mm}{-1.5mm}}(T)\equiv\sum C_Y(T)\Pm_Y(T)=
\Pm_{\two{-3mm}{-2mm}}(T)+(1+T)\Jm_{\two{-2mm}{-2mm}}(T).
\label{divlev2}
\ee
The simplest example below illustrates how such ``multilevel'' division may fix an ambiguous remainder.

Let $\Gm(T)=1+T$ and
\be
\begin{array}{|c||c|c|}
\hline\\[-4.5mm]
Y&1&2\\
\hline&&\\[-4.2mm]
\hline&&\\[-4mm]
C_Y&1&T^2\\
\hline&&\\[-4.5mm]
\Fm_Y&1+T&1\\
\hline
\end{array},
\ee
so that $\Fm(T)$ admits two positive integer decompositions, both with the monomial remainder,
\be
\Fm(T)=\Fm_1C_1+\Fm_2C_2=1+T+T^2=1+(1+T)\cdot T=T^2+1\cdot(1+T).
\ee
From the viewpoint of the multilevel division, one decomposition is preferred, namely,
\be
\mbox{I.}&&\Fm_1=0+1\cdot(1+T)\ \Rightarrow\ \Pm_1=0,\nn\\
&&\Fm_2=1+0\cdot(1+T)\ \Rightarrow\ \Pm_2=1.\nn\\
\mbox{II.}&&\Pm_{}=\Pm_1C_1+\Pm_2C_2=0\cdot 1+1\cdot T^2=T^2+0\cdot(1+T) %. \nn\\
%&&
\Rightarrow
\boxed{\Pm_{\two{-3mm}{-2mm}}=T^2}\, .
\ee

The suggested algorithm is motivated by the known properties of the
knot polynomials, see~\cite{AM} and sec.~\ref{sec:dexp}.

\section{Lie algebra structure in the complex and its breaking\label{sec:repth}}
In this section, we return to the particular complex that arises in
the context of knot invariants, associated with the resolution
hypercube (see sec.~\ref{sec:hyp}), and discuss in some further
details the spaces at the vertices and the morphsims along the edges.

Namely, we slightly touch the topic of representation theory
structures associated with the resolution hypercube, with the
constructed complex, and with the obtained knot invariants. Although
at the moment we are able neither to present a closed construction,
nor to extract any practical output, we see at least two reasons for
addressing this subject.

First, the description of the spaces and differentials in the original
KhR formalism~\cite{KhR, CarMuf} refers to the representation theory
(see ssec.~\ref{sec:hypsp}--\ref{sec:cmp}, especially footnote$^{\ref{foot:Dan}}$.).
In particular, it looks highly plausible that the representation
theory properties of the morphisms determine their particular
form to much extend. %???Dan

Second, the known properties of the resulting knot invariants, which
are the KhR invariants, as well as that of the closely related
superpolynomials~\cite{GSchV,DGN}, might have a natural representation
theory interpretation. Moreover, it seem that this structure becomes
more transparent, from the viewpoint of the differential expansion
technique~\cite{Art}, and the evolution method~\cite{MMM3} (see
sec.~\ref{sec:dexp}).

\

In this section, we use a number of representation theory notions and
statements without any comments, referring the reader
to~\cite{KlimSch}, whenever necessary.  The most necessary notions are
contained in the following

\paragraph{Quick reference. Standard basis in the Lie algebra and in the representation space.}
%For the sake of reference, now we briefly recall the standard description of the semi-simple Lie algebra and of its \textit{highest weight} representations (only such ones we consider here). For more details see, e.g.,~\cite{KlimSch}.

By definition, the Lie algebra $\mathfrak{g}$ is a linear space
(usually over complex numbers) with the additional operation (the
\textit{commutator}). I.e., any pair of elements $A_1$ and $A_2$ of
the algebra correspond an element $[A_1,A_2]$ of the algebra. Here we
consider only simple finite dimensional Lie algebras.

In this case all elements of an algebra $\mathfrak{g}$ can be obtained
from the %$3r$
$2r$ \textit{generators}, by taking successively the commutators and
linear combinations of the elements. The generators include lowering
operators $E_{\alpha}$ and the raising operators
$F_{\alpha}$, %and the Cartan operators $H_{\alpha}$
with $\alpha=\overline{1,r}$.

A representation of the algebra is a linear map of the algebra to the
space of linear operators consistent with the definition of the
commutator. These linear operators act in the \textit{representation
  space} $V$ of the algebra. Throughout the text we consider the
algebra together with its representation and do not make a difference
between the algebra elements and the corresponding linear operators.

%Let the spaces $\Vm_k$ in the complex be the spaces of 

We consider only the finite-dimensional representations of simple Lie
algebras. All these representations are \textit{highest weight}
representations. This means that the representation space contains the
special basis composed of the highest vector $X_{\emptyset}\equiv X$,
which is annihilated by all the raising operators ($F_{\alpha}X=0,\
\alpha=\overline{1,r}$) and the maximum linearly independent subset of
descendant vectors $X_{\{\alpha\}}=E_{\alpha_1}\ldots E_{\alpha_k}x$,
obtained from the successive action of the lowering operators on
$x$. The vectors $X_{\{\alpha\}}$ are the common eigenvectors of the
\textit{Cartan operators} $H_{\alpha}=[E_{\alpha},F_{\alpha}]$, i.e.,
$H_{\alpha}X_{\{\beta\}}=\Lambda_{\alpha\{\beta\}}X_{\{\beta\}}$ for
all possible $\alpha$ and $\{\beta\}$.

For further details see, e.g.,~\cite{KlimSch}.

\subsection{The idea}

As we discussed in sec.~\ref{sec:gen}, both the HOMFLY and the KhR
invariants can be considered as the generating functions for the
graded bases in the spaces in the Khovanov complex and in its
homologies, respectively. The HOMFLY invariant is explicitly expressed
through the $\Rm$-matrices, associated with certain Lie algebra
representations, and can be expanded over the characters of these
representations. This is not true for the KhR invariant. Yet, in
particular cases these invariants can be expressed via some remnants
of the representation theory structure that governs the HOMFLY
polynomial~\cite{DGN}, \cite{MMM3}, \cite{supsup}, \cite{Art},
\cite{MorKon2}, \cite{DM3}, \cite{AM}, \cite{AMev}. This phenomenon
finds a natural explanation, if the spaces $\Vm_k$ in the KhR complex
are representation spaces of the Lie algebra, while the homologies are
not. At the same time the latter ones can be invariant spaces of some
subalgebra. In turn, this happens if the morphisms in the complex
commute with this subalgebra, but not with the entire algebra.

\subsection{Spaces in the resolution hypercube as representation spaces\label{sec:repsp}}
In sec.~\ref{sec:gen} we describe the space $\Vm^*\subset V^{\circ m}$
at the hypercube vertex $*$ as a subspace of the tensor power of the
representation space $V$ initially associated with the knot (the power
$m$ is half the number of the turning points on the knot diagram, see
ssec.~\ref{sec:hypsp}--\ref{sec:grad} including the
footnote$^{\ref{foot:half}}$). Precisely, these spaces are defined as
invariant spaces of the grading operator $\Zbs$, which allows one to
introduce the graded basis $\Zbs$ of eigenvectors in each of these
spaces (see tab.~\ref{tab:sketch} for the basic formulas and for the
general plan of the construction). Each space $V_k$ in the complex is
in turn defined as a direct sum of several $\Vm^*$; hence, the $\Vm_k$
is an invariant space of $\Zbs$ and has a graded basis as well. In
this section, we consider an even stronger claim that
\begin{itemize}
\item{
The spaces %$\Vm^*$ defined in sec.~\ref{sec:repsp} 
in the complex constructed for a given knot are representation
spaces of the Lie algebra $\mathfrak{g}$ associated with the knot.
} 
\end{itemize}
Note that $\Vm^*$ is generally a \textbf{reducible} representation.

The above claim follows from the definition (\ref{mQm}--\ref{Zbcol})
of the operator $\Zbs$, which is a contraction of the operators $\Qm$
and $\Rm$, and from the representation theory properties of these
operators~\cite{KlimSch} (see also App.~\ref{app:RQ}). The derivation
is similar to the reasoning in footnote$^{\ref{foot:imsp}}$.

Since $\Qm$ and $\Rm$, and hence $\Zbs$ commute with all the elements of the algebra, i.e.,
\be
A\Zbs=\Zbs A,\ \ \forall A\in\mathfrak{g},\ \ *\in\big\{\circ,\bullet\big\}^n,\label{gcomm}
\ee
then 
%Relation~(\ref{gcomm}) implies that
\be
\forall x\in\Vm^*\ \Big(\!\Leftrightarrow\, x=\Zbs y\Big)\ \
\Rightarrow\ \ Ax=\Zbs(Ay)\in\Vm^*.
\ee
Hence, by definition, $\Vm^*$ is the representation space of $\mathfrak{g}$, and the same is true for the sums of $\Vm^*$ that yield $\Vm_k$.

\subsection{Are the homology spaces representation spaces?\label{sec:repdiff}}
If the original spaces $\Vm_k$ in the Khovanov complex are invariant spaces a linear operator $A$, then the homology spaces $H_k$ must as well be invariant space of $A$ \textit{if} the operator commutes with the differential, $A\hat d_k=d_k A$ (as we demonstrated in sec.~\ref{sec:grbas} for $A=\Zbs$). %Indeed, in this case 
%\be
%%x\in \ker\hat d_k\ \Leftrightarrow 
%\hat d_kx=0\ \Rightarrow\ \hat d_kAx=0\\
%Ax=\hat d_{k-1}y\ \Rightarrow\ ???
%\ee
However, the homology spaces may \textit{not} the invariant spaces of the operator $A$ generally.

The actual properties of the KhR complexes are nicely compatible with the following picture.
Assume that %the morphisms and hence 
the differentials in the complex satisfy
%\be
%\hat d_k \Zbs=\Zbs\hat d_k
%\ee %,\ \ \mathbf{and}\ \
%\be
$\hat d_k E_{\alpha}=E_{\alpha}\hat d_k$ %,\ \mathbf{and}\ 
and
$\hat d_k H_{\alpha}= H_{\alpha}\hat d_k$ for all $\alpha=\overline{1,r}$, but % \quad \alpha=\overline{1,r},\ k=\overline{1,n},
%\label{dE}
%\ee
%\textbf{but} 
%\be
$\hat d_k F_{\alpha}\ne F_{\alpha}\hat d_k$, at least for some alpha. 
%\ \ \mathbf{and}\ \%\hat d_k H_{\alpha}\ne H_{\alpha}\hat d_k.
%\label{dFdH}\ee
Applying these relations %(\ref{dFdH}) 
when the operators act on 
an arbitrary vector $x\in \Vm_k$, one concludes that
\begin{itemize}
\item{The image of the Cartan eigenvector \textbf{must} be a Catran eigenvector with the \textbf{eigenvalue},
\be H_{\alpha}x=\lambda x\ \Rightarrow\ H_{\alpha}\hat d_k x=\lambda\hat d_kx.\ee}
%\item{The image of a vector under $\hat d_k$ \textbf{must} have
%    the same grading as the original vector, 
%    \be \Zbs x = q^{\Delta}x\ \Rightarrow\ \Zbs \hat d_k x=q^{\Delta}\hat d_kx.\ee}
\item{The image of a descendant of a vector
    \textbf{must} be the same descendant of the image of the vector,
    \be \hat d_k \left(E_{\alpha_i}\ldots E_1 x\right)= E_{\alpha_i}\ldots E_1 \hat d_k x.\ee}
\item{The image of the highest vector may \textbf{not} be a
    highest vector,
    \be F_{\alpha}x=0\not\Rightarrow F_{\alpha}\hat d_kx=0.\ee}
%\item{The image of the Cartan eigenvector may \textbf{not} be a Catran eigenvector,
%\be H_{\alpha}x=\lambda x\ \not\Rightarrow\ H_{\alpha}\hat d_k x=\lambda^{\prime} \hat d_kx.\ee}
%\item{If both a vector and its image are the Cartan eigenvactors, they may \textbf{not} have the same eigenvalues,
%\be H_{\alpha}x=\lambda x,\ H_{\alpha}\hat d_kx=\lambda^{\prime} \hat d_kx\ \not\Rightarrow\ \lambda=\lambda^{\prime}.\ee}
\end{itemize}
% Thus, the subsets $\bar \Vm_m^{\prime} $ are generally
%\textbf{not} representations of $\mathfrak{g}$.
%The same is true in the case, when the differentials commute
The operators $E_{\alpha}$ and $H_{\alpha}$, which satisfy the commutation relations %for the $E$ and $H$ operators have the form 
$[E_{\alpha},H_{\beta}]=c_{\alpha\beta}E_{\alpha}$ (for some constants $c_{\alpha\beta}$)~\cite{KlimSch}, generate the subalgebra $\mathfrak{g}^{\prime}$ of the original Lie algebra $\mathfrak{g}$. Similarly, one consider the differentials that commute with all $F_{\alpha}$ and $H_{\alpha}$, but not 
with all $F_{\alpha}$.
In the both cases we obtain the following non-trivial property of the KhR construction,
\begin{itemize}
\item{The homology spaces are generally \textbf{not} representation
    spaces of $\mathfrak{g}$.}
\end{itemize}
Yet, 
\begin{itemize}
\item{The homology are %spaces may be 
invariant spaces of the subalgebra of $\mathfrak{g}^{\prime}\subset\mathfrak{g}$ that is respected by action of the differentials.}
\end{itemize}
In particular, the grading operator $\Zbs$ can be an element of the \textit{universal enveloping}~\cite{KlimSch} of $\mathfrak{g}^{\prime}$, e.g.,~\mbox{$\Zbs=q^{\sum_{\alpha=1}^r H_{\alpha}}$}. Then the operators $\Zbs$ commute with the differentials as well, and hence in full consistence with sec.~\ref{sec:grbas},
%E.g., all homologies could be invariant spaces of one particular element of the Cartan subalgebra $H=\sum_{\alpha=1}c_{\alpha}H_{\alpha}$. The grading operator $\Zbs$ then can be identified with the exponential of this element $q^{H}$, and
\begin{itemize}
\item{The homology spaces are graded spaces.}
\end{itemize}

The algebras generated by the $2r$ operators $E_{\alpha}$ (or $F_{\alpha}$) and $H_{\alpha}$ with $\alpha=\overline{1,r}$, are by the moment less studied, than simple Lie algebras
generated by all $3r$ operators $E_{\alpha}$, and $F_{\alpha}$, and $H_{\alpha}$. Yet, representations of the former  algebras already reveal some inspiring properties. An particular, the characters of certain representations reproduce the MacDonald polynomials at special point~\footnote{The author is indebted to S.Arthamonov for pointing out this reference.}
~\cite{BorAlg}. The same polynomials at an other special point often can be used as an expansion basis for the knot superpolynomials~\cite{DMMSS}.

\subsubsection{A toy example}

Below we present two explicit examples of differentials with the
properties described above. %~(\ref{dE}),~(\ref{dFdH}). 
A more involved example is given
in App.~\ref{app:repex}.

Diagram~(\ref{repex1}) illustrates two possible maps (the
differentials $\hat d^{(0)}$ and $\hat d^{(1)}$) of the first
symmetric (type $[2]$) representation of the $\mathfrak{su}_2$
algebra. The differential $\hat d^{(0)}$ is an exact map, commuting
with the whole Lie algebra. The differential $\hat d^{(1)}$ commutes
only with the lowering operator $E$ and has non-trivial
kernel and co-kernel (the underlined terms in the first and second
lines).
\be
\label{repex1}
\arraycolsep=0mm
\begin{array}{rcl}
Fx&=&0,\\
\hat\Delta y&=&2y
%,\ \Rightarrow\\
%\Delta_{y\otimes Ex}=\Delta_x
\end{array}
\arraycolsep=1mm
\begin{array}{ccccccccccccccc}
&&&&\hspace{-0.5mm}\scriptstyle{E}\hspace{0.5mm}\\[-2mm]
&&\hspace{1mm}\boxed{0}&\hspace{-2mm}\boxed{0}
&\rightarrow&
\hspace{2mm}\boxed{x}&\hspace{-4mm}\boxed{x}&\rightarrow&
\boxed{Ex}&\boxed{Ex}&\rightarrow&
\boxed{E^2x}&\boxed{\underline{E^2x}}&\rightarrow&
\boxed{0}\\
% % % % % % % % % % % % % % % % % % % % % % % % % % % % % % %
&\hspace{3mm}\hat d^{(0)}\hspace{-4mm}
&\Arrow{-1}{-4}{4}{3}{10}&\Arrow{1}{-4}{7}{-5}{10}
&\hspace{-9mm}\hat d^{(1)}\hspace{3mm}
&
\Arrow{-1}{-4}{4}{3}{10}&\Arrow{1}{-4}{7}{-6}{10}&&
\Arrow{-1}{-4}{3}{4}{9}&\Arrow{1}{-4}{6}{-5}{9}&&
\Arrow{-1}{-4}{3}{2}{9}&\Arrow{1}{-4}{7}{-5}{9}&&\\
% % % % % % % % % % % % % % % % % % % % % % % % % % % % % % % % % % % % % %
&&\boxed{0}&&&
\boxed{x}&&&
\boxed{Ex}&&&
\boxed{E^2x}\\[-2.5mm]
% % % % % % % % % % % % % % % % % % %
\boxed{0}&\rightarrow
&&&\rightarrow
&&&\rightarrow&
&&\rightarrow&
&&\rightarrow&\boxed{0}\\[-4.5mm]
% % % % % % % % % % % % % % % % % % % % % % % % % % % % % %
&&&\boxed{\underline{y\otimes\! x}}&
&&\boxed{y\otimes\! Ex}&&
&\boxed{y\otimes\! E^2x}&&
&\boxed{0}&&\\
\end{array}
\ee
The gradings of the vectors are respectively equal to
\be
%&Fx=0
\arraycolsep=1mm
\begin{array}{ccccccccccccccc}
&&&&\hspace{-0.5mm}\scriptstyle{E}\hspace{0.5mm}\\[-2mm]
&&\hspace{1mm}\boxed{\emptyset}&\hspace{-2mm}\boxed{\emptyset}
&\rightarrow&
\hspace{2mm}\boxed{\Delta}&\hspace{-4mm}\boxed{\Delta}&\rightarrow&
\boxed{\Delta-2}&\boxed{\Delta-2}&\rightarrow&
\boxed{\Delta-4}&\boxed{\underline{\Delta-4}}&\rightarrow&
\boxed{\emptyset}\\
% % % % % % % % % % % % % % % % % % % % % % % % % % % % % % %
&\hspace{3mm}\hat d^{(0)}\hspace{-4mm}
&\Arrow{-1}{-4}{4}{3}{12}&\Arrow{1}{-4}{7}{-6}{12}
&\hspace{-5mm}\hat d^{(1)}\hspace{3mm}
&
\Arrow{-1}{-4}{4}{3}{12}&\Arrow{1}{-4}{7}{-6}{12}&&
\Arrow{-1}{-4}{3}{4}{10}&\Arrow{1}{-4}{6}{-5}{10}&&
\Arrow{-1}{-4}{3}{2}{9}&\Arrow{1}{-4}{6}{-5}{9}&&\\
% % % % % % % % % % % % % % % % % % % % % % % % % % % % % % % % % % % % % %
&&\boxed{\emptyset}&&&
\boxed{\Delta}&&&
\boxed{\Delta-2}&&&
\boxed{\Delta-4}\\[-2.5mm]
% % % % % % % % % % % % % % % % % % %
\boxed{\emptyset}&\rightarrow
&&&\rightarrow
&&&\rightarrow&
&&\rightarrow&
&&\rightarrow&\boxed{\emptyset}\\[-4.5mm]
% % % % % % % % % % % % % % % % % % % % % % % % % % % % % %
&&&\boxed{\underline{\Delta+2}}&
&&\boxed{\Delta}&&
&\boxed{\Delta-2}&&
&\boxed{\emptyset}&&\\
\end{array}
\ee
If we construct the generating function for the special graded
basis vectors of the spaces in the complex, then the differential
$\hat d=\hat d^{(0)}+\hat d^{(1)}$ is associated with the
following positive integer decomposition (see sec.~\ref{sec:dec} for
the definitions and comments),
\be
\begin{array}{c}
\multicolumn{1}{l}{
\boxed{\Pfr=\Pm(q,T)+(1+T)\Qfr(q,T)}=}\\[2mm]
=\left(1+q^2T\right)
\left(q^{\Delta}+q^{\Delta-2}+q^{\Delta-4}\right)
=\\
=
\big(1+T\big)
\left(q^{\Delta}+q^{\Delta-2}+q^{\Delta-4}\right)+
\left(1+q^2T\right)
\left(q^{\Delta}+q^{\Delta-2}+q^{\Delta-4}\right)
=\\
=\big(1+T\big)
\underbrace{
\left(q^{\Delta}+q^{\Delta-2}+q^{\Delta-4}\right)
}  _{\textstyle\Im\hat d^{(0)}}
+
q^{\Delta+2}T+
\left(1+T\right)
\underbrace{
\left(q^{\Delta}+q^{\Delta-2}\right)
}_{\textstyle\Im\hat d^{(1)}}
+q^{\Delta-4},
\\\\
%\multicolumn{3}{l}
\multicolumn{1}{l}{\boxed{
\Jm=q^{\Delta}+2q^{\Delta-2}+q^{\Delta-4},\
\Pm=q^{\Delta+2}T+q^{\Delta-4}}\,.}
\end{array}
\ee

\subsubsection{Relation to the real case}
The simplest case appearing is the KhR construction is in fact reduced
to the example above, by means of the correspondence between the
$N$-th symmetric ($[N+1]$ type) representation of $\frak{su}_2$ and
the fundamental representation of
$\frak{su}_{N+1}$~\cite{KlimSch}. %???, Sham, Dan.
Namely, define $E=\sum_{\alpha=1}^{N}E_{\alpha}$ ($E\in
\mathfrak{su}_2$, $E_{\alpha} \in \mathfrak{su}_{N+1}$). Then, the
highest vector $x$ of the fundamental representation of
$\mathfrak{su}_{N+1}$ satisfies $F_{\alpha}x=0$ for
$\alpha={\overline{1,N}}$ and $E^kx=E_k\ldots E_2E_1 x$ for $0\le k\le
N$ and $E^kx=0$ for $k>N$.

The map corresponding to $\hat d^{(1)}$ in~(\ref{repex1}) for $N=2$ then generally has the form
\be
\arraycolsep=0mm
\begin{array}{lcrl}
F_{\alpha}x&=&0,&
\hspace{2mm}
\alpha={\overline{1,r}}\\
\hat\Delta y&=&2y
\end{array}
\arraycolsep=1mm
\hspace{-1cm}
\begin{array}{ccccccccccccccc}
&&\boxed{0}&
\rightarrow&\boxed{x}&
\rightarrow&\boxed{E_1x}&
\rightarrow&\boxed{E_2E_1x}&
\rightarrow&\ldots&
\rightarrow&\boxed{\underline{E_{n-1}\ldots E_2E_1x}}
&\rightarrow&\boxed{0}\\
&&\Arrow{0}{-1}{15}{0}{12}&
&\Arrow{0}{-1}{13}{0}{12}&
&\Arrow{0}{-1}{10}{0}{10}&
&\Arrow{0}{-1}{10}{0}{10}&
&\ldots
&&\Arrow{0}{-1}{10}{0}{10}\\
\boxed{0}&
\rightarrow&\boxed{\underline{y\otimes\! x}}&
\rightarrow&\boxed{y\otimes\! E_1x}&
\rightarrow&\boxed{y\otimes\! E_2E_1x}&
\rightarrow&\ldots&
\rightarrow&
\ldots
%\boxed{E_{n-1}\ldots E_2E_1x}
&\rightarrow&\boxed{0}
\end{array}.
\label{repex2}\ee
%??? Dan
The simplest of the KhR maps has the form of diagram~(\ref{repex2}),
the grading being equal to~\cite{KhR} %???
\be
%\arraycolsep=0mm
%\begin{array}{rcl}
%Fx&=&0,\\
%\hat\Delta y&=&2y
%\end{array}
\arraycolsep=1mm
\hspace{2mm}
\begin{array}{ccccccccccccccc}
&&\boxed{\emptyset}&
\rightarrow&\boxed{N-1}&
\rightarrow&\boxed{N-3}&
\rightarrow&\boxed{N-5}&
\rightarrow&\ldots&
\rightarrow&\boxed{\underline{-N+1}}
&\rightarrow&\boxed{\emptyset}\\
&&\Arrow{0}{-1}{15}{0}{12}&
&\Arrow{0}{-1}{13}{0}{12}&
&\Arrow{0}{-1}{10}{0}{10}&
&\Arrow{0}{-1}{10}{0}{10}&
&\ldots
&&\Arrow{0}{-1}{10}{0}{10}\\
\boxed{\emptyset}&
\rightarrow&\boxed{\underline{N-1}}&
\rightarrow&\boxed{N-3}&
\rightarrow&\boxed{N-5}&
\rightarrow&\ldots&
\rightarrow&
\ldots
%\boxed{E_{n-1}\ldots E_2E_1x}
&\rightarrow&\boxed{\emptyset}
\end{array}.
\ee

%\subsection{
\section{Differential expansion and evolution method\label{sec:dexp}}
\subsubsection{The sketch of the methods}
Now we briefly outline two more (highly interrelated) approaches to
the KhR (and super-) polynomials. These two approaches are the
\textit{evolution method}~\cite{DMMSS, MMM3, AntM2} and the
\textit{differential expansion}\footnote{See also discussion
  in~\cite{AM}, where, in particular, the relations between the
  variables used in different cited papers are explicitly written
  down.}~\cite{DGN, MMM3, supsup, Art, MorKon2}. 
Some grounds for these approaches are given by recently obtained
recurrent relations for the KR polynomials. In some cases these
relations are derived from the Khovanov/KhR construction (essentially,
from the exact skein triangle)~\cite{KhRrec1,KhRrec2}, while in other
cases they remain an empiric observation. Hence, at the moment neither
the differential expansion, nor the evolution method are based on any
rigorously formulated construction. Yet both methods are highly
effective as computational tools. % being in this sense much closer to
% the positive division approach considered here.

All three approaches\,---\,positive division, differential expansion and
evolution method\,---\,share more in common with each other than with
the original KhR construction. We include this issue in this section
to emphasize the possible (as yet mostly empiric) relation of these
approaches to Lie group representation theory. We
complete the discussion by comparing the two approaches with the
positive division approach (an interplay was already observed
in~\cite{AM}).
%, and with the CohFT approach~\cite{MG}, which is the subject of sec.~\ref{sec:Gal}.

\

In both the evolution and the differential expansion methods one
studies an entire family of knots instead of its single
representative. The family can be either relatively general one, such
as all knots that are the closures of three-strand braids, or a more
restricted one, e.g., the knots obtained from a given one by
performing subsequently a certain transformation. One then writes the
desired knot invariant in the form
\be
\mathrm{Inv}^{\Km}(q,T) = \sum_Y \mathrm{Inv}^{\Km}_Y(q,T)C_Y(q,T),\label{dexp}
\ee
where $Y$ is a summation index (e.g., an integer, or a partition) running over a finite (and ``small'') set. The quantities $C_Y$ are treated as the common ``expansion basis'' for the knot family, while the ``expansion coefficients'' $\mathrm{Inv}_Y$ are either (relatively) simple functions of parameter(s) inside the family, or at least have a much simpler form than the resulting sum.
Moreover, the ``expansion basis'' usually consists of the products of some simple functions of an integer $k$ (or, more generally, of a set of integers),
\be
C_Y(q,T)=\begin{array}{c}
\prod\limits_{k\in \phi\subset \mathbb{Z}} f^Y(k|q,T)\\
\hline\\[-4mm]
\prod\limits_{l\in \psi\subset \mathbb{Z}}  g^Y(l|q,T)
\end{array}\label{basfac}
\ee
By construction, expansion~(\ref{dexp}) for $T=-1$ must reproduce one
of the known similar expansions for the HOMFLY polynomial (e.g., the
character expansion~\cite{MMM2,Tur}). The factors in the ``expansion
basis'' in this limit often become certain representation theory
quantities (e.g., the quantum dimensions of the irreducible
representation spaces, see sec.~\ref{sec:repsp}), and the expansion
factors typically take form of the \textit{quantum numbers}
$[n]_q\equiv\frac{q^n-q^{-n}}{q-q^{-1}}$ in the variable $q$. If this
is the case, a typical ``deformation'' of the expansion for general $T$ consists in the substitution
\be
f^Y_k(q,T=-1)=[n_k]_q \equiv
\begin{array}{c}
q^{n_k}-q^{-n_k}\\
\hline\\[-4mm]
q-q^{-1}
\end{array}
\ \ \longrightarrow\ \
f^Y_k(q,T)=[n_k]_q \equiv
\begin{array}{c}
q^{n_k}+q^{-n_k}T\\
\hline\\[-4mm]
q+q^{-1}T,
\end{array}
\label{chdef}
\ee
for some integer $n_k$.
%The expansion coefficients are also subjected a certain (???case dependent) substitution,
%\be
%Inv_Y(q,T=-1)\ \ \stackrel{\mbox{something}}{\Arrow{1}{0}{60}{-30}{0}}\ \ Inv_Y(q,T)
%\ee
In many cases, the denominators cancel out, so that the answer takes
form of a polynomial expanded over the ring generated by
$\left\{(q^{n_k}+Tq^{-n_k})\right\}$ %_{k\in\phi}$ % ??? what is $\phi$,
%$n_k$ ??? 
(the generators are referred to as ``differentials''
in~\cite{Art, DGN, MMM3}).

\subsubsection{The interplay with the positive division method}
The observations on the structure of the KhR
polynomials, which we summarised above, in fact motivated us to
formulate the multi-level division algorithm
(sec.~\ref{sec:levdiv}). Namely, were the KhR invariants obtained
merely by the first-level division, one could just identify $C_Y$ and
$\Pm_Y$ in (\ref{divlev2}) with $c_Y(q,T)$ and Inv$_Y$ in
(\ref{dexp}), respectively. In the actual, more complicated case,
initial structure (\ref{Fdec}) is broken down by the second-level
division. Yet, one can observe some traces of this structure in the
resulting invariant, which still can be written in the form~(\ref{chdef}).

\subsubsection{The interplay with the group theory viewpoint}
%From the group theory standpoint, 
The quantities (\ref{chdef}) usually look like ``$T$-deformations'' of
some quantum dimensions. This observation might signal that they
(similarly to their undeformed counterparts at $T=-1$) are somehow
related to the Lie algebra $\mathfrak{g}$, which acts on the spaces in
the complex (see
sec.~\ref{sec:gen},~\ref{sec:repth}). %have certain representation theory sense, as their
%are somehow expressed in representation theory terms. 
The highest expectation would be that the $C_Y(q,T)$ are invariants of the subalgebra $\mathfrak{g}^{\prime}\subset\mathfrak{g}$ that commutes with the action of the differentials (see also discussion and examples in sec.~\ref{sec:repth}). In this case the KhR invariants could be explicitly computed as traces over the homology spaces of some linear operators, or at least somehow expressed in terms of such quantities.

%
%
%The resulting ``deformed'' coefficients often appear to be related to
%the first-level remainders from~\cite{AM}, while the ``deformation''
%of the characters (\ref{chdef}) coincides with the second-level
%reduction substitution (see sec.~\ref{sec:levdiv}). However, the
%general case is more complicated and yet far from being exhaustively
%studied.
%
%??? write more comments ??? From the Lie algebra viewpoint discussed
%above, the minimal positive division for separate coefficients
%in~(\ref{dexp}) can be associated with the action of the ``zero
%approximation'' components of the differentials, which commute with
%all the Lie algebra generators (and hence, map each subspace of an
%irreducible representation to a subspace of isomorphic representation,
%which contributes to the same term in (\ref{dexp}).), like $\hat
%d^{(0)}$ in fig.~\ref{repex1}.  In turn, the following second-level
%division (\ref{chdef}) may correspond to the ``symmetry breaking
%components'', like $\hat d^{(1)}$ in fig.~\ref{repex1}.

%\subsection{\label{sec:evm}}

\section{Minimal positive division approach and CohFT calculus\label{sec:Gal}}

In this section we turn to another point of view on the KhR calculus
which appeared recently. This new approach is the soliton counting
technique developed in~\cite{MG,Gal}, which is part of the general
framework of CohFT introduced in~\cite{Witt}. We do not go into the
field theory background, concentrating on the details of practical
computations instead. We mostly aim to compare this approach to the
KhR invariants with the approach discussed in
ssec.~\ref{sec:gen}--\ref{sec:repth}.

\subsection{Preliminary comments}
\subsubsection{CohFT diagram technique}
Generally, the CohFT diagram is a knot diagram (see the definition in
sec.~\ref{sec:kndef}), with a spin state ($+$ or $-$) assigned to each
edge, and with (several kinds of) solitons distributed over the
crossings and turning points of the knot diagram. Various possible
diagrams can then be organized into a graph, similar to the resolution
hypercube with various colourings in the vertices
(sec.~\ref{sec:gen}). However, one cannot distribute the solitons
(playing the roles of two colourings $\Wcr$ and $\Bcr$) at different
crossings arbitrarily (see the particular examples below). Hence, the
resulting graph is not simply a hypercube, and we will not describe it
explicitly.

In the original CohFT construction~\cite{Witt,MG,Gal}, the diagrams
span certain vector spaces. These spaces are used as building blocks
to construct a complex, the differentials being defined by their
matrix elements labelled by pairs of diagrams. The general idea is
thus similar to that of the Khovanov~\cite{Khov, BarNat} and
KhR~\cite{KhR, CarMuf} constructions.

\subsubsection{Selection rule for CohFT matrix elements\label{sec:Galiff}}
To each CohFT diagram one can associate an integer, which is preserved
by the differentials. Also in the basis associated with CohFT diagrams
\begin{itemize}
\item{All the differentials in the constructed complex have matrix
    elements equal either to~$0$, or to~$1$, or to~$-1$.}
\end{itemize}
In other words, the CohFT diagrams enumerate the basis vectors, and
play the role of grading considered in sec.~\ref{sec:grbas}. We recall
(see sec.~\ref{sec:dec}) that these bases give rise to the positive
integer decomposition for the generating function (on a deeper level
this relation follows from the structure of the differentials). The
generating function for the basis vectors of the homology spaces
(which gives the desired knot invariant) is then the remainder of
division of the original (polynomial) generating function by a certain
polynomial. Hence, a similar decomposition can be written down in
terms of the CohFT diagrams.

\subsubsection{Types of CohFT matrix elements and multilevel division\label{sec:Galmult}}
%(see sec.~\ref{sec:Gal} for the definitions).
General considerations together with intuition from case studies
result in further specification of the selection rules. In particular,
the differentials %preserving 
the %same
distribution of signs over the strands at the bottom of the
braid %initial spin state 
appear to be
especially simple. 
%??? give the definition of the spin state on the
%section of a braid ??? % For a knot diagram having the form of a
% braid closure,
%One can introduce the initial spin state as a
%distribution of signs over the strands at the bottom of the
%braid. Then
\begin{itemize}
\item{The morphisms responsible for the transitions between the
    soliton diagrams with the same initial spin state commute with the
    whole Lie algebra (associated with the gauge group),}
\end{itemize}
and hence
\begin{itemize}
\item{ %???
The matrix element of the differential between any pair of
    soliton diagrams with the same $q$ grading and the $T$ gradings
    differing by $1$ equals either 1 or -1. %is non-vanishing.
    %???
    }
\end{itemize}
In the language of the minimal positive division this implies (see sec.~\ref{sec:dec}) that, roughly speaking,
%\begin{itemize}
%\item{
the corresponding terms in the dividend polynomial do \textbf{not} enter the remainder, whenever possible.
%}
%\end{itemize}
The previous statement admits a rigorous formulation in the unambiguous division case (see sec.\ref{sec:dec}), namely,
\begin{itemize}
\item{Only the first-level (see sec.~\ref{sec:levdiv}) minimal
    remainders for certain initial spin states (but generally not all
    of them), if uniquely defined, enter the resulting remainder,
    which yields the knot invariant.}
\end{itemize}

\subsubsection{Our program}
In this section, we present a straightforward way to calculate the
explicit form of the generating function for all the CohFT
diagrams. The technique relies on the $\Rm$ and $\Qm$-matrix formalism
(described in sec.~\ref{sec:RQpol}), in principle being valid for any
knot diagram, with any associated Lie algebra and its representations
(see sec.~\ref{sec:kndef}). Here we study the case of fundamental
representation of the $\mathfrak{su}_N$ algebra, when the answer is
expected to reproduce the KhR polynomial. We restrict the set of our
knot diagrams to braid closures. More specifically, we consider
general two-strand braids and particular cases of three-strand braids.
Furthermore we require all crossings to be of type $\Pcr$. An $\Ncr$
type crossing gives rise to an extra set of solitons~\cite{MG}, which
is beyond the scope of the current paper.

We then apply the minimal positive division technique described in
sec.~\ref{sec:div} to the resulting generating function.

To determine the ambiguous remainder, we involve the idea of the
multilevel division from sec.~\ref{sec:levdiv}. Namely, we write a
decomposition of type~(\ref{divlev1}), with $Y$ running over the
various distributions of the solitons over the turning points, the
expansion term with index $Y$ generating all possible distributions of
the solitons over the crossings in the relevant case.

\

\subsection{Division algorithm for a generic two strand knot\label{sec:2str}}
\subsubsection{The CohFT calculus in the two-strand case}
Below we consider a particular case of knot presented as the closure
of a two strand positive braid with $n=2k+1$ crossings (all of type
$\Pcr$). A two strand braid involving negative crossings are
topologically equivalent either to a positive braid or to its
reflection depending on whether $\#\Pcr>\#\Ncr$ or $\#\Pcr<\#\Ncr$
respectively~(see
App.~\ref{app:RQmirr}).

A spin state, $\fp$ or $\fm$, is associated to a strand
of the braid. The corresponding label is placed in the picture near each strand segment connecting two crossings. A section of the braid is related then to a composite
spin state, $\fpp$, $\fpm$, $\fmp$, or $\fmm$. The states $\fpp$ and $\fmm$ are
conserved throughout the braid, while each of the states $\fpm$ and $\fmp$
can change into one another near crossings. The change happens if a
crossing is asscoiciated to a soliton. %, either ($-\!-$ or $=\!\!\!\bullet\!\!\!=$) is placed on the crossing (in
%the pictures the soliton is drawn slightly below the crossing). 
%The
%entire CohFT diagram then looks like a braid with 
Two kinds of the solitons are drown on the picture as a single ($-\!-$) and the double ($=\!\!\!\bullet\!\!\!=$) lines below the crossings. %Each segment of a
%strand between two crossings labeled by $-$ or $+$ (see examples in
%table (\ref{fig:2strinst})).
In addition, one can associate solitons to
cups and caps, which close the braid diagram at the
top and the bottom ends. The arrangement of the solitons encodes the initial
spin state of the system. All four possible arrangements are
illustrated in~(\ref{fig:2strinst}), where soliton carrying cups and caps with like $\cup\hspace{-0.7em}-$ and $\cap\hspace{-0.7em}-$, respectively. The obtained picture is called in~\cite{MG} a \textit{soliton diagram}.

We represent a soliton diagram as a
word, where each of $n$ letters indicates what happens at the corresponding crossing. Namely, $\times$ stands for the lack of soliton, while $\en$ and $\db$ stand for the solitons of the two kinds (following~\cite{Gal}, with put the arrows showing the transport of the spin state). We word put the word in the square brackets and write the initial spin states in front of them. The spin states in all other braid sections can be successively obtained as one knows if there is a soliton at each crossing. The final spin state of the system must coincide with the initial one.

The examples of the soliton diagrams together with the corresponding words are given in table~(\ref{fig:2strinst}).

%???Mir, Zen.
%(see~(\ref{fig:2strinst})).
\be
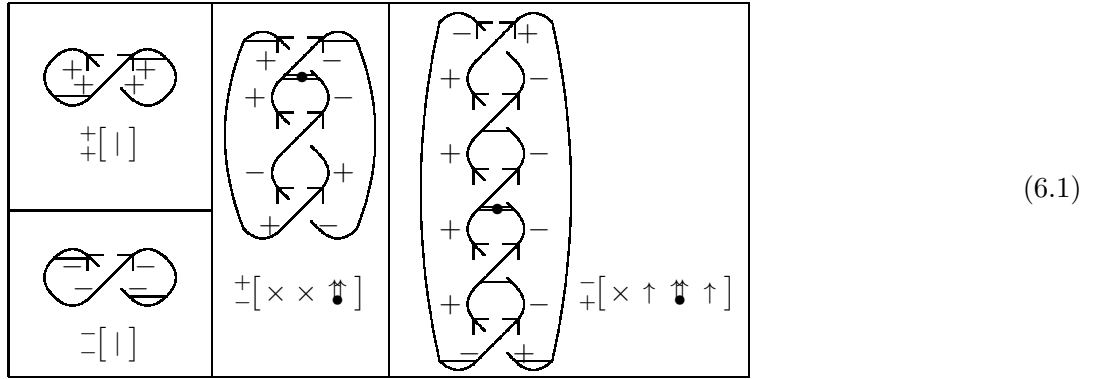
\begin{array}{|c|c|cc|}
\hline
\begin{array}{c}
\\[-40mm]
\unitlength=0.5mm
\begin{picture}(40,30)(-20,-30)
\figpcr{0}{0}
% % % % % % % % % % % % % % % % % % % % % % % %
%\put(-10,-5){\qbezier(-5,0)(0,-5)(5,0)}
\figsncup{-10}{-5}
\put(10,-5){\qbezier(-5,0)(0,-5)(5,0)}
%\figsncup{10}{-5}
\put(-10,5){\qbezier(-5,0)(0,5)(5,0)}
%\figsncap{-10}{5}
%\put(10,5){\qbezier(-5,0)(0,5)(5,0)}
\figsncap{10}{5}
% % % % % % % % % % % % % % % % % % % %
\put(-15,0){\qbezier(0,5)(-5,0)(0,-5)}
\put(15,0){\qbezier(0,5)(5,0)(0,-5)}
% % % % % % % % % % % % % % % % % % % % % % %
\put(-15,8){
\put(5,-12){$+$}\put(2,-8){$+$}
}
\put(8.5,8){
\put(-5,-12){$+$}\put(-2,-8){$+$}
}
\end{picture}\\[-10mm]
^+_+\hspace{-0.3em}
\left[
\arraycolsep=0.5mm
\begin{array}{cccc}
%\cap&\cup&
\en
\end{array}
\right]\\[2mm]
\end{array}
&
\unitlength=0.5mm
\begin{picture}(40,60)(-20,-15)
\figpcr{0}{0}\figpcr{0}{20}\figdbcr{0}{40}
\put(0,0){\qbezier(5,5)(10,10)(5,15)\qbezier(-5,5)(-10,10)(-5,15)}
\put(0,20){\qbezier(5,5)(10,10)(5,15)\qbezier(-5,5)(-10,10)(-5,15)}
% % % % % % % % % % % % % % % % % % % % % % % %
\put(-10,-5){\qbezier(-5,0)(0,-5)(5,0)}
%\figsncup{-10}{-5}
\put(10,-5){\qbezier(-5,0)(0,-5)(5,0)}
%\figsncup{-10}{-5}
%\put(-10,45){\qbezier(-5,0)(0,5)(5,0)}
\figsncap{-10}{45}
%\put(10,45){\qbezier(-5,0)(0,5)(5,0)}
\figsncap{10}{45}
\put(-15,0){\qbezier(0,45)(-10,20)(0,-5)}
\put(15,0){\qbezier(0,45)(10,20)(0,-5)}
% % % % % % % % % % % % % % % % % % % % % % %
\put(-15,8){
\put(4,-14){$+$}\put(0,0){$-$}\put(0,20){$+$}\put(3,30){$+$}
}
\put(8.5,8){
\put(-4,-14){$-$}\put(0,0){$+$}\put(0,20){$-$}\put(-3,30){$-$}
}
\end{picture}
&
\unitlength=0.5mm
\begin{picture}(40,70)(-25,20)
\figpcr{0}{0}\figsncr{0}{20}\figdbcr{0}{40}\figsncr{0}{60}\figvarpcr{0}{80}
\put(0,0){\qbezier(5,5)(10,10)(5,15)\qbezier(-5,5)(-10,10)(-5,15)}
\put(0,20){\qbezier(5,5)(10,10)(5,15)\qbezier(-5,5)(-10,10)(-5,15)}
\put(0,40){\qbezier(5,5)(10,10)(5,15)\qbezier(-5,5)(-10,10)(-5,15)}
\put(0,60){\qbezier(5,5)(10,10)(5,15)\qbezier(-5,5)(-10,10)(-5,15)}
% % % % % % % % % % % % % % % % % % % % % % % %
%\put(-10,-5){\qbezier(-5,0)(0,-5)(5,0)}
\figsncup{-10}{-5}
%\put(10,-5){\qbezier(-5,0)(0,-5)(5,0)}
\figsncup{10}{-5}
\put(-10,85){\qbezier(-5,0)(0,5)(5,0)}
%\figsncap{-10}{85}
\put(10,85){\qbezier(-5,0)(0,5)(5,0)}
%\figsncap{10}{85}
\put(-15,0){\qbezier(0,85)(-10,40)(0,-5)}
\put(15,0){\qbezier(0,85)(10,40)(0,-5)}
%\figsncap{15}{0}
% % % % % % % % % % % % % % % % % % % % % % %
\put(-15,8){
\put(5,-13){$-$}\put(0,0){$+$}\put(0,20){$+$}\put(0,40){$+$}\put(0,60){$+$}\put(3,72){$-$}
}
\put(8.5,8){
\put(-4,-13){$+$}\put(0,0){$-$}\put(0,20){$-$}\put(0,40){$-$}\put(0,60){$-$}\put(-3,72){$+$}
}
\end{picture}&\\[-9mm]
\cline{1-1}
\begin{array}{c}
\unitlength=0.5mm
\begin{picture}(40,30)(-20,-12)
\figpcr{0}{0}
% % % % % % % % % % % % % % % % % % % % % % % %
\put(-10,-5){\qbezier(-5,0)(0,-5)(5,0)}
%\figsncup{-10}{-5}
%\put(10,-5){\qbezier(-5,0)(0,-5)(5,0)}
\figsncup{10}{-5}
%\put(-10,5){\qbezier(-5,0)(0,5)(5,0)}
\figsncap{-10}{5}
\put(10,5){\qbezier(-5,0)(0,5)(5,0)}
%\figsncap{10}{5}
% % % % % % % % % % % % % % % % % % % %
\put(-15,0){\qbezier(0,5)(-5,0)(0,-5)}
\put(15,0){\qbezier(0,5)(5,0)(0,-5)}
% % % % % % % % % % % % % % % % % % % % % % %
\put(-15,8){
\put(5,-13){$-$}\put(2,-7){$-$}
}
\put(8.5,8){
\put(-4,-13){$-$}\put(-2,-7){$-$}
}
\end{picture}\\[-1mm]
^-_-\hspace{-0.3em}
\left[
\arraycolsep=0.5mm
\begin{array}{cccc}
%\cap&\cup&
\en
\end{array}
\right]\\[2mm]
\end{array}
&
^+_-\hspace{-0.3em}
\left[
\arraycolsep=0.5mm
\begin{array}{cccccc}
%\cap&\cap&
\c&\c&\db
\end{array}
\right]
&&
^-_+\hspace{-0.3em}
\left[
\arraycolsep=0.5mm
\begin{array}{cccccccc}
%\cup&\cup&
\c&\sn&\db&\sn
\end{array}
\right]\\[6mm]
\hline
\end{array}
\label{fig:2strinst}
\ee
 
Below we enumerate the soliton diagrams with help of the corresponding words, treating the latter ones as products of non-commutative variables
(the letters) and construct explicitly the generating function for all
the allowed words.

Then we search for an algorithm of presenting the generating function
as (\ref{gendec})-type decomposition with the desired knot invariant
as a remainder. For that we apply the ``multi-level'' division trick
from sec.~\ref{sec:levdiv} with 
%$Y$ running over the pairs of values of the first and last letters, 
$Y\in\big\{\fpp,\fmm,\fpm,\fmp\big\}$.

%???
We find that
\begin{itemize}
\item{The first-level remainders are unambiguous (see sec.~\ref{sec:amb}) for any number or crossings.}
\item{The second-level division can be performed algorithmically, by
    operating with the introduced non-commutative variables according
    to certain ``reduction rules''.}
\end{itemize}
The corresponding ``reduction rules'' are explicitly formulated below,
in ssec.~\ref{sec:div2},~\ref{sec:divN}.

\subsubsection{Khovanov ($N=2$) case\label{sec:div2}}
\begin{enumerate}
\item{Calculate
\be
\Pfr^{2,n}(q,T)=\Rfr^n_{=}\xi_{=}+\Tr\Big\{\Rfr^n_{\pm}
\Big({\footnotesize\arraycolsep=0mm
\begin{array}{cc}\xi_{\pm}\\&\xi_{\mp}
\end{array}}\Big)
\Big\}\label{Pfr2n}
\ee
with
\be
%\hspace{5mm}
\Rfr_==\left(
\begin{array}{c}
%\begin{array}{c|c}\\[-1.2cm]
%ii&\\
%\hline&\\[-4.5mm]
q %&ii
\end{array}\right),
\hspace{5mm}
\Rfr_{\pm}=\left(
\begin{array}{cc}
%\begin{array}{cc|c}\\[-1.2cm]
%ij&ji&\\
%\hline&&\\[-4.5mm]
q&T \\%&ij\\
T&q+Tq^{-1}\\ %&ji
\end{array}
%\hspace{-9mm}\right)\hspace{6mm},
\right).
%,\hspace{6mm}\mbox{and}\hspace{4mm}\hspace{5mm}\xi_==\xi_++\xi_-.
\label{PR2}\ee}
\item{Expand
\be\Pfr(\xi|q,T)=\sum_Y\Pfr^Y(q,T)\xi_Y,\ \
Y\in\{\fpp,\fmm,\fpm,\fmp\}.\label{2nexp}
\ee
\label{item:2}
}
\item{Perform the first-level reduction
\be
\Pfr^Y(q,T)=\sum_k q^k\sum_{\vartheta=\vartheta_{Y,k}^{\min}}^{\vartheta=\vartheta_{Y,k}^{\max}}C^Y_{\vartheta,u}T^{\vartheta}
\ \to\
\Pm^Y_{\one{-3mm}{-1mm}}\hspace{-1mm}(q,T)\equiv \sum_k q^kT^{\vartheta^{\min}_{Y,k}}.\label{2nred1}
\ee}
\item{Perform the second-level reduction %???
\be
\begin{array}{rrcrcc}
1.&q\xi_+&+&q^{-1}T\xi_{\pm}&\to&0.\\
2.&T^{2k+1}\xi_{\pm}&+&T^{2k+2}\xi_{\mp}&\to&0.\\
\end{array}
\label{2nred2}
\ee
\textbf{Do not reduce} $T^{2k}\xi_{\pm}+T^{2k+1}\xi_{\mp}\to0$ \textbf{!}\label{item:4}
}
\item{Substitute %??? %$\xi_-=q^{-6}$, $\xi_+=q^{-8}$, $\xi_{\pm}=q^{-8}$, $\xi_{\mp}=q^{-12}$
$\xi_+=q^2$, $\xi_-=q^{-2}$, $\xi_{\pm}=\xi_{\mp}=1$
.}
\item{The KhR polynomial equals
\be\boxed{\mbox{Kh}^{[2,n]}(q,T)=q^{-2n}\,\Pfr_{\two{-2mm}{-2mm}}(q,T)}\,.\label{fr2}\ee}
\end{enumerate}
For each particular odd $n$ the result reproduces the Khovanov
polynomial for the two-strand knot with $n$ crossings (torus knot
$T[2,n]$)~\cite{katlas}. The particular case $n=5$ is explicitly
worked out below. Even $n$ correspond to links, which we do not
consider here.

\subsubsection{KhR for generic $N$ \label{sec:divN}}
\begin{enumerate}
\item{Take the expansion~(\ref{2nexp}), reduced as in item~\ref{item:2} of sec.~\ref{sec:div2},
\be
\Pfr_{\one{-2mm}{-2mm}}(q,T)=
\sum_Y\Pfr_{\one{-2mm}{-3mm}}^Y\hspace{-1mm}(q,T)\xi_Y=\sum_Y\xi_Y\sum_k q^kT^{\vartheta^{\min}_{Y,k}}.\label{2nRed1}
\ee}
\item{Substitute whenever possible %???
\be
\begin{array}{rrcrcc}
1.&q^n\xi_{=}&+&q^{n-2}T\xi_{<}&\to&q^{N+n-1}[N].\\
2.&q^2T^{2k+1}\xi_{<}&+&q^{-2}T^{2k+2}\xi_{>}&\to&T^{2k+1}\left(q^N+Tq^{-N}\right)[N-1].\\
\end{array}
\label{2nRed2}
\ee
%\textbf{Do not reduce} $T^{2k}\xi_{\pm}+T^{2k+1}\xi_{\mp}\to0$ \textbf{!}
}
\item{The KhR polynomial of the torus knot $T[2,n]$ equals
    \be
    \boxed{\mbox{KhR}^{[2,n]}_N(q,T)=q^{-nN}\,
      \Pfr_{\two{-2mm}{-2mm}}(N|q,T)}\,.\label{frN}
    \ee
    \label{item:3}
}
\end{enumerate}
The result coincides with formula (4.55) from~\cite{AM} (earlier proposed in~\cite{DMMSS} for the superpolynomials).
\be
\mbox{KhR}^{[2,n]}_N(q,T)=
q^{-(n-1)(N-1)}\left([N]+[N-1]\left(1+q^{-2}T\right)
\sum_{j=0}^{2j\le n} q^{n-2j-4}T^{2j} \right).
\ee

\subsubsection{Primary polynomials as ``deformed'' HOMFLY polynomials\label{sec:Rm2n}}
Expression~(\ref{Pfr2n}) is the result of deformation~(\ref{Pfrgen})
of the $\Rm$-matrix representation for the HOMFLY
invariant~(\ref{HRm}), applied to the particular case of a two-strand
braid closure~\cite{MMM2, AM}. Namely, 
\be
\Pfr^{[2,n]}=\Qm_{i_1}^{i_{n+1}}\Qm_{j_1}^{j_{n+1}}\prod_{k=1^n}\Rfr^{i_kj_k}_{j_{k+1}i_{k+1}}=
\sum_{i=1}^N\left (\Qm^i_i\Qm^i_i %\left(\Rm_{=}^n\right)^{ii}_{ii} 
%\left (
\Rm^{ii}_{ii}\right)^n
+\sum_{i<j=1}^N \left(\Qm^i_i\Qm^j_j
%\sum_{{i_k,j_k}}
%\left(
%\Rfr^{ij}_{j_2i_2}\ldots\Rfr^{i_nj_n}_{ji}+\Rfr^{ji}_{j_2i_2}\ldots\Rfr^{i_nj_n}_{ji}
(\Rfr^N_{\pm})^{ij}_{ij}+\Qm^j_j\Qm^i_i(\Rfr^N_{\pm})^{ji}_{ji}
\right),
\label{2nexpl}
\ee
where
\be
\nn\\[-2mm]
\Rfr_= %(q,T)=
=\Rfr^{ii}_{ii}=
\begin{array}{c|c}\\[-1.2cm]
ii&\\
\hline&\\[-4.5mm]
q&ii
\end{array},\
1\le i\le
 N,
\hspace{5mm}
\Rfr_{\pm} %(q,T)
=\left(
\begin{array}{cc}
\Rfr^{ij}_{ij}&\Rfr^{ji}_{ij}\\
\Rfr^{ij}_{ji}&\Rfr^{ji}_{ji}
\end{array}\right)
\cong\left(
\begin{array}{cc|c}\\[-1.2cm]
ij&ji&\\
\hline&&\\[-4.5mm]
0&T&ij\\
T&q+Tq^{-1}&ji
\end{array}
\hspace{-9mm}\right)\hspace{6mm},\
1\le i<j\le N.
\label{Rfr2}\ee
%where $\Rfr_=$ and $\Rfr_{\pm}$
Note that $\Rfr_{\pm}$ is obtained from
the block of the ``deformed'' $\Rm$-matrices~(\ref{TRm}) by cancelling the
$q(1+T)$ entry.
%\be
%%\hspace{5mm}
%\Rfr_=(q,-1)=
%\begin{array}{c|c}\\[-1.2cm]
%ii&\\
%\hline&\\[-4.5mm]
%q&ii
%\end{array},\
%1\le i\le
% N,
%\hspace{5mm}
%\Rfr_{\pm}(q,-1)=\left(
%\begin{array}{cc|c}\\[-1.2cm]
%ij&ji&\\
%\hline&&\\[-4.5mm]
%q&-1&ij\\
%-1&q-q^{-1}&ji
%\end{array}
%\hspace{-9mm}\right)\hspace{6mm},\
%1\le i<j\le N,
%\ee
%???
%$\xi$'s do not enter the R-matrix?
%???
Since the blocks~(\ref{Rfr2}) of the $\Rfr$-matrix are independent of
$i$ and $j$, $\Qm$ enters (\ref{2nexpl}) only in combinations
\be
\xi_==\sum_{i=1}^{N}\left(Q^i_i\right)^2,\ \
\xi_{\pm}=\sum_{i<j=1}^{N}\Qm^i_i\Qm^j_j
,\ \ \xi_{\mp}=\sum_{j<i=1}^{N}\Qm^j_j\Qm^i_i.
\label{xi2}\ee
%Yet in
If one considers $\Qm^i_i$ as (generically non-commutative) operators,
this yields~(\ref{Pfr2n}). All the $\xi$'s must then be treated as
formal operators, independent of each other, as well as of $q$ and
$T$, until all the reduction steps are performed.  

The factors in~(\ref{fr2},~\ref{frN}) are the particular cases of the
framing factors (skipped in~(\ref{HRm}), but now restored; see
App.~\ref{app:Q}, where we should set $w=n$, and $N=2$ to
get~(\ref{frN})).

The substitutions in item~\ref{item:2} of sec.~\ref{sec:divN} are
motivated by the identities for the corresponding terms in the
HOMFLY polynomial, respectively,
\be
\begin{array}{cccccccc}
q\xi_{=}+q^{-1}T\xi_{<}&\stackrel{\mbox{HOMFLY}}{\Arrow{-1}{0}{28}{0}{0}\Arrow{1}{0}{28}{0}{0}}&
q&\frac{[2N]}{[2]}&-&q^{-1}\frac{[N][N-1]}{[2]}&=
q[N]^2-[N][N-1]=q^N[N]
\\[2mm]
&&&\parallel&\\
&&[N]^2\hspace{-4mm}&-&\hspace{-4mm}\frac{[N]{[N-1]}}{[2]}&
\end{array},
\ee
and
\be
q^2\xi_{>}+q^{-2}T\xi_{<}\hspace{2mm}
\stackrel{\mbox{HOMFLY}}{\Arrow{-1}{0}{28}{0}{0}\Arrow{1}{0}{28}{0}{0}}
\hspace{2mm}
\left(q^2-q^{-2}\right)\frac{[N][N-1]}{[2]}=\left(q-q^{-1}\right)[N][N-1]=\left(q^N-q^{-N}\right)[N-1].
\ee

\subsubsection{Comparing the $N=2$ and $N>2$ cases\label{sec:finN}}
\paragraph{The reduction rules: $N=2$ as a particular case of generic $N$.}
The level II reduction rules look differently for $N=2$
(sec.~\ref{sec:div2}, item 4) and for generic $N$ (sec.~\ref{sec:divN}, item 2).  Here we comment on their relationship.

The first of the $N=2$ rules, written in schematic form
\begin{equation}
  q^n\xi_{=}+q^{n-2}T\xi_{\pm}\to q^{n-1}[2],
\end{equation}
with $\xi_=\equiv\xi_++\xi_-$, just takes the form of the general
rule, with $N=2$ substituted.

The case of the second rule is different. Namely, the $N=2$ rule
(item~\ref{item:4} from sec.~\ref{sec:div2}) is manifestly asymmetric
between the odd and even $T$ powers (see the comment in the boldface),
while the general $N$ rule (item~\ref{item:3} from sec.\ref{sec:divN})
appear to be symmetric. In fact, there are several versions of the
reduction rule for $N\ge 3$, all yielding the same remainder. In
particular, one can take a straightforward generalization of the $N=2$
rule, which is illustrated in Tab.~\ref{redtab2},~\ref{redtab3}.II for
the $N=2$ and $N=3$ cases.  One should compare this with the rule in
Tab.~\ref{redtab3}.I, which we adhere to, and one more
possibility in Tab.~\ref{redtab3}.III.
\be
\arraycolsep=0.6mm
\begin{array}{|c|c|}
%\cline{2-3}\multicolumn{1}{c|}{}
\hline&\\[-4mm]
N=2&\min(i,j)\\
\hline&\\[-4mm]
\max(i,j)&0\\
\hline&\\[-4mm]
%q^{2p}\left(q^{-1}T\right)^{n-2p}&q^{2p+1}\left(q^{-1}T\right)^{n-2p-1}\\
1&q\cdot q^{-1}\\
\hline
\multicolumn{2}{c}{}\\[7mm]
\end{array}
%\right]
\times
\begin{array}{|cc|c|cc|c|}
\multicolumn{5}{c}{q^{-1}T^3+q^{-1}T^4\to\hspace{-4mm}/\hspace{4mm}0\,;}\\
%\cline{1-5}\multicolumn{6}{c}{}\\[-4mm]%\multicolumn{2}{c|}{}&&\multicolumn{2}{c}{}\\[-4mm]
\multicolumn{2}{c}{qT^2+qT^3\to 0,}&
\multicolumn{3}{c}{q^{-1}T^4+q^{-3}T^5\to 0.}\\
\hline&&&&&\\[-4mm]
%q^{2p}\left(q^{-1}T\right)^{n-2p}&q^{2p+1}\left(q^{-1}T\right)^{n-2p-1}\\
\Arrow{1}{-1}{15}{0}{10}
qT^2&
\boxed{q^{-1}T^3}&
&
\Arrow{2}{1}{32}{1}{-3}
q^{-3}T^4&
\boxed{q^{-5}T^5}&i<j\\[2.5mm]
\boxed{q^3T^2}&
\Arrow{1}{-1}{15}{0}{10}
qT^3&
&
\boxed{q^{-1}T^4}&
\Arrow{2}{1}{32}{1}{-3}
q^{-3}T^5&i>j\\[1mm]
\hline
\multicolumn{5}{c}{}\\
\multicolumn{5}{c}{}\\
\end{array}
\label{redtab2}
\ee

\be
\label{redtab3}
\arraycolsep=0.6mm
\begin{array}{|c|cc|}
%\cline{2-3}\multicolumn{1}{c|}{}
\hline&&\\[-4mm]
N=3&\multicolumn{2}{|c|}{\min(i,j)}\\
\hline&&\\[-4mm]
\max(i,j)&1&0\\
\hline&&\\[-4mm]
%q^{2p}\left(q^{-1}T\right)^{n-2p}&q^{2p+1}\left(q^{-1}T\right)^{n-2p-1}\\
2&q^2\cdot 1&q^2\cdot q^{-2}\\
1&&1\cdot q^{-2}\\
\hline
\multicolumn{3}{c}{}\\
\multicolumn{3}{c}{}
\end{array}
%\right]
\times
\begin{array}{|cc|c|}
\hline&&\\[-4mm]
%q^{2p}\left(q^{-1}T\right)^{n-2p}&q^{2p+1}\left(q^{-1}T\right)^{n-2p-1}\\
qT^2&q^{-1}T^3&i<j\\
q^3T^2&qT^3&i>j\\
\hline
\end{array}
\hspace{5mm}
\begin{array}{|c|c|}
\multicolumn{2}{c}{\mbox{I}}\\
\multicolumn{2}{c}{q[2]T^2\left(q+q^{-1}T\right)[3]\to qT^2[2]\left(q[2](q^3+q^{-3}T\right)}\\
\hline&\\[-4mm]
qT^2
\times\left(
\begin{array}{cc}
\boxed{q^2}
&\Arrow{1}{1}{10}{-1}{0}
1\\
&\Arrow{-1}{-1}{10}{7}{7}
q^{-2}\\[1mm]
\end{array}
\right)
&
q^{-1}T^3
\times\left(
\begin{array}{cc}
\Arrow{1}{1}{13}{-1}{-2}
q^2
&
\Arrow{-1}{-1}{10}{7}{8}
1\\
&\boxed{q^{-2}}\\[1mm]
\end{array}
\right)
\\
\hhline{|=|=|}&\\[-4mm]
q^3T^2
\times\left(
\begin{array}{cc}
\boxed{q^2}&
\Arrow{1}{-1}{10}{-1}{6}
1\\
&
\Arrow{-1}{1}{10}{7}{-2}
q^{-2}
\end{array}
\right)
&
qT^3
\times\left(
\begin{array}{cc}
\Arrow{1}{-1}{10}{-1}{6}
q^2
&
\Arrow{-1}{1}{10}{7}{-1}
1\\
&
\boxed{q^{-2}}\\[1mm]
\end{array}
\right)
\\
\hline
\end{array}
\\[-2mm]
\arraycolsep=0.6mm
\begin{array}{|c|c|}
\multicolumn{2}{c}{\mbox{II}}\\
\multicolumn{2}{c}{q^2T^2(q+q^{-1}T)[3]\to qT^2(q^2+q^{-2}T)[2]}\\
\multicolumn{2}{c}{qT^2(1+T)[3]\to 0}\\
\hline&\\[-4mm]
\Arrow{1}{-1}{15}{0}{10}
qT^2
\times\left(
\begin{array}{cc}
q^2&1\\
&q^{-2}
\end{array}
\right)
&
q^{-1}T^3
\times\left(
\begin{array}{cc}
\Arrow{-1}{1}{10}{7}{-3}
q^2&\boxed{1}\\[1mm]
&\boxed{q^{-2}}\\[1mm]
\end{array}
\right)
\\
\hhline{|=|=|}&\\[-4mm]
q^3T^2
\times\left(
\begin{array}{cc}
\boxed{q^2}&\boxed{1}\\
&\Arrow{-1}{1}{10}{7}{-3}
q^{-2}
\end{array}
\right)
&
\Arrow{1}{-1}{15}{0}{10}
qT^3
\times\left(
\begin{array}{cc}
q^2&1\\
&q^{-2}
\end{array}
\right)
\\
\hline
\end{array}
\hspace{5mm}
%\begin{array}{cc}
%q^3&1\\
%&q^{-2}
%\end{array}
\begin{array}{|c|c|}
\multicolumn{2}{c}{}\\
\multicolumn{2}{c}{}\\
\multicolumn{2}{c}{\mbox{III}}\\
\hline&\\[-4mm]
qT^2
\times\left(
\begin{array}{cc}
\Arrow{1}{-1}{10}{-1}{6}
q^2&
\Arrow{1}{1}{10}{-1.5}{-1}
1\\
&\Arrow{-1}{1}{10}{7}{-2}
q^{-2}
\end{array}
\right)
&
q^{-1}T^3
\times\left(
\begin{array}{cc}
\Arrow{1}{1}{13}{0}{-1}
q^2
&\boxed{1}\\[1mm]
&\boxed{q^{-2}}\\[1mm]
\end{array}
\right)
\\
\hhline{|=|=|}&\\[-4mm]
q^3T^2
\times\left(
\begin{array}{cc}
\boxed{q^2}
&
\boxed{1}\\
&
\Arrow{-1}{-1}{10}{7}{7}
q^{-2}
\end{array}
\right)
&
qT^3
\times\left(
\begin{array}{cc}
\Arrow{1}{-1}{10}{-1}{6}
q^2&
\Arrow{-1}{-1}{10}{6.5}{7.5}
1\\
&\Arrow{-1}{1}{10}{7}{-2}
q^{-2}
\end{array}
\right)
\\
\hline
\end{array}
%\\[5mm]
%\hspace{5mm}
%%\begin{array}{cc}
%%q^3&1\\
%%&q^{-2}
%%\end{array}
%\begin{array}{|c|c|}
%\hline&\\[-4mm]
%qT^2
%\times\left(
%\begin{array}{cc}
%q^3&1\\
%&q^{-2}
%\end{array}
%\right)
%&
%q^{-1}T^3
%\times\left(
%\begin{array}{cc}
%q^3&1\\
%&q^{-2}
%\end{array}
%\right)\\
%\hhline{|=|=|}&\\[-4mm]
%q^3T^2
%\times\left(
%\begin{array}{cc}
%q^3&1\\
%&q^{-2}
%\end{array}
%\right)
%&
%qT^3
%\times\left(
%\begin{array}{cc}
%q^3&1\\
%&q^{-2}
%\end{array}
%\right)
%\\
%\hline
%\end{array}
\nn\ee

%\be
%\begin{array}{|c|ccc|}
%\hline&&&\\[-4mm]
%N=4&\multicolumn{3}{|c|}{\min(i,j)}\\
%\hline
%\max(i,j)&2&1&0\\
%\hline&&&\\[-4mm]
%3&q^3\cdot q&q^3\cdot q^{-1}&q^3\cdot q^{-3}\\
%2&&q\cdot q^{-1}&q\cdot q^{-3}\\
%1&&&q^{-1}\cdot q^{-3}\\
%\hline
%\multicolumn{3}{c}{}\\
%\multicolumn{3}{c}{}
%\end{array}
%\times
%\begin{array}{|cc|c|}
%\hline&&\\[-4mm]
%qT^2&q^{-1}T^3&i<j\\
%qT^3&q^{-1}T^4&i>j\\
%\hline
%\end{array}
%\ee

In particular, one can see from the tables~(\ref{redtab3}) that the
odd-even asymmetry in fact is present in all versions of the reduction
rules. However, the resulting remainder does not include a $\sim(1+T)$
contribution in case $N\ge 3$. The difference is due to the additional
$q$-power factors (arranged below in the triangle matrices), which
account for various initial states giving rise to the same set of the
soliton diagrams. Hence,
\begin{itemize}
\item{The KhR polynomial of a two-stand knot is a minimal remainder for $N\ge 3$, unlike the Khovanov polynomial ($N=2$) of the same knot.
}
\end{itemize}

To summarize,
\begin{itemize}
\item{The $N=2$ reduction rules are a particular case of the generic $N$ reduction rules,}
\end{itemize}
in the considered case of the two-strand knots. However, \textbf{this is not so for a general knot}
because naive substitution of $N=2$ in the (large $N$) KhR polynomial does not always yield the correct Khovanov polynomial (compare the explicit examples in~\cite{CarMuf} with that in~\cite{katlas}, and see e.g., in~\cite{AM} for discussion). %??? Mor

%\paragraph{KhR and Khovanov polynomials as minimal and non-minimal remainders %\label{sec:termn}
%}
%In case of

\

\noindent
Now we return to the $N=2$ (Khvanov) case, which provides enough illustrations to the following discussion. 

\subsubsection{The primary polynomial as the generating function for the soliton diagrams\label{sec:inst2}}
Diagram rules~(A.33) from~\cite{MG} can be written in the matrix notations as
\be
\Pfr\to\hPfr:\hspace{5mm}
\Rfr_=\to\hRfr_==
\left(\begin{array}{c}\en\end{array}
\right),
\hspace{5mm}
\Rfr_{\pm}\to\hRfr_{\pm}={\footnotesize\arraycolsep=0.5mm
\bigg(\begin{array}{cc}0&\c\\[-2mm]\c&\sn+\db\\[-3mm]\end{array}
\bigg)},
\label{Rinst2}
\ee
(one must substitute $q\to qT$, $[0]\to 1$, $[f_0]\to 1$, $[f_0+1]\to T$  in (A.33)
and extract the common factor $q^{-\frac{1}{2}}$ to obtain our
formulas). 
The primary polynomial~(\ref{Pfr2n}) subjected to further ``deformation''~(\ref{Rinst2}) thus
becomes the generating function for the soliton diagrams
from~\cite{MG}. If we put $N=2$ 
and $i,j\in\left\{+,-\right\}$ in~(\ref{2nexpl}) (a more detailed
version of~(\ref{Pfr2n})), then the upper and the lower pair or indices of each $\Rfr$ corresponds to the composite spin state below and above the corresponding crossing, respectively. The braids with the composite spin state $\fpp$ and $\fmm$ in the first (and hence in any other) section are enumerated by the diagonal terms entering with the coefficient $\xi_=$, and all such diagrams yield the same ``amplitude'' $\Rfr^n_=$. In turn, the braids with composite spin states $\fpm$ or $\fmp$ in each section are enumerated by the off-diagonal terms. These states can exchange in each crossing in the way encoded in the $2\time 2$ matrix $\hat\Rfr_{\pm}$ in (\ref{Rinst2}), so that all possible soliton diagrams are enumerated by the expansion of the trace in (\ref{Pfr2n}), where the diagram with the initial spin state $+-$ or $-+$ enters with the coefficient $\xi_{\pm}$ or $\xi_{\mp}$, respectively. 

%
%Indeed, expression~(\ref{Rinst2}) is just 
%
%(where one
%must substitute $q\to qT$, $[0]\to 1$, $[f_0]\to 1$, $[f_0+1]\to T$
%and extract the common factor $q^{-\frac{1}{2}}$ to obtain our
%formulas). The coefficient of each $\xi$ in the resulting deformation
%of~(\ref{2nexp}) then generates all the soliton diagrams with the
%given initial state, while the label of $\xi$ runs over various
%initial states, $Y\in\left\{ {\tiny\begin{array}{c}+\\+\end{array}},\
%  {\tiny\begin{array}{c}-\\-\end{array}},\
%  {\tiny\begin{array}{c}+\\-\end{array}},\
%  {\tiny\begin{array}{c}-\\+\end{array}} \right\}$ with
%$\pm\sim{\tiny\begin{array}{c}+\\-\end{array}}$,
%$\mp\sim{\tiny\begin{array}{c}-\\+\end{array}}$, and $=\
%\sim{\tiny\begin{array}{c}+\\+\end{array}}\cup{\tiny\begin{array}{c}-\\-\end{array}}$
%(the two coefficients identically coincide). ??? clarify the
%notations, include in~(\ref{Rinst2}) ???

\paragraph{Combinatorial expressions for the numbers of the soliton diagrams.}
The number of the soliton diagrams with initial spin state $\pm$ or $\mp$ can as well be explicitly computed from combinatorial considerations, respectively, 
\be
\sharp\left\{^+_-\!\big[\big({\cc\!\!}^c {\sn\!\!}^w\db{\!\!}^{b}\big)\big]\right\} %_{2c+w+b=n} %=\nu_{n=w+b+2c|c,w,b}
  =\frac{(w+b+c)!}{w!\,b!\,c!},\ \ w+b+2c=n, %\ \ \mbox{and}\ \  \
\label{npm}
\ee
and
\be
\sharp\left\{^-_+\!\big[\!\c\! \big( {\cc}^c {\sn\!\!}^w \db{\!\!}^{b}\big)\c\!\big]\right\} %=\nu_{n=w+b+2c|c,w,b}
  =\frac{\left(w+b+c\right)!}{w!\,b!\,c!},\ \ w+b+2c=n-2, 
\label{nmp}
\ee
where the notations in the l.h.s. imply that we count distinct configurations obtained by the various permutations of $w$ letters $\en$, $b$ letters $\db$, and $c$ letters $\cc$ in the parentheses.
Denied that numbers (\ref{npm}) and (\ref{nmp}) are polynomial coefficients in the expansion 
of $\left(x+y+z\right)^{w+b+c}$, the proper generating functions
\be
f^{\pm}_n(x,y,z)=\sum_{w+b+2c=n}\frac{(w+b+c)!}{w!\,b!\,c!}x^wy^bz^{2c}=\Phi^n_{1,1},\label{genfun}\\
f^{\mp}_n(x,y,z)=f^{\pm}_{n-2}(x,y,z)=\Phi^n_{2,2},
\nn
\ee
are %\texit{not}
expressed for generic $n$ either as elements of powers of the matrix 
\be
\Phi=\left(\begin{array}{cc}0&c\\c&w+b\end{array}\right),\label{genmat}
\ee
or as rational functions of its eigenvalues
\be
\lambda_{1,2}=\frac{1}{2}\left(w+b\pm\sqrt{(b+w)^2+4c^2}\right).\label{genev}
\ee
These functions becomes the polynomials only if one substitutes the particular integer for $n$.

%\paragraph{Explicit expression for the primary polynomial.}

For the initial spin states $\fmm$ and $\fpp$ one gets trivially
\be
\sharp\left\{^+_+\!\big[\big(\en\!\!^n\big)\big]\right\}= \sharp\left\{^-_-\!\big[\big(\en\!\!^n\big)\big]\right\}=1 
\ee
with the generating function $f^=_n(x)=2x^n$.
\be
\ee
Primary polynomial~(\ref{Pfr2n},~\ref{PR2}) is then obtained from the entire generating function 
\be
f^n(x,y,z)=\xi_= f^n_=(x,y,z)+\xi_{\pm}f^n_{\pm}(x,y,z)+\xi_{mp}f^n_{\mp}(x,y,z)
\ee
as the result of the substitution
\be
\Pfr^{2,n}(\xi,q,T)=f^n(q,q^{-1}T,T).\label{Prf2f}
\ee

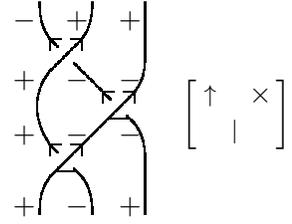
\begin{wrapfigure}{R}{112pt}
%\begin{array}{c}
\begin{picture}(60,45)(-20,15)
\figsncr{0}{0}\figsncr{20}{20}\figpcr{0}{40}
% % % % % % % % % % % % % % % % % % % % % % % % % % % % % % % % % %
\put(-6,6){\qbezier(0,0)(-10,15)(0,28)}
\put(5,5){\qbezier(0,0)(5,5)(10,10)}\put(15,25){\qbezier(0,0)(-5,5)(-10,10)}
\put(30,-10){\qbezier(0,-10)(0,5)(0,10)}\put(30,40){\qbezier(0,0)(0,5)(0,20)}
% % % % % % % % % % % % % % % % % % % % % % % % % % % % % % % % % % % % %
\put(30,30){\qbezier(-5,-5)(0,0)(0,10)}\put(10,50){\qbezier(-5,-5)(0,0)(0,10)}
\put(30,10){\qbezier(0,-10)(0,0)(-5,5)}\put(10,-10){\qbezier(0,-10)(0,0)(-5,5)}
\put(-10,-10){\qbezier(0,-10)(0,0)(5,5)}
\put(-10,50){\qbezier(5,-5)(0,0)(0,10)}
% % % % % % % % % % % % % % % % % % % % % % % % % % % % % % % % % % % %
\put(0,-20){\put(-20,0){$+$}\put(0,0){$-$}\put(20,0){$+$}}
\put(0,7){\put(-20,0){$+$}\put(0,0){$-$}\put(20,0){$-$}}
\put(0,27.5){\put(-20,0){$+$}\put(0,0){$-$}\put(20,0){$-$}}
\put(0,50){\put(-20,0){$-$}\put(0,0){$+$}\put(20,0){$+$}}
\end{picture} %\\
%\hspace{5mm}
$
\left[
\arraycolsep=0.5mm
\begin{array}{ccc}
\sn&&\c\\
&\en
\end{array}
\right]
$\\[5mm]
\caption{A piece of a three strand braid $\Bm=(\ldots, 1,2,1,\ldots)$, with a possible distribution of the spin states ($+$ and $-$) on the stands, and of the solitons in the crossings. On the right is our notation for the obtained diagram.\label{fig:3str}}
%\end{array}
\end{wrapfigure}

\paragraph{Employing the selection rules for the matrix elements in the division procedure.}

Deformation rule~(\ref{Rinst2}) sets a one-to-one correspondence
between the monomials in primary polynomial~(\ref{Pfr2n}) and the
soliton diagrams. Then,
\begin{itemize}
\item{The pairs of terms to delete in the reduction rules of
    sec.~\ref{sec:div2} correspond to the pairs of soliton
    diagrams of~\cite{MG} between which the transition is allowed\footnote{In fact, the allowed transitions in~\cite{MG} are
      related to a slightly different reduction rules, yet producing
      the same answer (see the explicit example below)}.}
\end{itemize}

\paragraph{Covariance under cyclic permutations.}
The suggested reduction rules are not invariant, but rather covariant
under the equivalence transformations of the knot diagram, namely
under the cyclic shift of the crossings (this is the only
transformation relevant for two-strand braids~\cite{KauffTB}).

The table below illustrates the behaviour of the various soliton
diagrams under the cyclic shift for two-strand braid with $n=3$
crossings (the closure is the trefoil knot, $3_1$
in~\cite{katlas}). Each diagram in the leftmost column is mapped to
itself. The diagrams arranged in triangles are mapped into each other
along the arrows. The cancellation of the diagrams arranged in the vertical pairs %(each one mapped along the arrows) ``cancel each other'' due to 
obeys the local rule $(\!\!\sn\!\db)+(\cc)\sim0$ (see~\cite{MG}) for detailed discussion of this rule in particular cases).  % (not discussed here???, see~\cite{MG}).
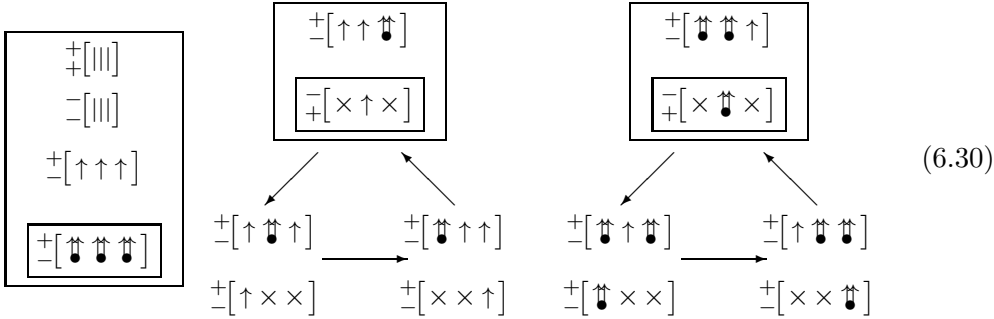
\be
\boxed{
\begin{array}{c}
\arraycolsep=0mm
{}^+_+\!\!\left[\begin{array}{ccc}\en&\en&\en\end{array}\right]\\[2mm]
% % %
\arraycolsep=0mm
{}^-_-\!\!\left[\begin{array}{ccc}\en&\en&\en\end{array}\right]
\\[3mm]
% % %
\arraycolsep=0.25mm
{}^+_-\!\!\left[\begin{array}{ccc}\sn&\sn&\sn\end{array}\right]\\[5mm]
% % %
\arraycolsep=0.25mm
\boxed{
{}^+_-\!\!\left[\begin{array}{ccc}\db&\db&\db\end{array}\right]
}
\end{array}
}
% % %
\begin{array}{ccc}
\multicolumn{3}{c}{\boxed{
\begin{array}{c}
\arraycolsep=0.25mm
{}^+_-\!\!\left[\begin{array}{ccc}\sn&\sn&\db\end{array}\right]\\[4mm]
\arraycolsep=0.25mm
\boxed{
^-_+\!\!\left[\begin{array}{ccc}\c&\sn&\c\end{array}\right]
}
\end{array}
}}\\
\Arrow{-1}{-1}{20}{20}{5}&&\Arrow{-1}{1}{20}{0}{-15}
\\[5mm]
\begin{array}{c}
\arraycolsep=0.25mm
{}^+_-\!\!\left[\begin{array}{ccc}\sn&\db&\sn\end{array}\right]\\[4mm]
\arraycolsep=0.25mm
{}^+_-\!\!\left[\begin{array}{ccc}\sn&\c&\c\end{array}\right]
\end{array}
&\Arrow{1}{0}{32}{-14}{5}&
\begin{array}{c}
\arraycolsep=0.25mm
{}^+_-\!\!\left[\begin{array}{ccc}\db&\sn&\sn\end{array}\right]\\[4mm]
\arraycolsep=0.25mm
{}^+_-\!\!\left[\begin{array}{ccc}\c&\c&\sn\end{array}\right]
\end{array}
\end{array}
% % %
\begin{array}{ccc}
\multicolumn{3}{c}{\boxed{
\begin{array}{c}
\arraycolsep=0.25mm
{}^+_-\!\!\left[\begin{array}{ccc}\db&\db&\sn\end{array}\right]\\[4mm]
\arraycolsep=0.25mm
\boxed{
^-_+\!\!\left[\begin{array}{ccc}\c&\db&\c\end{array}\right]
}
\end{array}
}}\\
\Arrow{-1}{-1}{20}{20}{5}&&\Arrow{-1}{1}{20}{0}{-15}
\\[5mm]
\begin{array}{c}
\arraycolsep=0.25mm
{}^+_-\!\!\left[\begin{array}{ccc}\db&\sn&\db\end{array}\right]\\[4mm]
\arraycolsep=0.25mm
{}^+_-\!\!\left[\begin{array}{ccc}\db&\c&\c\end{array}\right]
\end{array}
&\Arrow{1}{0}{32}{-14}{5}&
\begin{array}{c}
\arraycolsep=0.25mm
{}^+_-\!\!\left[\begin{array}{ccc}\sn&\db&\db\end{array}\right]\\[4mm]
\arraycolsep=0.25mm
{}^+_-\!\!\left[\begin{array}{ccc}\c&\c&\db\end{array}\right]
\end{array}
\end{array}
\ee

\subsubsection{Example: torus knot $T^{2,5}$.\label{sec:T25}}
The simplest example, when all the above features are non-trivial, is
presented in Tab.~\ref{tab:T25}. Nothing essentially new happens for a
two-stand knot with an arbitrary number of crossings.

The contributions to the first-level remainder are boxed. The
contributions to the second-level remainder are double boxed. The
terms marked by the same Roman numerals (uppercase) ``cancel each
other'' under the second level reduction. The lowercase labels mark
the pairs of terms ``cancelling'' in the~\cite{MG} version of the
reduction instead. The resulting remainders in the two cases contain
different sets of the soliton diagrams but are equal as polynomials in
$q$ and $T$, both yielding the correct value of the Khovanov
polynomial for $T[2,5]$ knot ($5_1$ in~\cite{katlas}).

\begin{table}
  \caption{Torus knot $[2,5]$. Soliton diagrams associated with all
    the terms of the primary polynomial, in the first-level remainders (boxed), and in the Khovanov polynomial (double boxed).\label{tab:T25}}
$
\begin{array}{|c|c|c|c|c|c|c|}
\hline&&\multicolumn{5}{|c|}{}\\[-3mm]
\small{\begin{array}{c}+\\[-1.5mm]+\end{array}}
&
\big[5\en\big]&\multicolumn{5}{|c|}{}\\[1mm]
&\boxed{\boxed{q^5}}&\multicolumn{5}{|c|}{}\\
&\mathbf{1}&\multicolumn{5}{|c|}{}\\
\hhline{|==|~~~~~|}&&\multicolumn{5}{|c|}{}\\[-3mm]
\small{\begin{array}{c}-\\[-2mm]-\end{array}}
&
\big[5\en\big]&\multicolumn{5}{|c|}{}\\[1mm]
&\boxed{q^5}&\multicolumn{5}{|c|}{}\\
&\mathbf{1}&\multicolumn{5}{|c|}{}\\
\cline{2-2}
&&\multicolumn{5}{|c|}{}\\[-4mm]
&\mathrm{I},i&\multicolumn{5}{|c|}{}\\
\hhline{|==|=====|}
\small{\begin{array}{c}+\\[-2mm]-\\[3mm]\end{array}}&
\big[(5\!\sn\!\big)]&
\big[(\!\sn\! 4\!\db\!\big)]&
\big[(2\!\sn\! 3\!\db\!\big)]&
\big[(3\!\sn\! 2\!\db\!\big)]&
\big[(\!\sn\! 4\!\db\!\big)]&
\big[(5\!\db\!\big)]\\[-2mm]
&\mathbf{1}&\mathbf{5}=\frac{5!}{1!4!}&\mathbf{10}=\frac{5!}{2!3!}&\mathbf{10}&\mathbf{5}&\mathbf{1}
\\[2mm]
&
\boxed{\boxed{q^5}}&
\boxed{q^3T}&
\boxed{qT^2}&
\boxed{\boxed{q^{-1}T^3}}&
\boxed{q^{-3}T^4}&
\boxed{\boxed{q^{-5}T^5}}\\
&\|&\|&\|&\|&\|&\|\\
&
q^5&
q^4\left(q^{-1}T\right)&
q^3\left(q^{-1}T\right)^2=&
q^2\left(q^{-1}T\right)^5&
q\left(q^{-1}T\right)^4&
\left(q^{-1}T\right)^5
\\[2mm]
\cline{2-7}&&&&&&\\[-4mm]
&&\mathrm{I},i&\mathrm{II}&ii&\mathrm{III}&iii\\
\cline{2-7}&&&&&&\\[-4mm]
&&\big[(3\!\sn\!\cc\big)]&
\big[(2\!\sn\! \!\db\!\cc\big)]&
\big[(\!\sn\! 2\!\db\!\cc\big)]&
\big[(3\!\db\!\cc\big)]
&\\[2mm]
&&
\mathbf{4}&\mathbf{12}=\frac{3!}{1!2!}\frac{(3+1)!}{3!1!}&\mathbf{12}&\mathbf{4}&\\[2mm]
&&
q^3T^2&
qT^3&
q^{-1}T^4&
q^{-3}T^5&\\
&&\|&\|&\|&\|&\\
&&
q^3\cdot T^2&
q^2\left(q^{-1}T\right)\cdot T^2&
q\left(q^{-1}T\right)^2\cdot T^2&
\left(q^{-1}T\right)^3\cdot T^2&
\\
\cline{2-7}&\multicolumn{2}{|c|}{}&&&\multicolumn{2}{|c|}{}\\[-3mm]
&\multicolumn{2}{|c|}{}&
\big[(\!\sn\!2\!\cc\big)]&
\big[(\!\db\! 2\!\cc\big)]&
\multicolumn{2}{|c|}{}\\[2mm]
&\multicolumn{2}{|c|}{}&
\mathbf{3}=\frac{(2+1)!}{2!1!}&\mathbf{3}&\multicolumn{2}{|c|}{}\\[2mm]
&\multicolumn{2}{|c|}{}&
\hspace{-4mm}\star\hspace{4mm}
qT^4&
\hspace{-4mm}\star\hspace{4mm}
q^{-1}T^5&
\multicolumn{2}{|c|}{}\\
&\multicolumn{2}{|c|}{}&
\|&\|&
\multicolumn{2}{|c|}{}\\
&\multicolumn{2}{|c|}{}&q\cdot T^2\cdot T^2&
\left(q^{-1}T\right)\cdot T^2\cdot T^2&
\multicolumn{2}{|c|}{}
\\
\hhline{|=======|}&&&&&&\\[-3mm]
\small{\begin{array}{c}-\\[-2.5mm]+\end{array}}&&
\big[\c (3\!\sn\!)\c\big]&
\big[\c (2\!\sn\! \!\db\!)\c\big]&
\big[\c (\!\!\sn\! 2\!\db\!)\c\big]&
\big[\c (3\!\db\!)\c\big]&\\[-2mm]
&&
\mathbf{1}&\mathbf{3}=\frac{3!}{1!2!}&\mathbf{3}&\mathbf{1}&\\[2mm]
&&
\boxed{\boxed{q^3T^2}}&
\boxed{qT^3}&
\boxed{\boxed{q^{-1}T^4}}&
\boxed{q^{-3}T^5}&\\
&&\|&\|&\|&\|&\\
&&
T\cdot q^3\cdot T&
T \cdot q^2\left(q^{-1}T\right)\cdot T&
T\cdot q\left(q^{-1}T\right)^2\cdot T&
T\cdot \left(q^{-1}T\right)^3\cdot T&
\\
\cline{2-7}&&&&&&\\[-4mm]
&&ii&\mathrm{II}&iii&\mathrm{III}&\\
\cline{2-7}\cline{2-7}&\multicolumn{2}{|c|}{}&&&\multicolumn{2}{|c|}{}\\[-4mm]
&\multicolumn{2}{|c|}{}&
\big[\c(\!\!\sn\!)\cc\c\big]&
\big[\c(\!\!\db\! \cc\!)\c\big]&
\multicolumn{2}{|c|}{}\\
&\multicolumn{2}{|c|}{}&
\mathbf{2}&\mathbf{2}&\multicolumn{2}{|c|}{}\\
&\multicolumn{2}{|c|}{}&
qT^4&
q^{-1}T^5&
\multicolumn{2}{|c|}{}\\
&\multicolumn{2}{|c|}{}&
\|&\|&\multicolumn{2}{|c|}{}\\
&\multicolumn{2}{|c|}{}&T\cdot q\cdot T^2\cdot T&
T\cdot\left(q^{-1}T\right)\cdot T^2\cdot T
&\multicolumn{2}{|c|}{}
\\[2mm]
\hline
\end{array}
$
\end{table}

\subsection{The first-level division for particular three-strand knots\label{sec:3str}}
Now we turn to the first attempt to treat a more general case using
our approach which is the hybrid between the CohFT calculus and positive division technique.

Below we restrict our considerations to the case of Khovanov
polynomials ($N=2$), which already appear to be rather involved.

\subsubsection{A draft of the three-strand reduction procedure\label{sec:div3str}}
The list below almost literally follows the two-stand algorithm, yet
differing at each step, with the most steps not being completely
worked out at the moment.
\begin{enumerate}
\item{For a three-strand braid $\Bm=\big(\Bm_i\big)_{i=1}^n$,\hspace{2mm} $\Bm_i=1,2$
(fig.~\ref{fig:3str}),
\be
\Pfr^{\Bm}(q,T)=
\sum_Y\Tr %\left\{
%{\Qm\rule{0mm}{3.5mm}\hspace{-0.5mm}}^Y\hspace{-2mm}(q)
\Qm^Y\hspace{-1.5mm}(q)
\prod_{i=1}^n %{\rule{0mm}{3.5mm}\hspace{-0.5mm}\Rfr}^Y_{\Bm_i}(q,T)%
\Rfr^Y_{\Bm_i}(q,T)
,\hspace{5mm} Y\in\big\{\mbox{\small(000)},\mbox{\small(001)},\mbox{\small(011)},\mbox{\small(111)}\big\}.
\label{Pfr3}
\ee
with
\be
\Qm^{(000)}={ %\small
\arraycolsep=0.2mm
\left(\begin{array}{c}\xi_{000}\end{array}\right)},
\
\Qm^{(001)}=
\mathrm{diag}{ %\small
\arraycolsep=0.2mm
\left(\begin{array}{ccc}\xi_{001}&\xi_{010}&\xi_{100}\end{array}\right)},
\
\Qm^{(011)}=
\mathrm{diag}{ %\small
\arraycolsep=0.2mm
\left(\begin{array}{ccc}\xi_{011}&\xi_{101}&\xi_{110}\end{array}\right)},
\
\Qm^{(111)}={ %\small
\arraycolsep=0.2mm
\left(\begin{array}{c}\xi_{111}\end{array}\right)},
\label{Qm3}
\ee
$\Rfr^Y_{-\Bm_i}(q,T)=\Rfr^Y_{\Bm_i}(q^{-1},T)$,\hspace{2mm} and $\Rfr^Y_{\Bm_i}(q,T)$ for $\Bm_i>0$ given by (\ref{Rfr3}).
}
\item{Expand
\be\Pfr(\xi|q,T)=\sum_Y\Pfr^Y\!(q,T)\,\xi_Y,\ \ Y\in\big\{\mbox{\small(000)},\mbox{\small(001)},\mbox{\small(011)},\mbox{\small(111)}\big\}.
\label{3exp}
\ee
}
\item{I. Reduce
\be
\Pfr^Y(q,T)=(1+T)\Jm^Y(q,T)+\Pm^Y_{\hspace{-1mm}\min}(q,T)
\ \to\
\Pm^Y_{\hspace{-1mm}\min}(q,T),\label{3red1}
\ee
to the residue containing the minimal possible number of $q$ and $T$ powers, \textbf{if it is unique}.}
\item{II. Reduce further
\be
\sum\Qm^Y\Pm^Y_{\min}(q,T)\to\ ?
\label{3nred2}
\ee
}
\item{Substitute
$\xi_{000}=q^3$,  $\xi_{001}=\xi_{010}=\xi_{100}=q$, $\xi_{011}=\xi_{101}=\xi_{110}=q^{-1}$, $\xi_{111}=q^{-3}$.
}
\item{The result must be proportional the the Khovanov polynomial of
    the knot $\Km=\overline{\Bm\rule{0mm}{3mm}}$ that is the closure
    of the braid $\Bm$ (the proportionality factor is $q^{-2w}$ with $w=\sum_{i=1}^N\Bm_i$, see sec.~\ref{sec:Rm2n}).}
\end{enumerate}

\subsubsection{Explicit formulas for the three-strand (deformed) $\Rm$-matrices}

From now on we switch to the notations $+\to 0$ and $-\to 1$, more
natural for the $\Rm$-matrix formalism (especially when aiming for the
general $N$ case). Just as in the two-strand case, we omit all the
$q(1+T)$ entries in the $\Rm$-matrices (as the sources of $\sim(1+T)$
contributions never enter the minimal remainders). All the needed
matrices are explicitly given in the table below.

\be
\begin{array}{ccccc}
&&\mathbf{I}&\mathbf{II}&\mathbf{III}\\
%\multicolumn{1}{p{5cm}}{HOMFLY}&
%\multicolumn{1}{p{5cm}}{Primary polynomial}&
%\multicolumn{1}{p{5cm}}{Generating function}\\
\multicolumn{2}{c}{\mbox{To compute}}&
\mbox{HOMFLY polynomial}&
\mbox{Primary polynomial}&
\mbox{Generating function}\\[1mm]
Y&k&\Rm_k^{Y}&\Rfr_k^{Y}&\hRfr_k^Y\\[4mm]
\begin{array}{l}(000),\\(111)\end{array}&
\begin{array}{l}1,\\2\end{array}&
\left(
\begin{array}{c|c}\\[-1.2cm]
iii&\\
\hline&\\[-3mm]
\hspace{-0.25cm}q&iii\end{array}\hspace{-1cm}\right)\hspace{1cm},
i=0,1
&
\left(\begin{array}{c}
q
\end{array}\right)
&
\left(
\begin{array}{ccc}
\en\\
\end{array}\right)\\[1.5cm]
(001)&1&
\left(
\begin{array}{ccc|c}\\[-1.2cm]
001&010&100&\\
\hline&&&\\[-4.5mm]
q&&&001\\
&q-q^{-1}&-1&010\\
&-1&&100
\end{array}\hspace{-1.2cm}\right)\hspace{1cm}
&
\left(\begin{array}{ccc}
q&&\\
&q+q^{-1}T&T\\
&T&
\end{array}\right)
&
\left(
\begin{array}{ccc}
\en&&\\
&\sn+\db&\c\\
&\c&
\end{array}\right)\\[1.5cm]
&2&
\left(
\begin{array}{ccc|c}\\[-1.5cm]
001&010&100&\\
\hline&&&\\[-4mm]
q-q^{-1}&-1&&001\\
-1&&&010\\
&&q&100
\end{array}\hspace{-1.2cm}\right)\hspace{1cm}
&
\left(
\begin{array}{ccc}
q+q^{-1}T&T&\\
T&&\\
&q&
\end{array}\right)
&
%\Rfr^{(001)}_2=
\left(
\begin{array}{ccc}
\en+\db&\c&\\
\c&&\\
&&\en
\end{array}\right)
\\[1.5cm]
(011)&1&
\left(
\begin{array}{ccc|c}\\[-1.2cm]
011&101&110&\\
\hline&&&\\[-4mm]
q-q^{-1}&-1&&101\\
-1&&&101\\
&&q&110
\end{array}
\hspace{-1.2cm}\right)\hspace{1cm}
&
\left(
\begin{array}{ccc}
q+q^{-1}T&T&\\
T&&\\
&&q
\end{array}\right)
&
\left(
\begin{array}{ccc}
\en+\db&\c&\\
\c&&\\
&&\en
\end{array}\right)\\[1.5cm]
&2&
\left(
\begin{array}{ccc|c}\\[-1.2cm]
011&101&110&\\
\hline&&&\\[-4.5mm]
q&&&011\\
&q-q^{-1}&-1&101\\
&-1&&110
\end{array}\hspace{-1.2cm}\right)\hspace{1cm}
&
\left(
\begin{array}{ccc}
q&&\\
&q+q^{-1}T&T\\
&T&
\end{array}\right)
&
\left(
\begin{array}{ccc}
\en&&\\
&\en+\db&\c\\
&\c&
\end{array}\right)
\end{array}
\label{Rfr3}
\ee

\subsection{The generating function for the three-strand soliton diagrams\label{sec:3inst}}
Similarly to the two-strand case, expression~(\ref{Pfr3}) for the three-strand primary polynomial turns into a generating function for the soliton diagrams associated with the braid $\Bm$, if one substitutes $\hat \Rfr$ from col.~\textbf{III} instead of $\Rfr$ from from col.~\textbf{II} of (\ref{Rfr3}), each monomial in the primary polynomial being then related to a certain soliton diagram.

We consider only the positive braids, i.e., all the crossings being
the $\Pcr$ type, and respectively all $\Bm_i>0$. The form of a three
strand braid is much more constrained by such assumption, than the
form of a two strand braid (see sec.~\ref{sec:2str}).

\subsubsection{Guide to the experimental data\label{sec:3strdata}}
The three-strand positive braids with no more than 8 crossings,
together with the knots or links being their closures, are enumerated
in App.~\ref{app:BrDet}. Examples of the unique remainders, with each
term related to a set of soliton diagrams, are given in
App.~\ref{app:inst}:
\begin{enumerate}
\item for the doubly twisted unknot diagram, presented
  as three-strand braid with 2 crossings (App.~\ref{app:un}),

\item for all possible diagrams of the trefoil knot having form of
  three-strand braids with 4 crossings and providing the unique first
  level remainders (App.~\ref{app:tref})

\item for torus knot $T[3,4]$ ($8_{19}$ in~\cite{katlas}), presented
  as a three-strand braid with 8 crossings~\footnote{This case
    deserves special interest, as the simplest case of the ``thick''
    knot, i.e., when the Khovanov polynomial is \textbf{not} recovered
    by setting $N=2$ in the KhR polynomial~\cite{AM}.
 %???Mor--cite--thick
  }.
\end{enumerate}
%???to do
%The program for producing these sets (in principle, for any three strand positive braid, in case of the unique first-level remainders) can be found in~\cite{knotebook}.

\section{Further directions\label{sec:summ}}
%\cite{KnHomLec} including \cite{SatNaw}
We complete our presentation by enumerating the most natural directions for the following research.
\begin{enumerate}
\item{\textbf{Systematic study of the $m\ge 3$ strand case.}
Of course, the attempt presented in sec.~\ref{sec:3str} is just the first and a very naive one.}

\item{\textbf{Relation to the standard Khovanov-Rozansky
      construction.}  It is mysterious that the KhR construction,
    apparently following just one of several possible options at many
    points of the procedure (in particular, see the footnotes in the
    beginning of sec.~\ref{sec:hypsp} and in the beginning
    of~\ref{sec:cmp}), is discussed in the mathematical literature
    only in its original literal formulation~\cite{KhR, CarMuf} ???
    $N=3$ ???. Any alternative (even not essentially different
    mathematically), if found, can be useful for various
    applications. A detailed comparison of the $\Rm$-matrix
    construction, schematically described in sec.~\ref{sec:gen} and in
    sec.~\ref{sec:repth}, with the rigorously formulated KhR
    construction, both on the general level and with explicit
    examples, provides some advances in this direction.}

\item{\textbf{Relation to the differential expansion.}  This issue
    touched in sec.~\ref{sec:dexp} definitely deserves a systematic
    treatment.}

\item{\textbf{The problem of finite $N$.}  The problem is in
    particular discussed in~\cite{AM} and mentioned in
    sec.~\ref{sec:finN}.  The definition of the KhR
    polynomial~\cite{KhR} initially refers to the same notions, as the
    $\Rm$-matrix definition of the HOMFLY polynomial, until the
    differentials are introduced. In particular, both invariants are
    associated with the $\mathfrak{su}_N$ series of the Lie algebras
    (we tried to present this point mostly explicit in
    sec.~\ref{sec:gen}). However, the HOMFLY invariants admit the most
    naive analytical continuation to arbitrary complex value of $N$,
    while the KhR polynomials admits one only for $N\le N_0$ (with
    $N_0$ depending on the knot). The ``phase transition'' at $N=N_0$,
    which definitely happens at the level of homologies, is still not
    well studied and attracts a lot of attention.}
  
%%%Mor

\item{\textbf{Higher representations and other groups.}  Finally, the
    $\Rm$-matrix based approach possesses a highly inspiring property,
    already mentioned in sec.~\ref{sec:gen}. Namely, in principle it
    can be extended to \textit{arbitrary} Lie groups and
    representations, on the ground of the already exiting $\Rm$-matrix
    approaches to other knot invariants~\cite{KauffTB}. The coloured
    (related to higher representations of the $\mathfrak{su}_N$ Lie
    algebra) analogs of the KhR invariants are widely studied in
    various (often semi-empiric) ways~\cite{DMMSS, MMM3, supsup, Art,
      IndSup, KnHomLec} (the list of references can be expanded). Yet
    the subject is even further from being exhausted, than the case of
    ordinary KhR invariants. In particular, even exact definition of
    the coloured KhR invariants (superpolynomials) faces essential
    difficulties, in particular, discussed in~\cite{KhovCol,Dan} (the
    second paper concerns a possible extension of the cabling
    approach~\cite{AMcab} to the superpolynomials).
%???Mir, cite
}
\item{\textbf{Can one use the CohFT calculus as a tool in the KhR calculus?} This is the most intriguing point of the story. In fact, this work was originally motivated by this question. Authors of~\cite{MG,Gal} follow the CohFT logic and associate the knot polynomials with the partition function for BPS states. They consider the resolutions of the knot diagram (representing the basis vectors in the complex) as the soliton states in the CohFT, and they relate the action of the differentials to the instantons responsible for the soliton scattering. They also discuss the wall-crossing phenomenon in this context. Yet, all their considerations remain rather a matter of art, than a regular technique. It is still unclear, whether the such technique can at all be developed in the CohFT framework.}
\end{enumerate}

%Never give up! -- as long as you are responsible for your generous RNF foundation...

\section*{Acknowledgements}
The author is deeply indebted to A.~Yu.~Morozov for scientific direction of this work, and to D.~Galakhov and I.~Danilenko for long inspiring discussions and patient explanations. The author truly appreciates the great work of Ye.~Zenkevich, who careffully read the draft and helped to prepare it for publishing, and also the work of G.~Aminov, who helped to correct the appendices. The author thanks S.~Arthamonov, And.~Morozov and A.~Popolitov for the valuable comments, and M.~Bishler, D.~Vasiliev, Ya.~Kononov, S.~Mironov, N.~Nemkov, A.~Sleptsov together with all other participants of the Mathematical physics group seminar for interest to this work.  

The author is also grateful to the secretary of the mathematical physics group E.~S.~Syslova, who makes the work of the author and his colleagues possible for many years.

This work was funded by the Russian Science Foundation (grant N~16-12-10344).

\bibliographystyle{gost2008}

\bibliography{Gal_bib}

\begin{thebibliography}{10}
\def\selectlanguageifdefined#1{
\expandafter\ifx\csname date#1\endcsname\relax
\else\selectlanguage{#1}\fi}
\providecommand*{\href}[2]{{\small #2}}
\providecommand*{\url}[1]{{\small #1}}
\providecommand*{\BibUrl}[1]{\url{#1}}
\providecommand{\BibAnnote}[1]{}
\providecommand*{\BibEmph}[1]{#1}
\ProvideTextCommandDefault{\cyrdash}{\hbox to.8em{--\hss--}}
\providecommand*{\BibDash}{\ifdim\lastskip>0pt\unskip\nobreak\hskip.2em\fi
\cyrdash\hskip.2em\ignorespaces}

\bibitem{MG}
\selectlanguageifdefined{english}
\BibEmph{Galakhov~D., Moore~G.} \BibDash
\newblock arXiv~: hep-th/1607.04222.

\bibitem{Gal}
\selectlanguageifdefined{english}
\BibEmph{Galakhov~D.} \BibDash
\newblock arXiv~: hep-th/1702.07086.

\bibitem{AM}
\selectlanguageifdefined{english}
\BibEmph{Anokhina~A., Morozov~A.}~// \BibEmph{JHEP}. \BibDash
\newblock 2014. \BibDash
\newblock Vol.~07, no. 063. \BibDash
\newblock arXiv~: hep-th/1403.8087.

\bibitem{KhR}
\selectlanguageifdefined{english}
\BibEmph{Khovanov~M., Rozansky~L.} Matrix factorizations and link homology~//
  \BibEmph{Fund. Math.} \BibDash
\newblock 2008. \BibDash
\newblock Vol. 199. \BibDash
\newblock P.~1--91. \BibDash
\newblock arXiv~: math.QA/0401268.

\bibitem{Khov}
\selectlanguageifdefined{english}
\BibEmph{Khovanov~M.} A categorification of the {Jones} polynomial~//
  \BibEmph{Duke Math. J.} \BibDash
\newblock 2000. \BibDash
\newblock Vol. 101. \BibDash
\newblock P.~359--426.

\bibitem{KauffTB}
\selectlanguageifdefined{english}
\BibEmph{Kauffman~L.~H.} The interface of knots and physics. \BibDash
\newblock Singapore~: World Scientific, 2001. \BibDash
\newblock P.~788.

\bibitem{katlas}
\selectlanguageifdefined{english}
\BibEmph{Bar-Natan~D., Scott~M., et~al.} The {K}not {A}tlas. \BibDash
\newblock URL: \BibUrl{http://katlas.org} (online; accessed: 22.10.17).

\bibitem{knbook}
\selectlanguageifdefined{english}
\BibEmph{Morozov~And., Sleptsov~A., et~al.} The knotebook. \BibDash
\newblock URL: \BibUrl{www.knotebook.org} (online; accessed: 22.10.17).

\bibitem{KnHomLec}
\selectlanguageifdefined{english}
\BibEmph{Gukov~Sergei, Khovanov~Mikhail, Walcher~Johannes}. Physics and
  Mathematics of Link Homology. \BibDash
\newblock Providence~: AMS, 2016. \BibDash
\newblock P.~177.

\bibitem{GSchV}
\selectlanguageifdefined{english}
\BibEmph{Gukov~S., Schwarz~A., Vafa~C.} {Khovanov-Rozansky} homology and
  topological strings~// \BibEmph{Lett. Math. Phys.} \BibDash
\newblock 2005. \BibDash
\newblock Vol.~74. \BibDash
\newblock P.~53--74. \BibDash
\newblock arXiv~: hep-th/0412243.

\bibitem{RasmKhR}
\selectlanguageifdefined{english}
\BibEmph{Rasmussen~Jacob}. \BibDash
\newblock 2006. \BibDash
\newblock arXiv~: math.GT/0607544.

\bibitem{DGN}
\selectlanguageifdefined{english}
\BibEmph{Dunfield~N.~M., Gukov~S., Rasmussen~J.} The superpolynomial for knot
  homologies~// \BibEmph{Experimental Math.} \BibDash
\newblock 2006. \BibDash
\newblock Vol.~15. \BibDash
\newblock P.~129--159. \BibDash
\newblock arXiv~: math/0505662.

\bibitem{GorLew}
\selectlanguageifdefined{english}
\BibEmph{Gorsky~Eugene, Lewark~Lukas}~// \BibEmph{Experimental Mathematics}.
  \BibDash
\newblock 2015. \BibDash
\newblock Vol.~24. \BibDash
\newblock P.~162--174. \BibDash
\newblock arXiv~: math.GT/1404.0623.

\bibitem{Witt}
\selectlanguageifdefined{english}
\BibEmph{Gaiotto~D., Witten~E.} \BibDash
\newblock arXiv~: hep-th/1106.4789.

\bibitem{Schw}
\selectlanguageifdefined{english}
\BibEmph{Schwarz~A.~S.} New topological invariants arising in the theory of
  quantized fields~// International Topological Conference. \BibDash
\newblock Vol.~2. \BibDash
\newblock Baku~: Institute of Mathematics and Mechanics of the Azerbaijan
  Academy of Sciences of USSR, 1987.

\bibitem{WittJ}
\selectlanguageifdefined{english}
\BibEmph{Witten~E.} Quantum field theory and the {Jones} polynomial~//
  \BibEmph{Comm. Math. Phys.} \BibDash
\newblock 1989. \BibDash
\newblock Vol. 121. \BibDash
\newblock P.~351--399.

\bibitem{WittH}
\selectlanguageifdefined{english}
\BibEmph{Witten~E.} Gauge theories and integrable lattice models~//
  \BibEmph{Nucl. Phys.} \BibDash
\newblock 1989. \BibDash
\newblock Vol. B322. \BibDash
\newblock P.~629--697.

\bibitem{BarNat}
\selectlanguageifdefined{english}
\BibEmph{Bar-Natan~D.} On {Khovanov's} categorification of the {Jones}
  polynomial~// \BibEmph{Algebr. Geom. Topol.} \BibDash
\newblock 2002. \BibDash
\newblock Vol.~2. \BibDash
\newblock P.~337--370. \BibDash
\newblock arXiv~: math.QA/0201043.

\bibitem{DM1}
\selectlanguageifdefined{english}
\BibEmph{Dolotin~V., Morozov~A.} Introduction to {Khovanov} homologies. {I}.
  {Unreduced} {Jones} superpolynomial~// \BibEmph{JHEP}. \BibDash
\newblock 2013. \BibDash
\newblock Vol. 1301, no. 065. \BibDash
\newblock arXiv~: hep-th/1208.4994.

\bibitem{DM2}
\selectlanguageifdefined{english}
\BibEmph{Dolotin~V., Morozov~A.} Introduction to {Khovanov} homologies. {II}.
  {Reduced} {Jones} superpolynomials~// \BibEmph{J. Phys.: Conf. Ser.} \BibDash
\newblock 2013. \BibDash
\newblock Vol. 411, no. 012013. \BibDash
\newblock arXiv~: hep-th/1209.5109.

\bibitem{SatNaw}
\selectlanguageifdefined{english}
\BibEmph{Nawata~Satoshi, Oblomkov~Alexei}. Proceedings of.

\bibitem{CarMuf}
\selectlanguageifdefined{english}
\BibEmph{Carqueville~N., Murfet~D.} Computing {Khovanov-Rozansky} homology and
  defect fusion~// \BibEmph{Algebr. Geom. Topol.} \BibDash
\newblock 2014. \BibDash
\newblock Vol.~14. \BibDash
\newblock P.~489--537. \BibDash
\newblock arXiv~: hep-th/1108.1081.

\bibitem{DM3}
\selectlanguageifdefined{english}
\BibEmph{Dolotin~V., Morozov~A.} Introduction to {Khovanov} homologies. {III.}
  {A} new and simple tensor-algebra construction of khovanov-rozansky
  invariants~// \BibEmph{Nucl. Phys.} \BibDash
\newblock 2014. \BibDash
\newblock Vol. B878. \BibDash
\newblock P.~12--81. \BibDash
\newblock arXiv~: hep-th/1308.5759.

\bibitem{DMMSS}
\selectlanguageifdefined{english}
Superpolynomials for torus knots from evolution induced by cut-and-join
  operators~/ P.~Dunin-Barkowski, A.~Mironov, A.~Morozov et~al.~//
  \BibEmph{JHEP}. \BibDash
\newblock 2013. \BibDash
\newblock Vol.~03, no. 021. \BibDash
\newblock arXiv~: hep-th/1106.4305.

\bibitem{MMM3}
\selectlanguageifdefined{english}
\BibEmph{Mironov~A., Morozov~A., Morozov~An.} Evolution method and
  ``differential hierarchy'' of colored knot polynomials~// \BibEmph{AIP Conf.
  Proc.} \BibDash
\newblock 2013. \BibDash
\newblock Vol. 1562. \BibDash
\newblock arXiv~: hep-th/1306.3197.

\bibitem{AntM2}
\selectlanguageifdefined{english}
Evolution method and {HOMFLY} polynomials for virtual knots~/ L.~Bishler,
  A.~Morozov, And.~Morozov, Ant.~Morozov~// \BibEmph{Int. J. of Mod. Phys.}
  \BibDash
\newblock 2015. \BibDash
\newblock Vol. A30, no. 1550074. \BibDash
\newblock arXiv~: hep-th/1411.2569.

\bibitem{supsup}
\selectlanguageifdefined{english}
\BibEmph{Gorsky~E., Gukov~S., Stosic~M.} Quadruply-graded colored homology of
  knots. \BibDash
\newblock 2014. \BibDash
\newblock arXiv~: math.QA/1304.3481.

\bibitem{Art}
\selectlanguageifdefined{english}
\BibEmph{S.Arthamonovв, A.Mironov, A.Morozov}~// \BibEmph{Theor.Math.Phys.}
  \BibDash
\newblock 2014. \BibDash
\newblock Vol. 179. \BibDash
\newblock P.~509--542. \BibDash
\newblock arXiv~: hep-th/1306.5682.

\bibitem{MorKon2}
\selectlanguageifdefined{english}
\BibEmph{Kononov~Ya., Morozov~A.}~// \BibEmph{JETP Letters}. \BibDash
\newblock 2015. \BibDash
\newblock Vol. 101. \BibDash
\newblock P.~831--834. \BibDash
\newblock arXiv~: hep-th/1504.07146.

\bibitem{KhovCol}
\selectlanguageifdefined{english}
\BibEmph{Khovanov~Mikhail}~// \BibEmph{Knot theory and its Ramifications}.
  \BibDash
\newblock 2005. \BibDash
\newblock Vol.~14, no.~1. \BibDash
\newblock P.~111--130. \BibDash
\newblock arXiv~: math.QA/0302060.

\bibitem{IndSup}
\selectlanguageifdefined{english}
\BibEmph{Nawata~S., Ramadevi~P., Zodinmawia}~// \BibEmph{JHEP}. \BibDash
\newblock 2014. \BibDash
\newblock Vol. 1401, no. 126. \BibDash
\newblock arXiv~: hep-th/1310.2240.

\bibitem{Dan}
\selectlanguageifdefined{english}
\BibEmph{Danilenko~I.} \BibDash
\newblock arXiv~: hep-th/1405.0884.

\bibitem{Tur}
\selectlanguageifdefined{english}
\BibEmph{Turaev~V.~G.} The {Yang-Baxter} equation and invariants of links~//
  \BibEmph{Invent. Math.} \BibDash
\newblock 1988. \BibDash
\newblock Vol.~92. \BibDash
\newblock P.~527--533.

\bibitem{ReshTur}
\selectlanguageifdefined{english}
\BibEmph{Reshetikhin~N.~Yu., Turaev~V.~G.} Ribbon graphs and their invariants
  derived from quantum groups~// \BibEmph{Commun. Math. Phys.} \BibDash
\newblock 1990. \BibDash
\newblock Vol. 127. \BibDash
\newblock P.~1--26.

\bibitem{MorSm}
\selectlanguageifdefined{english}
\BibEmph{Morozov~A., Smirnov~A.} Chern-simons theory in the temporal gauge and
  knot invariants through the universal quantum {R-matrix}~// \BibEmph{Nucl.
  Phys.} \BibDash
\newblock 2010. \BibDash
\newblock Vol. B835. \BibDash
\newblock P.~284--313. \BibDash
\newblock arXiv~: hep-th/1001.2003.

\bibitem{MMM2}
\selectlanguageifdefined{english}
\BibEmph{Mironov~A., Morozov~A., Morozov~And.} Character expansion for {HOMFLY}
  polynomials. {II.} fundamental representation. up to five strands in braid~//
  \BibEmph{JHEP}. \BibDash
\newblock 2012. \BibDash
\newblock Vol.~03. \BibDash
\newblock arXiv~: hep-th/1112.2654.

\bibitem{Ano}
\selectlanguageifdefined{english}
\BibEmph{Anokhina~A.} On {R-matrix} approaches to knot invariants. \BibDash
\newblock 2014. \BibDash
\newblock arXiv~: hep-th/1412.8444v2.

\bibitem{KlimSch}
\selectlanguageifdefined{english}
\BibEmph{Klimyk~A., Schm{\"u}dgen~K.} Quantum groups and their representations.
  \BibDash
\newblock Berlin~: Springer, 2012. \BibDash
\newblock P.~552.

\bibitem{PrSig}
\selectlanguageifdefined{english}
\BibEmph{Groups~Loop}. Andrew Pressley and Graeme Segal. \BibDash
\newblock Oxford~: Clarendon Press, Oxford Mathematical Monographs, 1988.
  \BibDash
\newblock P.~316.

\bibitem{GelMan}
\selectlanguageifdefined{english}
\BibEmph{S.I.~Gelfand, Yu.I.~Manin}. Homological Algebra. \BibDash
\newblock Berlin~: Springer, 1994. \BibDash
\newblock P.~222.

\bibitem{BorAlg}
\selectlanguageifdefined{english}
\BibEmph{G.Fourier, P.Littelmann}~// \BibEmph{Advances in Mathematics}.
  \BibDash
\newblock 2007. \BibDash
\newblock Vol. 211, no.~2. \BibDash
\newblock P.~566--593. \BibDash
\newblock arXiv~: hep-th/1702.07086.

\bibitem{KhRrec1}
\selectlanguageifdefined{english}
\BibEmph{Elias~Ben, Hogancamp~Matthew}. \BibDash
\newblock 2016. \BibDash
\newblock arXiv~: math.GT/1603.00407.

\bibitem{KhRrec2}
\selectlanguageifdefined{english}
\BibEmph{Hogancamp~Matthew}. \BibDash
\newblock 2017. \BibDash
\newblock arXiv~: math.GT/1704.01562.

\bibitem{AMcab}
\selectlanguageifdefined{english}
\BibEmph{A.Anokhina, An.Morozov}~// \BibEmph{Theor.Math.Phys.} \BibDash
\newblock 2014. \BibDash
\newblock Vol. 178. \BibDash
\newblock P.~1--58. \BibDash
\newblock arXiv~: hep-th/1307.2216.

\bibitem{GukSt}
\selectlanguageifdefined{english}
\BibEmph{Sergei~Gukov~Marko~Stosic}~// \BibEmph{Geometry \& Topology
  Monographs}. \BibDash
\newblock 2012. \BibDash
\newblock Vol.~18. \BibDash
\newblock P.~309--367. \BibDash
\newblock arXiv~: hep-th/1112.0030.

\end{thebibliography}

\appendix

\section{Basic properties of the special point operators\label{app:RQ}}

\subsection{General constraints on the special point operators}

\subsubsection{Continuous space transformations and transformations of the knot diagram.}
The knot $\Km$ as a curve in the space is considered up to arbitrary continuous space transformations. However, the corresponding transformations of the knot  diagram $\Dm(\Km)$ (see sec.~\ref{sec:gen}) include, apart from arbitrary continuous plane transformations, certain singular transformations generated by the \textit{Reidemester moves}~\cite{KauffTB}. The latter ones are the elementary transformations of the special points on the knot diagram. Two of the moves, RII and RIII~(fig.~\ref{fig:R23}), involve only the crossings, and one more move, RI~(\ref{fig:R1}), involves the turning points as well. In addition, one should consider the transformations generated by the elementary transformations in fig.~\ref{fig:QQ}, which involve only the turning points on the knot diagram~\cite{KauffTB}.

\subsubsection{Topological invariance constraints on the crossing and turning point operators.}
By definition, contraction (\ref{HRm})  ignores the continuous, turning point-preserving planar transformations. Hence, the obtained quantity is a topological invariant 
%(ignores the entire group of the transformations representing the continuous space transformations, see above)
provided that the operators $\Rm$ and $\Qm$ satisfy a finite set of constraints explicitly presented below.

\subsection{Constraints on the $\Rm$-matrices\label{app:R}}
Transformations in fig.~\ref{fig:R23} are the planar projections of the continuous space transformations. Yet they affect non-trivially on the knot diagram and can identified with the generators of the vertices changing subgroup of the entire equivalence transformations group. The corresponding space constraints on the vertex ($\Rm$) operators are presented below each figure.

\be
\begin{array}{rclrclr}
\unitlength=0.5mm
\begin{picture}(20,40)(0,-18)
\figncr{0}{0}\figpcr{0}{20}
%\qbezier(-10,10)(5,-5)(10,-10)\qbezier(30,10)(15,-5)(10,-10)
%\qbezier(0,0)(5,5)(10,10)\qbezier(20,0)(15,5)(10,10)
\qbezier(5,5)(10,10)(5,15)\qbezier(-5,5)(-10,10)(-5,15)
% % % % % % % % % % % % % % % % % % % % % % % % % % % % %
\put(-10,-12){$i$}\put(8,-12){$j$}
\put(-13,10){$a$}\put(10,10){$b$}
\put(-10,27){$k$}\put(8,27){$l$}
\end{picture}
&\begin{picture}(10,40)\put(0,40){$=$}\end{picture}&
\begin{picture}(20,20)(-15,-42)
\qbezier(10,-10)(0,0)(10,10)
\qbezier(-10,-10)(0,0)(-10,10)
\put(10,10){\qbezier(0,0)(-2,0)(-4,0)\qbezier(0,0)(0,-2)(0,-4)}
% % % % % % % % % % % % % % % % % % % % % % % % % % % % %
\put(-10,10){\qbezier(0,0)(2,0)(4,0)\qbezier(0,0)(0,-2)(0,-4)}
% % % % % % % % % % % % % % % % % % % % % % % % % % % % % %
\put(-10,-18){$i$}\put(8,-18){$j$}
\put(-10,12){$k$}\put(8,12){$l$}
% % % % % % % % % % % % % % % % % % % % % % % % % % % % % %
\end{picture}
&
%\label{R3}\ee
%\be
\unitlength=0.5mm
\begin{picture}(75,60)(-60,-25)
\figpcr{0}{0}\figpcr{0}{20}\figpcr{-10}{10}
%\figpcr{0}{-20}\figpcr{10}{-10}
\qbezier(6,6)(10,10)(6,14)
\put(0,-20){\qbezier(-6,6)(-10,10)(-6,14)}
\put(-20,0){\qbezier(-6,-6)(0,0)(6,6)}
% % % % % % % % % % % % % % % % % % % % % % % % %
\put(0,-20){\qbezier(-6,6)(0,0)(6,-6)}
\put(10,-10){\qbezier(-6,6)(0,0)(6,-6)}
% % % % % % % % % % % % % % % % % % % % %
\put(10,-6){$j$}\put(10,9){$a$}
\put(-4,-12){$i$}\put(-10,-2){$c$}\put(-10,16){$b$}
\put(-24,16){$m$}
\put(6,27){$l$}\put(-6,27){$k$}
\put(-27,0){$p$}
\end{picture}
&\begin{picture}(10,40)\put(0,40){$=$}\end{picture}&
\unitlength=0.5mm
\begin{picture}(40,60)(-20,-30)
\figpcr{0}{0}
%\figpcr{0}{20}\figpcr{-10}{10}
\figpcr{0}{-20}\figpcr{10}{-10}
\qbezier(6,6)(10,10)(6,14)
\put(0,-20){\qbezier(-6,6)(-10,10)(-6,14)}
%\put(-20,0){\qbezier(-6,-6)(0,0)(6,6)}
% % % % % % % % % % % % % % % % % % % % % % % % %
\put(0,20){\qbezier(-6,6)(0,0)(6,-6)}
\put(-10,10){\qbezier(-6,6)(0,0)(6,-6)}
% % % % % % % % % % % % % % % % % % % % %
\put(18,-21){$j$}
\put(12,-30){$i$} \put(7,-3){$c$}\put(-5,-12){$a$}
\put(-21,18){$m$}
\put(16,-3){$l$}\put(7,-21){$b$}
\put(-12,24){$k$}
\put(-12,-30){$p$}
\end{picture}
\\
\Rn^{ji}_{ab}\Rp^{al}_{kb}&=&\delta^i_k\delta^j_l&
\Rs^{ia}_{jc}\Rs^{pb}_{cm}\Rs^{bl}_{ak}&=&\Rs^{pb}_{ia}\Rs^{bl}_{jc}\Rs^{ak}_{cm},
&
*\in\left\{{\scriptstyle+},{\scriptstyle-}\right\}.
\end{array}
\label{fig:R23}
\ee

\subsection{Relations between the $\Rm$ and $\Qm$-matrices\label{app:Q}}
\paragraph{Extremum point operators.}
The described approach to the knot invariants also requires %not only to 
%After for projecting the knot on a plane, %but for 
%and 
for selecting a direction in the plane %as well
~\cite{ReshTur, KauffTB, MorSm}. As a result, one must treat the turning points as two-valent vertices on the knot diagram %(apart from the crossings treated as four-valent vertices)
of the four kinds, $\Rmin$, $\Rmax$, $\Lmin$, and $\Lmax$.

The equivalence transformations of the knot diagram that change the number of the turning points must be considered separately. These transformations form a finite subgroup, and their generators give rise to the relations between different $\Qm$, as well as between $\Rm$ and $\Qm$ matrices (both presented below).
%\be
%\begin{array}{p{5cm}|c|cccc|}
%\hline&&\multicolumn{4}{c}{}\\[-4mm]
%\mbox{Preferred direction}&\mbox{yes}&\multicolumn{4}{c}{\mbox{yes}}
%\Pcr&\Pcr&\QPcr&\PcrQ&\QPcrQ\\
%\Nc&\PNr&\NPcr&\NcrQ&\QNcrQ\\
%\end{array}
%\ee

\be
\begin{array}{rclcrcl}
\unitlength=0.4mm
\begin{picture}(15,15)(3,-6)
\figlmin{-6}{-6}\figlmax{6}{6}
\put(13,2){$i$}\put(-6,2){$j$}\put(-15,2){$k$}
\end{picture}
&
\cong
&
\unitlength=0.4mm
\begin{picture}(15,3)(-15,-3)
\put(-15,0){\qbezier(0,0)(1,-1)(3,-3)\qbezier(0,0)(1,1)(3,3)}
\qbezier(0,0)(-10,0)(-15,0)
\put(-3,2){$i$}\put(-18,2){$k$}
\end{picture}
&\hspace{1cm}&
\unitlength=0.4mm
\begin{picture}(18,15)(0,-6)
\figrmin{-6}{-6}\figrmax{6}{6}
\put(-12,2){$i$}\put(-5,2){$j$}\put(15,2){$k$}
\end{picture}
&
\cong
&
\unitlength=0.4mm
\begin{picture}(15,3)(0,-3)
\put(15,0){\qbezier(0,0)(-1,-1)(-3,-3)\qbezier(0,0)(-1,1)(-3,3)}
\qbezier(0,0)(10,0)(15,0)
\put(-1,2){$i$}\put(15,2){$k$}
\end{picture}\\[2mm]
\Qlcap^i_j\ \, \Qlcup^j_k&=&\delta^i_k&&
\Qrcap^i_j\ \, \Qrcup^j_k&=&\delta^i_k
\end{array}
\label{fig:QQ}
\ee

\be
\begin{array}{rcccl}
\unitlength=0.4mm
\begin{picture}(18,12)(9,-5)
\figpcr{0}{0}
\qbezier(6,6)(11,10)(12,0)
\qbezier(6,-6)(11,-10)(12,0)
\put(8.5,6){\circle*{3}}\put(8.5,-8){\circle*{3}}
\put(15,-2){$p$}
\put(-6,7){$l$}\put(-6,-13){$i$}
\put(1,7){$k$}\put(1,-13){$j$}
\end{picture}
&
\cong
&
\begin{picture}(18,12)(-9,-5)
\figpcr{0}{0}
\qbezier(-6,6)(-11,10)(-12,0)
\qbezier(-6,-6)(-11,-10)(-12,0)
\put(-8.5,6){\circle*{3}}\put(-8.5,-8){\circle*{3}}
\put(-20,-2){$p$}
\put(6,7){$l$}\put(6,-13){$j$}
\put(-5,7){$k$}\put(-5,-13){$i$}
\end{picture}
&
\cong
&
\unitlength=0.4mm
\begin{picture}(8,12)(-3,3)
\put(0,15){\qbezier(0,0)(-1,-1)(-3,-3)\qbezier(0,0)(1,-1)(3,-3)}
\qbezier(0,0)(0,10)(0,15)
\put(5,0){$i$}\put(5,10){$k$}
\end{picture}
\\[5mm]
\Qlcup^j_p\ \, \Qrcap^p_l\ \Rm^{ij}_{kl}\ &=&
\ \Qlcap^j_p\ \, \Qrcup^p_l\Rm^{ij}_{kl}
&=&\varkappa\delta^i_k
\end{array}
\label{fig:R1}
\ee
The factor $\varkappa$ here depends on the particular solutions of (\ref{fig:R23}) for the $\Rm$ operators.
In particular, $\kappa=q^{-N}$ for solution (\ref{Rm}).

\paragraph{Framing factor.}
%Factor 
Multiplied on $\varkappa^w=q^{-wN}$, where $w\equiv\#\Pcr-\#\Ncr$ is the algebraic number of crossings on the knot diagram, contraction (\ref{HRm}) becomes topologically invariant. One can set $\varkappa=1$ by using homogeneity property of constraints (\ref{fig:R23}) to rescale the $\Rm\to\varkappa\Rm$. Yet the resulting $\Rm$ explicitly depends on the chosen Lie algebra $su_N$ via $N$, and this is why we do not make such rescaling here.

\paragraph{Types of crossings in presence of the selected direction.}
%As the second consequence of the direction selected, 
Once the ???preferred direction in the projection plane is chosen, one must consider 8 kinds of crossings instead of 2, 4 variants of each $\Pcr$ and $\Ncr$, differing by orientations of arrows w.r.t. the preferred direction. The topological invariance constraints then relate the corresponding 8 crossing operators to each other via the extremum point operators. The corresponding relations are given below.

\vspace{-2cm}
\be
\begin{array}{c}
\\[-3mm]
{\unitlength=0.6mm\pcrQ}
\end{array}
\hspace{5mm}
{\large\mbox{$\cong$}}
\hspace{5mm}
{\unitlength=0.45mm
\begin{picture}(48,48)(-27,6)
\qbezier(-6,-6)(-4,-4)(-3,-3)\qbezier(6,6)(4,4)(3,3)
\qbezier(-6,-6)(-4,-6)(-2,-6)\qbezier(-6,-6)(-6,-4)(-6,-2)
% % % % % % % % % % % % % % % % % % % % % % % % % % % % %
\qbezier(-6,6)(0,0)(6,-6)
\qbezier(-6,6)(-4,6)(-2,6)\qbezier(-6,6)(-6,4)(-6,2)
% % % % % % % % % % % % % % % % % % % % % % % % % % % % % % % % % % %
\put(0,12){
\qbezier(-6,-6)(0,0)(6,6)
\qbezier(6,6)(4,6)(2,6)\qbezier(6,6)(6,4)(6,2)}
% % % % % % % % % % % % % % % % % % % % % % % % % % % % %
\put(0,24){
\qbezier(-6,6)(0,0)(6,-6)
\qbezier(-6,6)(-4,6)(-2,6)\qbezier(-6,6)(-6,4)(-6,2)}
% % % % % % % % % % % % % % % % % % % % % % % % % % % % % % % % %
%\put(0,-12){
%\qbezier(-6,-6)(0,0)(6,6)
%\qbezier(6,6)(4,6)(2,6)\qbezier(6,6)(6,4)(6,2)}
% % % % % % % % % % % % % % % % % % % % % % % % % % % % % % % %
\put(-12,0){
\qbezier(-6,6)(0,0)(6,-6)
\qbezier(-6,6)(-4,6)(-2,6)\qbezier(-6,6)(-6,4)(-6,2)}
% % % % % % % % % % % % % % % % % % % % % % % % % % % % % % % %
\put(-24,0){
\qbezier(-6,-6)(0,0)(6,6)
\qbezier(-6,-6)(-4,-6)(-2,-6)\qbezier(-6,-6)(-6,-4)(-6,-2)
}
% % % % % % % % % % % % % % % % % % % % % % % % % % % % % %
%\put(12,0){
%\qbezier(-6,6)(0,0)(6,-6)
%\qbezier(-6,6)(-4,6)(-2,6)\qbezier(-6,6)(-6,4)(-6,2)}
% % % % % % % % % % % % % % % % % % % % % % % % % % % % %
%\put(-11,-9){$l$}\put(-13,2){$k$}\put(8,-9){$i$}\put(8,2){$j$}
\end{picture}
}
{\large\mbox{$\cong$}}
{\unitlength=0.6mm
\begin{picture}(60,48)(-30,0)
\qbezier(-6,-6)(0,0)(6,6)
\qbezier(6,6)(4,6)(2,6)\qbezier(6,6)(6,4)(6,2)
% % % % % % % % % % % % % % % % % % % % % % % % % % % % % % % %
\qbezier(-6,6)(-4,4)(-3,3)\qbezier(6,-6)(4,-4)(3,-3)
\qbezier(-6,6)(-4,6)(-2,6)\qbezier(-6,6)(-6,4)(-6,2)
% % % % % % % % % % % % % % % % % % % % % % % % % % % % %
\put(0,12)
{\qbezier(-6,6)(0,0)(6,-6)
\qbezier(-6,6)(-4,6)(-2,6)\qbezier(-6,6)(-6,4)(-6,2)}
% % % % % % % % % % % % % % % % % % % % % % % % % % % % %
\put(0,-12)
{\qbezier(-6,6)(0,0)(6,-6)
%\qbezier(-6,6)(-4,6)(-2,6)\qbezier(-6,6)(-6,4)(-6,2)
}
% % % % % % % % % % % % % % % % % % % % % % % % % % % % % %
%\put(-12,12){
%\qbezier(-6,-6)(0,0)(6,6)
%\qbezier(-6,-6)(-4,-6)(-2,-6)\qbezier(-6,-6)(-6,-4)(-6,-2)}
% % % % % % % % % % % % % % % % % % % % % % % % % % % % % %
\put(-12,0){
\qbezier(-6,-6)(0,0)(6,6)
\qbezier(-6,-6)(-4,-6)(-2,-6)\qbezier(-6,-6)(-6,-4)(-6,-2)
}
% % % % % % % % % % % % % % % % % % % % % % % % % % % % % %
\put(12,0){
\qbezier(-6,-6)(0,0)(6,6)
%\qbezier(-6,-6)(-4,-6)(-2,-6)\qbezier(-6,-6)(-6,-4)(-6,-2)
}
% % % % % % % % % % % % % % % % % % % % % % % % % % % % %
%\put(0,24){
%\qbezier(-6,-6)(0,0)(6,6)
%\qbezier(6,6)(4,6)(2,6)\qbezier(6,6)(6,4)(6,2)}
% % % % % % % % % % % % % % % % % % % % % % % % % % % % % % % %
%\put(-24,0){
%\qbezier(-6,6)(0,0)(6,-6)
%\qbezier(-6,6)(-4,6)(-2,6)\qbezier(-6,6)(-6,4)(-6,2)}
% % % % % % % % % % % % % % % % % % % % % % % % % % % % % %
%\put(5,-16){$i$}
\put(-5,7){$l^{\prime}$}\put(5,-3){$j^{\prime}$}
%\put(0,14){$k$}
\put(-7,-4){$i$}\put(-16,0){$l$}\put(15,-5){$j$}\put(6,7){$k$}
\end{picture}
}
\Qlcup^j_{j^{\prime}}\ \Qlcap_l^{l^{\prime}}\
\Rp^{i^{\prime}j^{\prime}}_{k^{\prime}l^{\prime}}
\ee

\vspace{-1.5cm}

\be
\begin{array}{c}
\\[-3mm]
{\unitlength=0.6mm\Qpcr}
\end{array}
\hspace{5mm}
{\large\mbox{$\cong$}}
\hspace{5mm}
{\unitlength=0.45mm
\begin{picture}(48,48)(-12,6)
\qbezier(-6,-6)(-4,-4)(-3,-3)\qbezier(6,6)(4,4)(3,3)
\qbezier(6,6)(4,6)(2,6)\qbezier(6,6)(6,4)(6,2)
% % % % % % % % % % % % % % % % % % % % % % % % % % % % %
\qbezier(-6,6)(0,0)(6,-6)
\qbezier(6,-6)(4,-6)(2,-6)\qbezier(6,-6)(6,-4)(6,-2)
% % % % % % % % % % % % % % % % % % % % % % % % % % % % % % % % % % %
%\put(-12,0){
%\qbezier(-6,-6)(0,0)(6,6)
%\qbezier(6,6)(4,6)(2,6)\qbezier(6,6)(6,4)(6,2)}
% % % % % % % % % % % % % % % % % % % % % % % % % % % % % % % % % % % %
\put(12,0){
\qbezier(-6,-6)(0,0)(6,6)
\qbezier(6,6)(4,6)(2,6)\qbezier(6,6)(6,4)(6,2)}
% % % % % % % % % % % % % % % % % % % % % % % % % % % % % % % % % % % %
\put(0,24){
\qbezier(-6,-6)(0,0)(6,6)
\qbezier(6,6)(4,6)(2,6)\qbezier(6,6)(6,4)(6,2)}
% % % % % % % % % % % % % % % % % % % % % % % % % % % % % % % % % % % %
%\put(0,-12){
%\qbezier(-6,6)(0,0)(6,-6)
%\qbezier(-6,6)(-4,6)(-2,6)\qbezier(-6,6)(-6,4)(-6,2)}
% % % % % % % % % % % % % % % % % % % % % % % % % % % % % %
\put(0,12){
\qbezier(-6,6)(0,0)(6,-6)
\qbezier(-6,6)(-4,6)(-2,6)\qbezier(-6,6)(-6,4)(-6,2)}
% % % % % % % % % % % % % % % % % % % % % % % % % % % % % %
\put(24,0){
\qbezier(-6,6)(0,0)(6,-6)
\qbezier(6,-6)(4,-6)(2,-6)\qbezier(6,-6)(6,-4)(6,-2)}
% % % % % % % % % % % % % % % % % % % % % % % % % % % % % % % % % % %
%\put(-11,-9){$j$}\put(-10,2){$i$}\put(8,-9){$k$}\put(8,2){$l$}
\end{picture}
}
{\large\mbox{$\cong$}}
{\unitlength=0.6mm
\begin{picture}(60,48)(-27,0)
\qbezier(-6,-6)(0,0)(6,6)
\qbezier(6,6)(4,6)(2,6)\qbezier(6,6)(6,4)(6,2)
% % % % % % % % % % % % % % % % % % % % % % % % % % % % % % % %
\qbezier(-6,6)(-4,4)(-3,3)\qbezier(6,-6)(4,-4)(3,-3)
\qbezier(-6,6)(-4,6)(-2,6)\qbezier(-6,6)(-6,4)(-6,2)
% % % % % % % % % % % % % % % % % % % % % % % % % % % % %
\put(12,0){
\qbezier(-6,6)(0,0)(6,-6)
\qbezier(6,-6)(4,-6)(2,-6)\qbezier(6,-6)(6,-4)(6,-2)}
% % % % % % % % % % % % % % % % % % % % % % % % % % % % % %
\put(-12,0){
\qbezier(-6,6)(0,0)(6,-6)
%\qbezier(6,-6)(4,-6)(2,-6)\qbezier(6,-6)(6,-4)(6,-2)
}
% % % % % % % % % % % % % % % % % % % % % % % % % % % % % %
%\put(-12,12){
%\qbezier(-6,-6)(0,0)(6,6)
%\qbezier(-6,-6)(-4,-6)(-2,-6)\qbezier(-6,-6)(-6,-4)(-6,-2)}
% % % % % % % % % % % % % % % % % % % % % % % % % % % % % %
\put(0,12){
\qbezier(-6,-6)(0,0)(6,6)
\qbezier(6,6)(4,6)(2,6)\qbezier(6,6)(6,4)(6,2)
}
% % % % % % % % % % % % % % % % % % % % % % % % % % % % %
\put(0,-12){
\qbezier(-6,-6)(0,0)(6,6)
%\qbezier(6,6)(4,6)(2,6)\qbezier(6,6)(6,4)(6,2)
}
% % % % % % % % % % % % % % % % % % % % % % % % % % % % % % % %
%\put(24,0){
%\qbezier(-6,-6)(0,0)(6,6)
%\qbezier(6,6)(4,6)(2,6)\qbezier(6,6)(6,4)(6,2)}
% % % % % % % % % % % % % % % % % % % % % % % % % % % % % % % %
%\put(0,24){
%\qbezier(-6,6)(0,0)(6,-6)
%\qbezier(-6,6)(-4,6)(-2,6)\qbezier(-6,6)(-6,4)(-6,2)}
% % % % % % % % % % % % % % % % % % % % % % % % % % % % % %
\put(-6,-3){$i^{\prime}$}
%\put(-10,3){$l^{\prime}$}\put(5,-3){$j^{\prime}$}
\put(2,7){$k^{\prime}$}
\put(-12,-8){$i$}\put(-7,7){$l$}\put(6,-3){$j$}\put(12,2){$k$}
\end{picture}
}
\Qrcup^i_{i^{\prime}}\ \Qrcap_k^{k^{\prime}}
\Rp^{i^{\prime}j^{\prime}}_{k^{\prime}l^{\prime}}
\ee

\be
\begin{array}{clcl}
\Rp^{ij}_{kl}
&
\begin{array}{c}
\\[-3mm]
{\unitlength=0.4mm\pcr}
\end{array}
\hspace{1.5cm}
&
\Rn^{ij}_{kl}
&
\begin{array}{c}
\\[-3mm]
{\unitlength=0.4mm\ncr}
\end{array}
\\
\parallel&&\parallel
\\
\Qlcup^j_{j^{\prime}}\ \Qlcap_l^{l^{\prime}}\ \
\Qrcup^i_{i^{\prime}}\ \Qrcap_k^{k^{\prime}}
\Rp^{i^{\prime}j^{\prime}}_{k^{\prime}l^{\prime}}
&
\begin{array}{c}
\\[-3mm]
{\unitlength=0.4mm\QpcrQ}
\end{array}
&
\Qlcup^i_{i^{\prime}}\ \Qlcap_k^{k^{\prime}}\ \
\Qrcup^j_{j^{\prime}}\ \Qrcap_l^{l^{\prime}}
\Rn^{i^{\prime}j^{\prime}}_{k^{\prime}l^{\prime}}
&
\begin{array}{c}
\\[-3mm]
{\unitlength=0.4mm\QncrQ}
\end{array}
\\\\
{\Qlcup}^j_{j^{\prime}}\ {\Qlcap}_l^{l^{\prime}}\Rp^{ij^{\prime}}_{kl^{\prime}}
&
\begin{array}{c}
\\[-3mm]
{\unitlength=0.4mm\Qpcr}
\end{array}
&
{\Qlcup}^i_{i^{\prime}}\ {\Qlcap}_k^{k^{\prime}}\Rn^{i^{\prime}l}_{k^{\prime}l}
&

\begin{array}{c}
\\[-3mm]
{\unitlength=0.4mm\Qncr}
\end{array}
\\
\parallel&&\parallel
\\
\Qrcup^i_{i^{\prime}}\ \Qrcap_k^{k^{\prime}}\Rp^{i^{\prime}j}_{k^{\prime}l}
&
\begin{array}{c}
\\[-3mm]
{\unitlength=0.4mm\pcrQ}
\end{array}
&
\Qrcup^k_{k^{\prime}}\ \Qrcap_l^{l^{\prime}}\Rn^{ij^{\prime}}_{kl^{\prime}}
&
\begin{array}{c}
\\[-3mm]
{\unitlength=0.4mm\ncrQ}
\end{array}
\end{array}
\label{Rvar}\ee

\paragraph{The commutation relations.}
%Finally, 
Both $\Rm$ operators must commute with the tensor squares of all the four $\Qm$ operators (we skip the corresponding equivalence transformations),
\be
{\rule{0mm}{3.5mm}}^{\times}\hspace{-2mm}
\Qm_{i^{\prime}}^i\,
{\rule{0mm}{3.5mm}}^{\times}\hspace{-2mm}
\Qm_{j^{\prime}}^j\,
\Rs^{i^{\prime}j^{\prime}}_{kl}=
{\rule{0mm}{3.5mm}}^{\times}\hspace{-2mm}\Qm^{k^{\prime}}_k\,
{\rule{0mm}{3.5mm}}^{\times}\hspace{-2mm}\Qm^{l^{\prime}}_l\,
\Rs_{k^{\prime}l^{\prime}}^{ij},
\hspace{1cm}
*\in\left\{+,-\right\},
\hspace{5mm}
\times\in\left\{
\Lcap,\ \Rcap,\ \Lcup,\ \Rcup
\right\}.
\label{RQcomm}\ee
Relations (\ref{RQcomm}) follow from relations (\ref{fig:R1}, \ref{fig:QQ}) and provide the vertical equalities
in relations (\ref{Rvar}).

\paragraph{Freedom in the definition of the turning point operators.}
%In fact, 
Topological invariance constraints (\ref{fig:R1}, \ref{fig:QQ}) fix only the pairwise products. Any four $\Qm$ operators satisfying  equations~(\ref{fig:QQ}, \ref{fig:R1}) yield the same expression for the knot invariant~\cite{ReshTur, KauffTB, MorSm}. In particular, one can select these operators as in (\ref{Qm}), where the two operators are the unity operators, so that the related turning point can be ignored (just as we do through the text).

\subsection{Properties of the particular solution\label{app:skein}}
The $\Rp$- and $\Rn$-operators given by particular solution (\ref{Rm}) of the topological invariance constraints also satisfy the 
%additional relations called
\textit{skein} relations, namely~\cite{Tur, KlimSch}
\be
\begin{array}{cccccc}
%q^N&
\unitlength=0.5mm\begin{picture}(25,20)(0,8)\pcr\end{picture}
&-&
%q^{-N}&
\unitlength=0.5mm\begin{picture}(25,20)(0,8)\ncr\end{picture}
&=&
\left(q-q^{-1}\right)&
\begin{picture}(20,20)(-5,-3)
\qbezier(10,-10)(0,0)(10,10)
\qbezier(-10,-10)(0,0)(-10,10)
\put(10,10){\qbezier(0,0)(-2,0)(-4,0)\qbezier(0,0)(0,-2)(0,-4)}
% % % % % % % % % % % % % % % % % % % % % % % % % % % % %
\put(-10,10){\qbezier(0,0)(2,0)(4,0)\qbezier(0,0)(0,-2)(0,-4)}
% % % % % % % % % % % % % % % % % % % % % % % % % % % % % %
\put(-10,-18){$i$}\put(8,-18){$j$}
\put(-10,12){$k$}\put(8,12){$l$}
% % % % % % % % % % % % % % % % % % % % % % % % % % % % % %
\end{picture}\\[5mm]
\Rp^{ij}_{kl}&-&\Rn^{ij}_{kl}&=&
\multicolumn{2}{c}{\left(q-q^{-1}\right)\delta^i_k\delta^j_l}.
\label{fig:skein}
\end{array}
\ee
These relations together with (\ref{fig:R23}) lead to the characteristic equations
\be
\left(\Rp-q\right)\left(\Rp+q^{-1}\right)=0,\ \
\left(\Rn-q^{-1}\right)\left(\Rn+q\right)=0.
\label{Rev}
\ee
Relations (\ref{Rev}) imply that each of the crossing operators has just two distinct eigenvalues: $q$, $\left(-q^{-1}\right)$) for $\Rp$, and $q^{-1}$, $\left(-q\right)$ for $\Rn$.
The operators $\Rm$ and $\Qm$ given by (\ref{Rm},\ref{Qm}) (so that $\Rm$ satisfies (\ref{Rev}))
correspond to relating the space of fundamental representation of the $\mathfrak{gl}_N$ algebra %as the space $V$ related 
to each edge of the knot diagram. %Hence, 
The choice of this space %$V$ 
is essential for the entire construction, which in all versions~\cite{Khov, KhR, DM3,AM} heavily relies on decomposition (\ref{Rdec}), and hence on (\ref{Rev}) whence the decomposition follows.

Although an analog of (\ref{Rdec}) is explicitly known for any representation of any simple Lie~\cite{KlimSch}), there are hardly any advances in extending the KhR construction beyond the fundamental representation of $\mathfrak{gl}_N$.

%One can explicitly verify that the relation $\Rp\Rn=\Rn\Rp=\IdId$ holds as well.

\subsection{Inverting of all crossings and mirror symmetry of the knot invariants\label{app:RQmirr}}
Many important knot invariants, among them are
the Jones, HOMFLY, Khovanov, and the KhR invariants, possess the so called \textit{mirror symmetry}~\cite{GukSt}. The symmetry relates the expressions for the invariant for a knot and its mirror image. Generally these two knots are topologically
distinct~\cite{KauffTB}, and all the enumerated invariants do distinguish them. Still, the invariants for the knot and for its mirror image are related by simple change of the variables, e.g.,~$q\to 1/q, T\to 1/T$ for the KhR invariants.

The mirror transformation corresponds to inverting all the crossings $\Pcr\leftrightarrow\Ncr$ on the knot diagram. By this reason, only the diagrams with $\#\Pcr>\#\Ncr$ are usually considered. In particular, in case of two-strand braids it is enough to consider only $\Pcr$ type crossings, just as we do in sec.~\ref{sec:2str}.

\section{The ``graded'' basis respected by the differentials\label{app:spbas}}
Here we describe in more details of the ``graded'' basis, selected in sec.~\ref{sec:grbas} in the Khovanov complex and used in~\ref{sec:dec} to derive the Positive integer decomposition for the generating function $\Pfr(q,T)$.

\

Recall that by definition
\be
\Vm_k=\Im\hat d_k\oplus\coIm \hat d_{k+1}\oplus H_k
\ \Leftrightarrow\
x=\hat d_ky+z+h,\ \  \forall x\in \Vm_k.
\ee
In particular, $x\in\coIm\hat d\stackrel{\mathrm{def}}{\Leftrightarrow}\hat d x\ne 0$.

To simplify the subsequent formulas, we introduce the grading operator $\hat \Delta$ such that $\Zbs=q^{\hat \Delta}$.
Then we chose the basis in the space $\Vm_k$ composed of three groups of the vectors
\be
%\left\{x_{k,i}\right\}_{i=1}^{\dim\Vm_k}=
\begin{array}{lclcl}
\left\{y_{k,i}:\ \hat d_{k+1}y_{k,i}\ne 0\right\}_{i=1}^{\dim\Im\hat d_{k+1}}
&\cup&
\left\{z_{k,i}=\hat d_ky_{k-1,i}\right\}_{i=1}^{\dim\Im \hat d_k}
&\cup&
\Big\{h_{k,i}\Big\}_{i=1}^{\dim\Vm_k-\dim\Im \hat d_k-\dim\Im\hat d_{k+1}},\\[4mm]
\hat\Delta y_{k,i}=\Delta_{k,i}y_{k,i}
&&
\hat\Delta z_{k,i}=\Delta_{k-1,i}z_{k,i}
&&
\hat\Delta h_{i,k}=\Delta^{\prime}_{i,k}h_{i,k}
\end{array}
%\label{basdec}
\ee
The groups are selected successively for all spaces $\Vm_k$ ($k=\overline{0,n}$) in the complex, and the transition $k\to k+1$ is performed as follows.
First, the vectors $\left\{z_{k,i}=\hat d_ky_{k-1,i}\right\}_{i=1}^{\dim\Im \hat d_k}$ are linearly independent and graded by construction. Second, these vectors $z$ span the space $\Im\hat d_k$, which is an invariant subspace of the grading operator $\hat\Delta$  (because $x=\hat d_ky\Rightarrow \hat\Delta x=\hat d_k\Delta y\in\Im\hat d$ for any $y\in\hat d_k$). The factor space $\coIm \hat d_k\equiv\Vm_k/\Im\hat d_k$. Similarly, the factor space is also an invariant subspace of $\hat\Delta$ and contains the graded basis %of the $\hat\Delta$ eigenvectors,
\be
\left\{x_{k,i}\in \coIm \right\}_{i=1}^{\coIm =\dim\Vm_k-\dim\Im \hat d_k}.
\ee
%Enumerate the remaining basis vectors
Next, %if $Y\in\Im \hat d_{k+1}$, then
%\be
%Y=\hat d_{k+1}X=\sum_{i,k}\hat d_{k+1} x_{i,k},
%\ee
%so
the number of the linearly independent vectors among $\left\{\hat d_{k+1} x_{i,k}\right\}_{i=1}^{\dim\Vm}$ is by definition $\dim\Im\hat d_{k+1}$, and because $\hat d_{k+1}z_{k,i}=0$ for all $i$ (due to the nilpotency condition $\hat d_{k+1}\hat d_k=0$), the basis vectors in $\Im\hat d_{k+1}$ must be %found 
among the vectors $\hat d_{k+1}x_{k,i}$. %Take then the %??? pre-images of these basis vectors as the vectors $y$, i.e., $y_{i,k}=x_{i^{\prime},k}$ for $i=\overline{1,\dim\Im\hat d_{k+1}}$.

%Set then $y_{i,k}=x_{i^{\prime},k}$ for all $\{i^{\prime}\}$ such that the vectors $\{dx_{k,i^{\prime}}\}$
%form the basis in $\dim\Im\hat d_{k+1} \Vm_k$.
Images of all the remaining basis vectors can be expanded as $\hat d_{k+1}x_{k,j}=\sum_{i=1}^{\dim\Im\hat d_{k+1}}\alpha_i \hat d_{k+1}y_{k,i}$ for some $\alpha$ (in the particular case $\hat d_{k+1}x_{k,l+1}=0$ all $\alpha=0$).
The vectors $h_{k,i}\equiv x_{k,j}-\sum_{i=1}^{\dim\Im\hat d_{k+1}}\alpha_i y_{k,i}$ form %then
 the basis in the homology space.
Indeed, by definition $\hat d_{k+1}h_{k,i}=0$, hence $h_{k,i}\in\ker\hat d_{k+1}$, and $h_{k,i}\not\in\Im\hat d_k$.
These vectors are linearly independent thanks to the presence of $x_{k,j}$, and the number of them equals to the homology dimension.

The bases in $\Vm_0$ 
and in $\Vm_n$ can be chosen just in same way, if we assume that $\Vm_{-1}\equiv\emptyset$ and $\Vm_{n+1}\equiv\emptyset$, respectively.

\section{Morphisms of the representation spaces. A more involved example\label{app:repex}}
Here we provide one more illustration of the representation theory standpoint on the KhR formalism discussed in sec.~\ref{sec:repth}. The figure below demonstrates a non-trivial (not exact) map between two $[2,1]$ representations of the $su_3$ algebra~\cite{KlimSch}. This map has the same properties as the morphisms %discussed
in sec.~\ref{sec:repth}.
\be
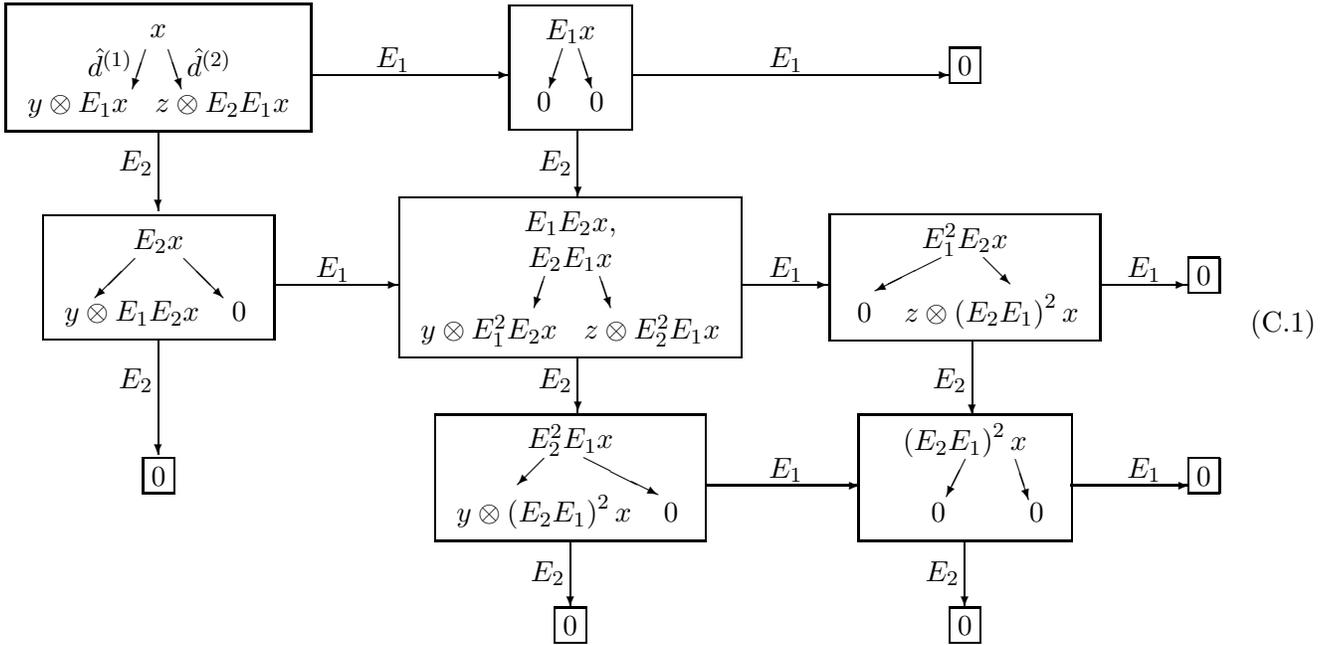
\begin{array}{cccccccc}
\boxed{
\begin{array}{cccc}
\multicolumn{4}{c}{x}\\
\hspace{8mm}\hat d^{(1)}\hspace{-8mm}
&\Arrow{-1}{-3}{5}{12}{10} &\Arrow{1}{-3}{5}{5}{10}&
\hspace{-3.5mm}\hat d^{(2)}\hspace{3.5mm}\\
\multicolumn{2}{c}{y\otimes E_1x}&
\multicolumn{2}{c}{z\otimes E_2E_1x}
\end{array}
}
&
\hspace{5mm}
\stackrel{\textstyle{E_1}}{\Arrow{1}{0}{75}{-31}{0}}
\hspace{-5mm}
&
\boxed{
\begin{array}{cccc}
\multicolumn{4}{c}{E_1x}\\
%\hspace{8mm}\hat d^{(1)}\hspace{-8mm}
&\Arrow{-1}{-3}{5}{2}{10} &\Arrow{1}{-3}{5}{-2}{10}&
%\hspace{-3.5mm}\hat d^{(2)}\hspace{3.5mm}
\\
\multicolumn{2}{c}{0}&
\multicolumn{2}{c}{0}
\end{array}
}
& %\hspace{5mm}
\stackrel{\textstyle{E_1}}{\Arrow{1}{0}{120}{-58}{0}}
%\hspace{-5mm}
&\boxed{0}\\[8mm]
\hspace{-6mm}
E_2\Arrow{0}{-1}{30}{2}{15}&&
\hspace{-4mm}
E_2\Arrow{0}{-1}{25}{2}{15}\\[2mm]
\boxed{
\begin{array}{cccc}
\multicolumn{4}{c}{E_2x}\\
%\hspace{8mm}\hat d^{(1)}\hspace{-8mm}
&\Arrow{-1}{-1}{15}{-4}{10} &\Arrow{1}{-1}{15}{-16}{10}&
%\hspace{-3.5mm}\hat d^{(2)}\hspace{3.5mm}
\\
\multicolumn{2}{c}{y\otimes E_1E_2x}&
\multicolumn{2}{c}{0}
\end{array}
}
&
\hspace{-3mm}
\stackrel{\textstyle{E_1}}{\Arrow{1}{0}{47}{-22}{0}}
\hspace{3mm}
&
\boxed{
\begin{array}{cccc}
\multicolumn{4}{c}{
\begin{array}{c}
E_1E_2x,\\
E_2E_1x
\end{array}}\\
%\hspace{8mm}\hat d^{(1)}\hspace{-8mm}
&\Arrow{-1}{-3}{4}{16}{10} &\Arrow{1}{-3}{4}{6}{10}&
%\hspace{-3.5mm}\hat d^{(2)}\hspace{3.5mm}
\\
\multicolumn{2}{c}{y\otimes E_1^2E_2x}&
\multicolumn{2}{c}{z\otimes E_2^2E_1x}
\end{array}
}
&
%\hspace{5mm}
\stackrel{\textstyle{E_1}}{\Arrow{1}{0}{32}{-16}{0}}
%\hspace{-5mm}
&
\boxed{
\begin{array}{cccc}
\multicolumn{4}{c}{E_1^2E_2x}\\
%\hspace{8mm}\hat d^{(1)}\hspace{-8mm}
&\Arrow{-2}{-1}{25}{24}{10}
&\Arrow{1}{-1}{10}{30}{10}&
%\hspace{-3.5mm}\hat d^{(2)}\hspace{3.5mm}
\\
\multicolumn{2}{c}{0}&
\multicolumn{2}{c}{z\otimes \left(E_2E_1\right)^2x}
\end{array}
}
&
%\hspace{5mm}
\stackrel{\textstyle{E_1}}{\Arrow{1}{0}{32}{-16}{0}}
%\hspace{-5mm}
&\boxed{0}\\[9mm]
\hspace{-6mm}
E_2\Arrow{0}{-1}{43}{2}{18}&&
\hspace{-4mm}
E_2\Arrow{0}{-1}{21}{2}{12}&&
\hspace{-4mm}
E_2\Arrow{0}{-1}{28}{2}{18}
\\[2mm]
\boxed{0}
&&
\boxed{
\begin{array}{cccc}
\multicolumn{4}{c}{E_2^2E_1x}\\
%\hspace{8mm}\hat d^{(1)}\hspace{-8mm}
&\Arrow{-1}{-1}{10}{-5}{10} &\Arrow{2}{-1}{28}{-28}{10}&
%\hspace{-3.5mm}\hat d^{(2)}\hspace{3.5mm}
\\
\multicolumn{2}{c}{y\otimes \left(E_2E_1\right)^2x}&
\multicolumn{2}{c}{0}
\end{array}
}
&
%\hspace{5mm}
\stackrel{\textstyle{E_1}}{\Arrow{1}{0}{58}{-30}{0}}
%\hspace{-5mm}
&
\boxed{
\begin{array}{cccc}
\multicolumn{4}{c}{\left(E_2E_1\right)^2x}\\
%\hspace{8mm}\hat d^{(1)}\hspace{-8mm}
\hspace{12mm}&\Arrow{-1}{-2}{7}{-12}{10} &\Arrow{1}{-3}{5}{-3}{10}&
%\hspace{-3.5mm}\hat d^{(2)}\hspace{3.5mm}
\\
\multicolumn{2}{c}{0}&
\multicolumn{2}{c}{0}
\end{array}
}
&
%\hspace{5mm}
\stackrel{\textstyle{E_1}}{\Arrow{1}{0}{45}{-28}{0}}
%\hspace{-5mm}
&
\boxed{0}\\[8mm]
&&
\hspace{-6mm}
E_2\Arrow{0}{-1}{25}{2}{15}&&
\hspace{-6mm}
E_2\Arrow{0}{-1}{25}{2}{15}\\[2mm]
&&
\boxed{0}
&&
\boxed{0}
\end{array}
\ee

\begin{landscape}
%\phantom{fu}

\section{List of braids providing the unique level I reminders\label{app:BrDet}}
\footnotesize
\vspace{-0.5cm}
%\phantom{ha}

%\vspace{-2cm}
\be
%\hspace{-2cm}
\arraycolsep=0.5mm
%\begin{array}{|p{9cm}|p{9cm}|}
\begin{array}{|p{12.5cm}|p{12.5cm}|}
\hline\multicolumn{2}{|c|}{}\\[-3mm]
\multicolumn{2}{|c|}{\mbox{The first-level minimal positive integer divition of the primary polynomial, the type}}\\
\hline&\\[-3mm]
Uniquely determined&Ambigious\\
\hline\multicolumn{2}{|c|}{}\\[-3mm]
\multicolumn{2}{|c|}{0_1\cup 0_1, \mbox{the pair of unlinked Unknots}}\\\hline&\\[-3mm]
\mbox{$[  1]$,}  \mbox{$[  2]$.} &\\\hline\multicolumn{2}{|c|}{}\\[-3mm]
\multicolumn{2}{|c|}{0_1, \mbox{the Unknot}, \mbox{Torus}\,[2,1]}\\\hline&\\[-3mm]
\mbox{$[  2 1]$,}  \mbox{$[  1 2]$.} &\\\hline\multicolumn{2}{|c|}{}\\[-3mm]
\multicolumn{2}{|c|}{0_1\cup 2_1^1,\ \mbox{the Hopf link unlinked with the Unknot} }\\\hline&\\[-3mm]
\mbox{$[  2 2]$,}  \mbox{$[  1 1]$.} &\\\hline\multicolumn{2}{|c|}{}\\[-3mm]
\multicolumn{2}{|c|}{2_1^1,\ \mbox{the Hopf link}, \mbox{Torus}\,[2,2]}\\\hline&\\[-3mm]
\mbox{$[  2 2 1]$,}  \mbox{$[  1 2 2]$,}  \mbox{$[  1 2 1]$,}  \mbox{$[  1 1 2]$,}  \mbox{$[  2 1 1]$,}  \mbox{$[  2 1 2]$.} &\\\hline\multicolumn{2}{|c|}{}\\[-3mm]
\multicolumn{2}{|c|}{0_1\cup 3_1, \mbox{the Trefoil knot as Tourus}\,[2,3]\mbox{ unlinked with the Unknot}}\\\hline&\\[-3mm]
\mbox{$[  2 2 2]$,}  \mbox{$[  1 1 1]$.} &\\\hline\multicolumn{2}{|c|}{}\\[-3mm]
\multicolumn{2}{|c|}{3_1, \mbox{the Trefoil knot}, \mbox{as Torus}\,[3,2]}\\\hline&\\[-3mm]
\mbox{$[  1 2 1 1]$,}  \mbox{$[  2 2 2 1]$,}  \mbox{$[  2 1 2 2]$,}  \mbox{$[  1 2 2 2]$,}  \mbox{$[  1 1 1 2]$,}  \mbox{$[  1 2 1 2]$,}  \mbox{$[  2 1 2 1]$,}  \mbox{$[  2 2 1 2]$,}  \mbox{$[  2 1 1 1]$,}  \mbox{$[  1 1 2 1]$.}& \\\hline\multicolumn{2}{|c|}{}\\[-3mm]
\multicolumn{2}{|c|}{4^2_1,\ L4a1(1), \mbox{the Solomon link}, \mbox{Torus}\,[2,4]}\\\hline&\\[-3mm]
\mbox{$[  2 2 1 2 2]$,}  \mbox{$[  2 1 2 2 2]$,}  \mbox{$[  2 1 2 1 1]$,}  \mbox{$[  2 2 1 2 1]$,}  \mbox{$[  1 2 1 1 1]$,}  \mbox{$[  1 2 1 1 2]$,}  \mbox{$[  1 1 2 1 2]$,}  \mbox{$[  1 2 1 2 2]$,}  \mbox{$[  1 1 1 2 1]$,}  \mbox{$[  2 2 2 1 2]$,}  \mbox{$[  2 1 2 2 1]$,}  \mbox{$[  1 2 2 1 2]$,}  \mbox{$[  1 1 2 1 1]$,}  \mbox{$[  2 1 2 1 2]$,}  \mbox{$[  1 2 1 2 1]$,}  \mbox{$[  2 1 1 2 1]$.}&
\mbox{$[ 2 2 2 2 1 ]$},  \mbox{$[ 2 1 1 1 1 ]$},  \mbox{$[ 1 2 2 2 2 ]$},  \mbox{$[ 1 1 1 1 2 ]$}.
\\\hline\multicolumn{2}{|c|}{}\\[-3mm]
\multicolumn{2}{|c|}{5_1, \mbox{the Fivefoil knot}, \mbox{Torus}\,[2,5]}\\\hline&\\[-3mm]
\mbox{$[  1 2 1 1 2 2]$,}  \mbox{$[  1 2 1 2 1 1]$,}  \mbox{$[  2 2 2 2 1 2]$,}  \mbox{$[  2 1 2 1 1 1]$,}  \mbox{$[  2 1 2 1 2 2]$,}  \mbox{$[  1 2 1 1 1 1]$,}  \mbox{$[  1 1 1 2 1 2]$,}  \mbox{$[  2 2 2 1 2 2]$,}  \mbox{$[  1 2 2 1 2 1]$,}  \mbox{$[  2 1 2 2 2 2]$,}  \mbox{$[  2 1 1 2 1 2]$,}  \mbox{$[  2 1 2 2 1 1]$,}  \mbox{$[  2 1 2 1 1 2]$,}  \mbox{$[  1 1 2 1 2 2]$,}  \mbox{$[  1 2 1 1 1 2]$,}  \mbox{$[  1 2 2 1 1 2]$,}  \mbox{$[  2 1 2 2 2 1]$,}  \mbox{$[  1 2 1 2 2 1]$,}  \mbox{$[  2 2 1 2 2 2]$,}  \mbox{$[  1 2 2 2 1 2]$,}  \mbox{$[  1 1 2 1 2 1]$,}  \mbox{$[  2 1 1 2 2 1]$,}  \mbox{$[  1 2 1 2 2 2]$,}  \mbox{$[  2 2 1 1 2 1]$,}  \mbox{$[  1 1 1 2 1 1]$,}  \mbox{$[  2 2 1 2 1 2]$,}  \mbox{$[  2 2 2 1 2 1]$,}  \mbox{$[  2 1 1 1 2 1]$,}  \mbox{$[  1 1 2 2 1 2]$,}  \mbox{$[  1 1 1 1 2 1]$,}  \mbox{$[  2 2 1 2 1 1]$,}  \mbox{$[  1 1 2 1 1 1]$.}&
\mbox{$[ 2 2 2 2 2 1 ]$},  \mbox{$[ 1 2 2 2 2 2 ]$},  \mbox{$[ 1 1 1 1 1 2 ]$},  \mbox{$[ 2 1 1 1 1 1 ]$}.
 \\\hline\multicolumn{2}{|c|}{}\\[-3mm]
\multicolumn{2}{|c|}{6_3^3,\ L6n1(0,1), \mbox{Torus}\,[3,3]}
\\\hline&\\[-3mm]
\mbox{$[  2 1 1 2 1 1]$,}  \mbox{$[  2 2 1 2 2 1]$,}  \mbox{$[  1 2 1 1 2 1]$,}  \mbox{$[  1 2 1 2 1 2]$,}  \mbox{$[  2 1 2 1 2 1]$,}  \mbox{$[  1 2 2 1 2 2]$,}  \mbox{$[  2 1 2 2 1 2]$,}  \mbox{$[  1 1 2 1 1 2]$.}&
 \\\hline\multicolumn{2}{|c|}{}\\[-3mm]
\multicolumn{2}{|c|}{7^2_7,\ L7n1(0)}\\\hline&\\[-3mm]
\mbox{$[  1 2 2 1 2 1 2]$,}  \mbox{$[  1 2 2 1 2 2 1]$,}  \mbox{$[  1 2 1 1 1 2 1]$,}  \mbox{$[  2 1 2 2 1 2 1]$,}  \mbox{$[  1 1 2 1 1 1 2]$,}  \mbox{$[  2 2 1 1 2 2 1]$,}  \mbox{$[  2 1 2 1 2 2 1]$,}  \mbox{$[  1 2 1 2 1 1 2]$,}  \mbox{$[  1 1 1 2 1 1 2]$,}  \mbox{$[  2 1 2 1 1 2 1]$,}  \mbox{$[  2 1 2 2 1 2 2]$,}  \mbox{$[  1 2 2 2 1 2 2]$,}  \mbox{$[  1 2 2 1 2 2 2]$,}  \mbox{$[  2 1 2 1 2 1 2]$,}  \mbox{$[  1 1 2 1 2 1 2]$,}  \mbox{$[  2 1 1 2 1 1 1]$,}  \mbox{$[  2 2 1 2 2 2 1]$,}  \mbox{$[  2 1 2 1 2 1 1]$,}  \mbox{$[  2 1 1 2 1 2 1]$,}  \mbox{$[  2 1 1 1 2 1 1]$,}  \mbox{$[  2 1 2 2 2 1 2]$,}  \mbox{$[  1 1 2 2 1 1 2]$,}  \mbox{$[  2 1 1 2 1 1 2]$,}  \mbox{$[  1 2 1 2 1 2 2]$,}  \mbox{$[  1 2 2 1 1 2 2]$,}  \mbox{$[  2 1 1 2 2 1 1]$,}  \mbox{$[  1 2 1 2 1 2 1]$,}  \mbox{$[  1 2 1 2 2 1 2]$,}  \mbox{$[  2 2 1 2 1 2 1]$,}  \mbox{$[  2 2 2 1 2 2 1]$,}  \mbox{$[  1 2 1 1 2 1 2]$,}  \mbox{$[  2 2 1 2 2 1 2]$,}  \mbox{$[  1 1 2 1 1 2 1]$,}  \mbox{$[  1 2 1 1 2 1 1]$.}&
\mbox{$[ 2 2 1 1 2 1 1 ]$},  \mbox{$[ 1 1 2 1 1 2 2 ]$},  \mbox{$[ 2 2 1 2 2 1 1 ]$},  \mbox{$[ 1 2 2 1 1 2 1 ]$},  \mbox{$[ 1 1 2 2 1 2 2 ]$},  \mbox{$[ 1 2 1 1 2 2 1 ]$},  \mbox{$[ 2 1 1 2 2 1 2 ]$},  \mbox{$[ 2 1 2 2 1 1 2 ]$}.
\\\hline\multicolumn{2}{|c|}{}\\[-3mm]
\multicolumn{2}{|c|}{8_{19}, \mbox{Torus}\,[3,4],\ \mbox{the first ``thick'' knot}}\\\hline&\\[-3mm]
\mbox{$[  1 2 2 1 2 2 1 2]$,}  \mbox{$[  2 1 2 1 1 2 1 1]$,}  \mbox{$[  1 1 1 2 1 1 2 2]$,}  \mbox{$[  2 1 2 2 1 2 2 1]$,}  \mbox{$[  2 2 1 2 2 1 2 1]$,}  \mbox{$[  2 2 2 1 2 2 1 1]$,}  \mbox{$[  2 1 1 2 1 2 1 1]$,}  \mbox{$[  2 1 1 2 1 1 2 1]$,}  \mbox{$[  1 1 2 2 1 2 2 2]$,}  \mbox{$[  1 2 1 1 2 1 1 2]$,}  \mbox{$[  2 1 2 1 2 1 2 1]$,}  \mbox{$[  1 1 2 1 2 1 2 2]$,}  \mbox{$[  1 2 1 2 2 1 2 2]$,}  \mbox{$[  1 2 2 1 2 1 2 2]$,}  \mbox{$[  2 2 1 1 2 1 1 1]$,}  \mbox{$[  2 2 1 2 1 2 1 1]$,}  \mbox{$[  1 1 2 1 2 1 1 2]$,}  \mbox{$[  1 1 2 1 1 2 1 2]$,}  \mbox{$[  2 2 1 2 1 2 2 1]$,}  \mbox{$[  1 2 1 2 1 2 1 2]$.} &
\mbox{$[ 1 2 1 1 2 2 1 1 ]$},  \mbox{$[ 2 1 2 2 1 1 2 2 ]$},  \mbox{$[ 2 2 2 1 1 2 2 1 ]$},  \mbox{$[ 1 2 2 1 1 2 1 2 ]$},  \mbox{$[ 1 1 2 2 1 1 1 2 ]$},  \mbox{$[ 1 1 2 1 1 1 2 1 ]$},  \mbox{$[ 1 2 1 1 2 1 2 1 ]$},  \mbox{$[ 2 1 1 2 1 1 1 2 ]$},  \mbox{$[ 1 2 1 2 1 1 2 1 ]$},  \mbox{$[ 2 2 1 1 2 2 2 1 ]$},  \mbox{$[ 2 2 1 2 2 2 1 2 ]$},  \mbox{$[ 1 1 2 2 2 1 2 2 ]$},  \mbox{$[ 2 1 1 2 2 1 1 1 ]$},  \mbox{$[ 2 1 1 2 2 2 1 2 ]$},  \mbox{$[ 1 2 1 2 1 1 2 2 ]$},  \mbox{$[ 1 2 1 1 1 2 2 1 ]$},  \mbox{$[ 2 2 1 1 2 1 2 1 ]$},  \mbox{$[ 1 2 1 1 1 2 1 1 ]$},  \mbox{$[ 2 1 2 1 1 2 2 1 ]$},  \mbox{$[ 2 1 1 1 2 1 1 1 ]$},  \mbox{$[ 1 2 2 1 1 1 2 1 ]$},  \mbox{$[ 1 1 1 2 1 1 1 2 ]$},  \mbox{$[ 2 1 2 2 2 1 2 2 ]$},  \mbox{$[ 2 1 2 1 2 2 1 2 ]$},  \mbox{$[ 1 2 1 2 2 1 1 2 ]$},  \mbox{$[ 2 1 1 2 2 1 2 2 ]$},  \mbox{$[ 2 1 1 1 2 1 1 2 ]$},  \mbox{$[ 2 1 1 1 2 2 1 1 ]$},  \mbox{$[ 1 2 2 2 1 2 2 2 ]$},  \mbox{$[ 1 1 2 2 1 1 2 1 ]$},  \mbox{$[ 2 1 2 1 2 2 1 1 ]$},  \mbox{$[ 2 1 2 2 1 2 1 2 ]$},  \mbox{$[ 1 1 2 1 1 2 2 1 ]$},  \mbox{$[ 2 2 2 1 2 2 2 1 ]$},  \mbox{$[ 1 1 2 1 1 1 2 2 ]$},  \mbox{$[ 1 1 1 2 2 1 1 2 ]$},  \mbox{$[ 2 2 1 2 2 1 1 2 ]$},  \mbox{$[ 1 2 2 2 1 1 2 2 ]$},  \mbox{$[ 2 1 1 2 2 1 2 1 ]$},  \mbox{$[ 1 2 2 1 2 1 2 1 ]$},  \mbox{$[ 1 2 2 1 2 2 2 1 ]$},  \mbox{$[ 2 1 1 2 1 2 1 2 ]$},  \mbox{$[ 1 2 1 1 2 2 1 2 ]$},  \mbox{$[ 2 2 1 1 1 2 1 1 ]$},  \mbox{$[ 1 1 2 2 1 2 1 2 ]$},  \mbox{$[ 2 2 1 1 2 2 1 2 ]$},  \mbox{$[ 1 2 2 1 1 2 1 1 ]$},  \mbox{$[ 2 1 2 2 1 1 2 1 ]$},  \mbox{$[ 2 1 2 2 2 1 1 2 ]$},  \mbox{$[ 1 2 2 2 1 2 2 1 ]$},  \mbox{$[ 2 1 2 1 2 1 1 2 ]$},  \mbox{$[ 1 2 2 1 1 2 2 2 ]$},  \mbox{$[ 2 2 1 2 2 2 1 1 ]$},  \mbox{$[ 1 2 1 2 1 2 2 1 ]$}.
\\\hline
\end{array}
\nn\ee

\end{landscape}

\section{Examples of the unique minimal remainders related to the CohFT diagrams in the case of three strands\label{app:inst}}
%\cite{knbook}
\subsection{The doubly twisted diagram of the unknot\label{app:un}}
\be
\begin{array}{|c|c|c|c|c|}
\hline
\multicolumn{5}{|c|}{[1,2],\mbox{ the doubly twisted unknot}}\\
\hline
T\mbox{-grading}&q\mbox{-grading}&\chi&\mbox{Diagrams}&
\begin{array}{c}
\mbox{Khovanov} \\\mbox{polynomial,}\\
q\mbox{-gradings}\\\mbox{of the terms}
\end{array}\\
\hline
&&&&\\[-4mm]
0&
2+\Delta_{111}&1&
\left[\arraycolsep=0.5mm\begin   {array}{cc}\en&\\&\en\end{array}\right]&
\\[2.5mm]\cline{2-4}&&&&\\[-4mm]
 &2+\Delta_{001}&1&
\left[\arraycolsep=0.5mm\begin{array}{cc}\en&\\&\sn\end{array}\right]&
3,
\\[2.5mm]\cline{2-4}&&&&\\[-4mm]
 &2+\Delta_{011}&1 &
\left[\arraycolsep=0.5mm\begin{array}{cc}\sn&\\&\en\end{array}\right]&
1
\\[2.5mm]\cline{2-4}&&&&\\[-4mm]
&2+\Delta_{000}&1 &
\left[\arraycolsep=0.5mm\begin{array}{cc}\en&\\&\en\end{array}\right]& \\
\hline&&&&\\[-4mm]
1&
\Delta_{011}&-1 &
\left[\arraycolsep=0.5mm\begin{array}{cc}\db&\\&\en\end{array}\right]&
\\[2.5mm]\cline{2-4}&&&&\\[-4mm]
&\Delta_{001}&-1 &
\left[\arraycolsep=0.5mm\begin{array}{cc}\en&\\&\db\end{array}\right]& \\
\hline
\end{array}
\ee

\begin{landscape}

\subsection{The trefoil knot in different three strand presentations\label{app:tref}}
\be
\begin{array}{|c|c|c|c|c|}
\hline
\multicolumn{5}{|c|}{1.\ [1112],\mbox{ the Trefoil knot}}\\
\hline
T\mbox{-grading}&q\mbox{-grading}&|\chi|&\mbox{Diagrams}&
\begin{array}{c}
\mbox{Khovanov} \\\mbox{polynomial,}\\
q\mbox{-gradings}\\\mbox{of the terms}
\end{array}\\
\hline&&&&\\[-4mm]0
&4+\Delta_{111}&1&
\left[\arraycolsep=0.2mm\begin{array}{cccc}\en & \en & \en &\\& & & \en\end{array}\right]&
\\\cline{2-4}&&&&\\[-4mm]

 &4+\Delta_{000}&1&
\left[\arraycolsep=0.2mm\begin{array}{cccc}\en & \en & \en &\\& & & \en\end{array}\right]&
-1,
\\\cline{2-4}&&&&\\[-4mm]

 &4+\Delta_{001}&1&
\left[\arraycolsep=0.2mm\begin{array}{cccc}\en & \en & \en &\\& & & \sn\end{array}\right]& -3\\\cline{2-4}&&&&\\[-4mm]

 &4+\Delta_{011}&1& \left[\arraycolsep=0.2mm\begin{array}{cccc}\sn & \sn & \sn &\\& & & \en\end{array}\right]&
\\\hline&&&&\\[-4mm]
1
&2+\Delta_{011}&1&
\left[\arraycolsep=0.2mm\begin{array}{cccc}\db & \sn & \sn &\\& & & \en\end{array}\right],
 \left[\arraycolsep=0.2mm\begin{array}{cccc}\sn & \db & \sn &\\& & & \en\end{array}\right],
 \left[\arraycolsep=0.2mm\begin{array}{cccc}\sn & \sn & \db &\\& & & \en\end{array}\right]& \\\cline{2-4}&&&&\\[-4mm]

 &2+\Delta_{001}&1&
\left[\arraycolsep=0.2mm\begin{array}{cccc}\en & \en & \en &\\& & & \db\end{array}\right]& \\

 \hline&&&&\\[-4mm]2&
2+\Delta_{100}&1&
\left[\arraycolsep=0.2mm\begin{array}{cccc}\c & \sn & \c &\\& & & \en\end{array}\right]& \\\cline{2-4}&&&&\\[-4mm]

 &2+\Delta_{101}&1&
\left[\arraycolsep=0.2mm\begin{array}{cccc}\c & \sn & \c &\\& & & \sn\end{array}\right]&
-5
\\\cline{2-4}&&&&\\[-4mm]

 &\Delta_{011}&1&
\left[\arraycolsep=0.2mm\begin{array}{cccc}\db & \db & \sn &\\& & & \en\end{array}\right],
 \left[\arraycolsep=0.2mm\begin{array}{cccc}\db & \sn & \db &\\& & & \en\end{array}\right],
 \left[\arraycolsep=0.2mm\begin{array}{cccc}\sn & \db & \db &\\& & & \en\end{array}\right]& \\

 \hline&&&&\\[-4mm]3
&2+\Delta_{011}&1&
\left[\arraycolsep=0.2mm\begin{array}{cccc}\db & \db & \db &\\& & & \en\end{array}\right]&
\\\cline{2-4}&&&&\\[-4mm]

 &\Delta_{101}&2&
\left[\arraycolsep=0.2mm\begin{array}{cccc}\c & \db & \c &\\& & & \sn\end{array}\right],
 \left[\arraycolsep=0.2mm\begin{array}{cccc}\c & \sn & \c &\\& & & \db\end{array}\right]& -9

\\\cline{2-4}&&&&\\[-4mm]

 &\Delta_{100}&1&
\left[\arraycolsep=0.2mm\begin{array}{cccc}\c & \db & \c &\\& & & \en\end{array}\right]& \\

 \hline&&&&\\[-4mm]4
&2+\Delta_{101}&1&
\left[\arraycolsep=0.2mm\begin{array}{cccc}\c & \db & \c &\\& & & \db\end{array}\right]&\\
\hline
\end{array}\
%\ee
%\be
\begin{array}{|c|c|c|c|c|}
\hline
\multicolumn{5}{|c|}{2.\ [1121],\mbox{ the Trefoil knot}}\\
\hline
T\mbox{-grading}&q\mbox{-grading}&|\chi|&\mbox{Diagrams}&
\begin{array}{c}
\mbox{Khovanov} \\\mbox{polynomial,}\\
q\mbox{-gradings}\\\mbox{of the terms}
\end{array}
\\[2.5mm]\hline&&&&\\[-4mm]0
&4+\Delta_{111}&1&
\left[\arraycolsep=0.2mm\begin{array}{cccc}\en&\en&&\en\\&&\en&\end{array}\right]&

 \\[2.5mm]\cline{2-4}&&&&\\[-4mm]&4+\Delta_{000}&1&
\left[\arraycolsep=0.2mm\begin{array}{cccc}\en&\en&&\en\\&&\en&\end{array}\right]&
-1,

 \\[2.5mm]\cline{2-4}&&&&\\[-4mm]&4+\Delta_{011}&1&
\left[\arraycolsep=0.2mm\begin{array}{cccc}\sn&\sn&&\sn\\&&\en&\end{array}\right]&
-3

 \\[2.5mm]\cline{2-4}&&&&\\[-4mm]&4+\Delta_{001}&1&
\left[\arraycolsep=0.2mm\begin{array}{cccc}\en&\en&&\en\\&&\sn&\end{array}\right]&

 \\[2.5mm]\hline&&&&\\[-4mm]1
&2+\Delta_{001}&1&
\left[\arraycolsep=0.2mm\begin{array}{cccc}\en&\en&&\en\\&&\db&\end{array}\right]&

 \\[2.5mm]\cline{2-4}&&&&\\[-4mm]&2+\Delta_{011}&1&
\left[\arraycolsep=0.2mm\begin{array}{cccc}\db&\sn&&\sn\\&&\en&\end{array}\right],
 \left[\arraycolsep=0.2mm\begin{array}{cccc}\sn&\db&&\sn\\&&\en&\end{array}\right],
 \left[\arraycolsep=0.2mm\begin{array}{cccc}\sn&\sn&&\db\\&&\en&\end{array}\right]&

 \\[2.5mm]\hline&&&&\\[-4mm]2&
2+\Delta_{101}&1&
\left[\arraycolsep=0.2mm\begin{array}{cccc}\c&\sn&&\c\\&&\en&\end{array}\right]&
-5

 \\[2.5mm]\cline{2-4}&&&&\\[-4mm]&2+\Delta_{010}&1&
\left[\arraycolsep=0.2mm\begin{array}{cccc}\sn&\c&&\c\\&&\en&\end{array}\right]&

 \\[2.5mm]\hline&&&&\\[-4mm]3
&\Delta_{101}&1&
\left[\arraycolsep=0.2mm\begin{array}{cccc}\c&\db&&\c\\&&\en&\end{array}\right]&
-9

 \\[2.5mm]\cline{2-4}&&&&\\[-4mm]&\Delta_{010}&1&
\left[\arraycolsep=0.2mm\begin{array}{cccc}\db&\c&&\c\\&&\en&\end{array}\right]&

\\
\hline
\end{array}\
\nn\ee
\be
\begin{array}{|c|c|c|c|c|}
\hline
\multicolumn{5}{|c|}{3.\ [1211],\mbox{ the Trefoil knot}}\\
\hline
T\mbox{-grading}&q\mbox{-grading}&|\chi|&\mbox{Diagrams}&
\begin{array}{c}
\mbox{Khovanov} \\\mbox{polynomial,}\\
q\mbox{-gradings}\\\mbox{of the terms}
\end{array}
\\[2.5mm]\hline&&&&\\[-4mm]0
&4+\Delta_{001}&1&
\left[\arraycolsep=0.2mm\begin{array}{cccc}\en&&\en&\en\\&\sn&&\end{array}\right]&

 \\[2.5mm]\cline{2-4}&&&&\\[-4mm]&4+\Delta_{011}&1&
\left[\arraycolsep=0.2mm\begin{array}{cccc}\sn&&\sn&\sn\\&\en&&\end{array}\right]&
-1,

 \\[2.5mm]\cline{2-4}&&&&\\[-4mm]&4+\Delta_{111}&1&
\left[\arraycolsep=0.2mm\begin{array}{cccc}\en&&\en&\en\\&\en&&\end{array}\right]&
-3

 \\[2.5mm]\cline{2-4}&&&&\\[-4mm]&4+\Delta_{000}&1&
\left[\arraycolsep=0.2mm\begin{array}{cccc}\en&&\en&\en\\&\en&&\end{array}\right]&

 \\[2.5mm]\hline&&&&\\[-4mm]1
&2+\Delta_{001}&1&
\left[\arraycolsep=0.2mm\begin{array}{cccc}\en&&\en&\en\\&\db&&\end{array}\right]&

\\[2.5mm]\cline{2-4}&&&&\\[-4mm]&2+\Delta_{011}&1&
\left[\arraycolsep=0.2mm\begin{array}{cccc}\db&&\sn&\sn\\&\en&&\end{array}\right],
 \left[\arraycolsep=0.2mm\begin{array}{cccc}\sn&&\db&\sn\\&\en&&\end{array}\right],
 \left[\arraycolsep=0.2mm\begin{array}{cccc}\sn&&\sn&\db\\&\en&&\end{array}\right]&

 \\[2.5mm]\hline&&&&\\[-4mm]2&
2+\Delta_{101}&1&
\left[\arraycolsep=0.2mm\begin{array}{cccc}\c&&\sn&\c\\&\en&&\end{array}\right]&
-5

 \\[2.5mm]\cline{2-4}&&&&\\[-4mm]&2+\Delta_{010}&1&
\left[\arraycolsep=0.2mm\begin{array}{cccc}\c&&\c&\sn\\&\en&&\end{array}\right]&

 \\[2.5mm]\hline&&&&\\[-4mm]3
&\Delta_{010}&1&
\left[\arraycolsep=0.2mm\begin{array}{cccc}\c&&\c&\db\\&\en&&\end{array}\right]&
-9

\\[2.5mm]\cline{2-4}&&&&\\[-4mm]&\Delta_{101}&1&
\left[\arraycolsep=0.2mm\begin{array}{cccc}\c&&\db&\c\\&\en&&\end{array}\right]&
\\
\hline
\end{array}\
%\ee
%\be
\begin{array}{|c|c|c|c|c|}
\hline
\multicolumn{5}{|c|}{4.\ [2111],\mbox{ the Trefoil knot}}\\
\hline
T\mbox{-grading}&q\mbox{-grading}&|\chi|&\mbox{Diagrams}&
\begin{array}{c}
\mbox{Khovanov} \\\mbox{polynomial,}\\
q\mbox{-gradings}\\\mbox{of the terms}
\end{array}
\\[2.5mm]\hline&&&&\\[-4mm]0
&4+\Delta_{111}&1&
\left[\arraycolsep=0.2mm\begin{array}{cccc}&\en&\en&\en\\\en&&&\end{array}\right]&

 \\[2.5mm]\cline{2-4}&&&&\\[-4mm]&4+\Delta_{000}&1&
\left[\arraycolsep=0.2mm\begin{array}{cccc}&\en&\en&\en\\\en&&&\end{array}\right]&
-1,

 \\[2.5mm]\cline{2-4}&&&&\\[-4mm]&4+\Delta_{011}&1&
\left[\arraycolsep=0.2mm\begin{array}{cccc}&\sn&\sn&\sn\\\en&&&\end{array}\right]&
-3,

 \\[2.5mm]\cline{2-4}&&&&\\[-4mm]&4+\Delta_{001}&1&
\left[\arraycolsep=0.2mm\begin{array}{cccc}&\en&\en&\en\\\sn&&&\end{array}\right]&

 \\[2.5mm]\hline&&&&\\[-4mm]1
&2+\Delta_{001}&1&
\left[\arraycolsep=0.2mm\begin{array}{cccc}&\en&\en&\en\\\db&&&\end{array}\right]&

 \\[2.5mm]\cline{2-4}&&&&\\[-4mm]&2+\Delta_{011}&1&
\left[\arraycolsep=0.2mm\begin{array}{cccc}&\db&\sn&\sn\\\en&&&\end{array}\right],
 \left[\arraycolsep=0.2mm\begin{array}{cccc}&\sn&\db&\sn\\\en&&&\end{array}\right],
 \left[\arraycolsep=0.2mm\begin{array}{cccc}&\sn&\sn&\db\\\en&&&\end{array}\right]&

 \\[2.5mm]\hline&&&&\\[-4mm]2&
2+\Delta_{101}&1&
\left[\arraycolsep=0.2mm\begin{array}{cccc}&\c&\sn&\c\\\sn&&&\end{array}\right]&
-5

 \\[2.5mm]\cline{2-4}&&&&\\[-4mm]&\Delta_{011}&1&
\left[\arraycolsep=0.2mm\begin{array}{cccc}&\db&\db&\sn\\\en&&&\end{array}\right],
 \left[\arraycolsep=0.2mm\begin{array}{cccc}&\db&\sn&\db\\\en&&&\end{array}\right],
 \left[\arraycolsep=0.2mm\begin{array}{cccc}&\sn&\db&\db\\\en&&&\end{array}\right]&

 \\[2.5mm]\cline{2-4}&&&&\\[-4mm]&2+\Delta_{100}&1&
\left[\arraycolsep=0.2mm\begin{array}{cccc}&\c&\sn&\c\\\en&&&\end{array}\right]&

 \\[2.5mm]\hline&&&&\\[-4mm]3
&\Delta_{100}&1&
\left[\arraycolsep=0.2mm\begin{array}{cccc}&\c&\db&\c\\\en&&&\end{array}\right]&
-9

 \\[2.5mm]\cline{2-4}&&&&\\[-4mm]&\Delta_{101}&2&
\left[\arraycolsep=0.2mm\begin{array}{cccc}&\c&\sn&\c\\\db&&&\end{array}\right],
 \left[\arraycolsep=0.2mm\begin{array}{cccc}&\c&\db&\c\\\sn&&&\end{array}\right]&

 \\[2.5mm]\cline{2-4}&&&&\\[-4mm]&2+\Delta_{011}&1&
\left[\arraycolsep=0.2mm\begin{array}{cccc}&\db&\db&\db\\\en&&&\end{array}\right]&

 \\[2.5mm]\hline&&&&\\[-4mm]4
&2+\Delta_{101}&1&
\left[\arraycolsep=0.2mm\begin{array}{cccc}&\c&\db&\c\\\db&&&\end{array}\right]&

\\
\hline
\end{array}
\nn\ee

\be
\begin{array}{|c|c|c|c|c|}
\hline
\multicolumn{5}{|c|}{5.\ [1222],\mbox{ the Trefoil knot}}\\
\hline
T\mbox{-grading}&q\mbox{-grading}&|\chi|&\mbox{Diagrams}&
\begin{array}{c}
\mbox{Khovanov} \\\mbox{polynomial,}\\
q\mbox{-gradings}\\\mbox{of the terms}
\end{array}
\\[2.5mm]\hline&&&&\\[-4mm]0
&4+\Delta_{001}&1&
\left[\arraycolsep=0.2mm\begin{array}{cccc}\en&&&\\&\sn&\sn&\sn\end{array}\right]&

 \\[2.5mm]\cline{2-4}&&&&\\[-4mm]&4+\Delta_{011}&1&
\left[\arraycolsep=0.2mm\begin{array}{cccc}\sn&&&\\&\en&\en&\en\end{array}\right]&
-1,

 \\[2.5mm]\cline{2-4}&&&&\\[-4mm]&4+\Delta_{111}&1&
\left[\arraycolsep=0.2mm\begin{array}{cccc}\en&&&\\&\en&\en&\en\end{array}\right]&
-3

 \\[2.5mm]\cline{2-4}&&&&\\[-4mm]&4+\Delta_{000}&1&
\left[\arraycolsep=0.2mm\begin{array}{cccc}\en&&&\\&\en&\en&\en\end{array}\right]&

 \\[2.5mm]\hline&&&&\\[-4mm]1
&2+\Delta_{001}&1 &
\left[\arraycolsep=0.2mm\begin{array}{cccc}\en&&&\\&\db&\sn&\sn\end{array}\right],
 \left[\arraycolsep=0.2mm\begin{array}{cccc}\en&&&\\&\sn&\db&\sn\end{array}\right],
 \left[\arraycolsep=0.2mm\begin{array}{cccc}\en&&&\\&\sn&\sn&\db\end{array}\right]&

 \\[2.5mm]\cline{2-4}&&&&\\[-4mm]&2+\Delta_{011}&1&
\left[\arraycolsep=0.2mm\begin{array}{cccc}\db&&&\\&\en&\en&\en\end{array}\right]&

 \\[2.5mm]\hline&&&&\\[-4mm]2&
2+\Delta_{110}&1&
\left[\arraycolsep=0.2mm\begin{array}{cccc}\en&&&\\&\c&\sn&\c\end{array}\right]&

 \\[2.5mm]\cline{2-4}&&&&\\[-4mm]&\Delta_{001}&1&
\left[\arraycolsep=0.2mm\begin{array}{cccc}\en&&&\\&\db&\db&\sn\end{array}\right],
 \left[\arraycolsep=0.2mm\begin{array}{cccc}\en&&&\\&\db&\sn&\db\end{array}\right],
 \left[\arraycolsep=0.2mm\begin{array}{cccc}\en&&&\\&\sn&\db&\db\end{array}\right]&
-5

 \\[2.5mm]\cline{2-4}&&&&\\[-4mm]&2+\Delta_{010}&1&
\left[\arraycolsep=0.2mm\begin{array}{cccc}\sn&&&\\&\c&\sn&\c\end{array}\right]&

 \\[2.5mm]\hline&&&&\\[-4mm]3
&\Delta_{110}&1&
\left[\arraycolsep=0.2mm\begin{array}{cccc}\en&&&\\&\c&\db&\c\end{array}\right]&

 \\[2.5mm]\cline{2-4}&&&&\\[-4mm]&\Delta_{010}&2&
\left[\arraycolsep=0.2mm\begin{array}{cccc}\db&&&\\&\c&\sn&\c\end{array}\right],
 \left[\arraycolsep=0.2mm\begin{array}{cccc}\sn&&&\\&\c&\db&\c\end{array}\right]&
-9

 \\[2.5mm]\cline{2-4}&&&&\\[-4mm]&2+\Delta_{001}&1&
\left[\arraycolsep=0.2mm\begin{array}{cccc}\en&&&\\&\db&\db&\db\end{array}\right]&

 \\[2.5mm]\hline&&&&\\[-4mm]4
&2+\Delta_{010}&1&
\left[\arraycolsep=0.2mm\begin{array}{cccc}\db&&&\\&\c&\db&\c\end{array}\right]&

\\
\hline
\end{array}\
%\ee
%\be
\begin{array}{|c|c|c|c|c|}
\hline
\multicolumn{5}{|c|}{6.\ [2122],\mbox{ the Trefoil knot}}\\
\hline
T\mbox{-grading}&q\mbox{-grading}&|\chi|&\mbox{Diagrams}&
\begin{array}{c}
\mbox{Khovanov} \\\mbox{polynomial,}\\
q\mbox{-gradings}\\\mbox{of the terms}
\end{array}
\\[2.5mm]\hline&&&&\\[-4mm]0
&4+\Delta_{111}&1&
\left[\arraycolsep=0.2mm\begin{array}{cccc}&\en&&\\\en&&\en&\en\end{array}\right]&

 \\[2.5mm]\cline{2-4}&&&&\\[-4mm]&4+\Delta_{000}&1&
\left[\arraycolsep=0.2mm\begin{array}{cccc}&\en&&\\\en&&\en&\en\end{array}\right]&
-1,

 \\[2.5mm]\cline{2-4}&&&&\\[-4mm]&4+\Delta_{011}&1&
\left[\arraycolsep=0.2mm\begin{array}{cccc}&\sn&&\\\en&&\en&\en\end{array}\right]&
-3

 \\[2.5mm]\cline{2-4}&&&&\\[-4mm]&4+\Delta_{001}&1&
\left[\arraycolsep=0.2mm\begin{array}{cccc}&\en&&\\\sn&&\sn&\sn\end{array}\right]&

 \\[2.5mm]\hline&&&&\\[-4mm]1
&2+\Delta_{001}&1&
\left[\arraycolsep=0.2mm\begin{array}{cccc}&\en&&\\\db&&\sn&\sn\end{array}\right],
 \left[\arraycolsep=0.2mm\begin{array}{cccc}&\en&&\\\sn&&\db&\sn\end{array}\right],
 \left[\arraycolsep=0.2mm\begin{array}{cccc}&\en&&\\\sn&&\sn&\db\end{array}\right]&

 \\[2.5mm]\cline{2-4}&&&&\\[-4mm]&2+\Delta_{011}&1&
\left[\arraycolsep=0.2mm\begin{array}{cccc}&\db&&\\\en&&\en&\en\end{array}\right]&

 \\[2.5mm]\hline&&&&\\[-4mm]2&
2+\Delta_{101}&1&
\left[\arraycolsep=0.2mm\begin{array}{cccc}&\en&&\\\c&&\c&\sn\end{array}\right]&
-5

 \\[2.5mm]\cline{2-4}&&&&\\[-4mm]&2+\Delta_{010}&1&
\left[\arraycolsep=0.2mm\begin{array}{cccc}&\en&&\\\c&&\sn&\c\end{array}\right]&

 \\[2.5mm]\hline&&&&\\[-4mm]3
&\Delta_{101}&1&
\left[\arraycolsep=0.2mm\begin{array}{cccc}&\en&&\\\c&&\c&\db\end{array}\right]&
-9
 \\[2.5mm]\cline{2-4}&&&&\\[-4mm]&\Delta_{010}&1&
\left[\arraycolsep=0.2mm\begin{array}{cccc}&\en&&\\\c&&\db&\c\end{array}\right]&

\\
\hline
\end{array}
\nn\ee

\be
\begin{array}{|c|c|c|c|c|}
\hline
\multicolumn{5}{|c|}{7.\ [2212],\mbox{ the Trefoil knot}}\\
\hline
T\mbox{-grading}&q\mbox{-grading}&|\chi|&\mbox{Diagrams}&
\begin{array}{c}
\mbox{Khovanov} \\\mbox{polynomial,}\\
q\mbox{-gradings}\\\mbox{of the terms}
\end{array}
\\[2.5mm]\hline&&&&\\[-4mm]0
&4+\Delta_{001}&1&
\left[\arraycolsep=0.2mm\begin{array}{cccc}&&\en&\\\sn&\sn&&\sn\end{array}\right]&

 \\[2.5mm]\cline{2-4}&&&&\\[-4mm]&4+\Delta_{011}&1&
\left[\arraycolsep=0.2mm\begin{array}{cccc}&&\sn&\\\en&\en&&\en\end{array}\right]&
-1,

 \\[2.5mm]\cline{2-4}&&&&\\[-4mm]&4+\Delta_{111}&1&
\left[\arraycolsep=0.2mm\begin{array}{cccc}&&\en&\\\en&\en&&\en\end{array}\right]&
-3

 \\[2.5mm]\cline{2-4}&&&&\\[-4mm]&4+\Delta_{000}&1&
\left[\arraycolsep=0.2mm\begin{array}{cccc}&&\en&\\\en&\en&&\en\end{array}\right]&

 \\[2.5mm]\hline&&&&\\[-4mm]1
&2+\Delta_{001}&1&
\left[\arraycolsep=0.2mm\begin{array}{cccc}&&\en&\\\db&\sn&&\sn\end{array}\right],
 \left[\arraycolsep=0.2mm\begin{array}{cccc}&&\en&\\\sn&\db&&\sn\end{array}\right],
 \left[\arraycolsep=0.2mm\begin{array}{cccc}&&\en&\\\sn&\sn&&\db\end{array}\right]&

 \\[2.5mm]\cline{2-4}&&&&\\[-4mm]&2+\Delta_{011}&1&
\left[\arraycolsep=0.2mm\begin{array}{cccc}&&\db&\\\en&\en&&\en\end{array}\right]&

 \\[2.5mm]\hline&&&&\\[-4mm]2&
2+\Delta_{101}&1&
\left[\arraycolsep=0.2mm\begin{array}{cccc}&&\en&\\\sn&\c&&\c\end{array}\right]&
-5

 \\[2.5mm]\cline{2-4}&&&&\\[-4mm]&2+\Delta_{010}&1&
\left[\arraycolsep=0.2mm\begin{array}{cccc}&&\en&\\\c&\sn&&\c\end{array}\right]&

 \\[2.5mm]\hline&&&&\\[-4mm]3
&\Delta_{010}&1&
\left[\arraycolsep=0.2mm\begin{array}{cccc}&&\en&\\\c&\db&&\c\end{array}\right]&
-9

 \\[2.5mm]\cline{2-4}&&&&\\[-4mm]&\Delta_{101}&1&
\left[\arraycolsep=0.2mm\begin{array}{cccc}&&\en&\\\db&\c&&\c\end{array}\right]&

\\
\hline
\end{array}\
%\ee
%\be
\begin{array}{|c|c|c|c|c|}
\hline
\multicolumn{5}{|c|}{8.\ [2221],\mbox{ the Trefoil knot}}\\
\hline
T\mbox{-grading}&q\mbox{-grading}&|\chi|&\mbox{Diagrams}&
\begin{array}{c}
\mbox{Khovanov} \\\mbox{polynomial,}\\
q\mbox{-gradings}\\\mbox{of the terms}
\end{array}
\\[2.5mm]\hline&&&&\\[-4mm]0
&4+\Delta_{111}&1&
\left[\arraycolsep=0.2mm\begin{array}{cccc}&&&\en\\\en&\en&\en&\end{array}\right]&

 \\[2.5mm]\cline{2-4}&&&&\\[-4mm]&4+\Delta_{000}&1&
\left[\arraycolsep=0.2mm\begin{array}{cccc}&&&\en\\\en&\en&\en&\end{array}\right]&
-1,

 \\[2.5mm]\cline{2-4}&&&&\\[-4mm]&4+\Delta_{011}&1&
\left[\arraycolsep=0.2mm\begin{array}{cccc}&&&\sn\\\en&\en&\en&\end{array}\right]&
-3

 \\[2.5mm]\cline{2-4}&&&&\\[-4mm]&4+\Delta_{001}&1&
\left[\arraycolsep=0.2mm\begin{array}{cccc}&&&\en\\\sn&\sn&\sn&\end{array}\right]&

 \\[2.5mm]\hline&&&&\\[-4mm]1
&2+\Delta_{001}&1&
\left[\arraycolsep=0.2mm\begin{array}{cccc}&&&\en\\\db&\sn&\sn&\end{array}\right],
 \left[\arraycolsep=0.2mm\begin{array}{cccc}&&&\en\\\sn&\db&\sn&\end{array}\right],
 \left[\arraycolsep=0.2mm\begin{array}{cccc}&&&\en\\\sn&\sn&\db&\end{array}\right]&

 \\[2.5mm]\cline{2-4}&&&&\\[-4mm]&2+\Delta_{011}&1&
\left[\arraycolsep=0.2mm\begin{array}{cccc}&&&\db\\\en&\en&\en&\end{array}\right]&

 \\[2.5mm]\hline&&&&\\[-4mm]2&
2+\Delta_{110}&1&
\left[\arraycolsep=0.2mm\begin{array}{cccc}&&&\en\\\c&\sn&\c&\end{array}\right]&

 \\[2.5mm]\cline{2-4}&&&&\\[-4mm]&2+\Delta_{010}&1&
\left[\arraycolsep=0.2mm\begin{array}{cccc}&&&\sn\\\c&\sn&\c&\end{array}\right]&
-5

 \\[2.5mm]\cline{2-4}&&&&\\[-4mm]&\Delta_{001}&1&
\left[\arraycolsep=0.2mm\begin{array}{cccc}&&&\en\\\db&\db&\sn&\end{array}\right],
 \left[\arraycolsep=0.2mm\begin{array}{cccc}&&&\en\\\db&\sn&\db&\end{array}\right],
 \left[\arraycolsep=0.2mm\begin{array}{cccc}&&&\en\\\sn&\db&\db&\end{array}\right]&

 \\[2.5mm]\hline&&&&\\[-4mm]3
&\Delta_{110}&1&
\left[\arraycolsep=0.2mm\begin{array}{cccc}&&&\en\\\c&\db&\c&\end{array}\right]&

 \\[2.5mm]\cline{2-4}&&&&\\[-4mm]&\Delta_{010}&2&
\left[\arraycolsep=0.2mm\begin{array}{cccc}&&&\sn\\\c&\db&\c&\end{array}\right],
 \left[\arraycolsep=0.2mm\begin{array}{cccc}&&&\db\\\c&\sn&\c&\end{array}\right]&
-9

 \\[2.5mm]\cline{2-4}&&&&\\[-4mm]&2+\Delta_{001}&1&
\left[\arraycolsep=0.2mm\begin{array}{cccc}&&&\en\\\db&\db&\db&\end{array}\right]&

 \\[2.5mm]\hline&&&&\\[-4mm]4
&2+\Delta_{010}&1&
\left[\arraycolsep=0.2mm\begin{array}{cccc}&&&\db\\\c&\db&\c&\end{array}\right]&

\\
\hline
\end{array}
\nn\ee

\be
\begin{array}{|c|c|c|c|c|}
\hline
\multicolumn{5}{|c|}{9.\ [1212].  ,\mbox{ the Trefoil knot}}\\
\hline
T\mbox{-grading}&q\mbox{-grading}&|\chi|&\mbox{Diagrams}&
\begin{array}{c}
\mbox{Khovanov} \\\mbox{polynomial,}\\
q\mbox{-gradings}\\\mbox{of the terms}
\end{array}

\\[2.5mm]\hline&&&&\\[-4mm]0
&4+\Delta_{111}&1&
\left[\arraycolsep=0.2mm\begin{array}{cccc}\en&&\en&\\&\en&&\en\end{array}\right]&

 \\[2.5mm]\cline{2-4}&&&&\\[-4mm]&4+\Delta_{000}&1&
\left[\arraycolsep=0.2mm\begin{array}{cccc}\en&&\en&\\&\en&&\en\end{array}\right]&
-1,

 \\[2.5mm]\cline{2-4}&&&&\\[-4mm]&4+\Delta_{011}&1&
\left[\arraycolsep=0.2mm\begin{array}{cccc}\sn&&\sn&\\&\en&&\en\end{array}\right]&
-3

 \\[2.5mm]\cline{2-4}&&&&\\[-4mm]&4+\Delta_{001}&1&
\left[\arraycolsep=0.2mm\begin{array}{cccc}\en&&\en&\\&\sn&&\sn\end{array}\right]&

 \\[2.5mm]\hline&&&&\\[-4mm]1&
2+\Delta_{001}&1&
\left[\arraycolsep=0.2mm\begin{array}{cccc}\en&&\en&\\&\db&&\sn\end{array}\right],
 \left[\arraycolsep=0.2mm\begin{array}{cccc}\en&&\en&\\&\sn&&\db\end{array}\right]&

 \\[2.5mm]\cline{2-4}&&&&\\[-4mm]&2+\Delta_{011}&1&
\left[\arraycolsep=0.2mm\begin{array}{cccc}\db&&\sn&\\&\en&&\en\end{array}\right],
 \left[\arraycolsep=0.2mm\begin{array}{cccc}\sn&&\db&\\&\en&&\en\end{array}\right]&

 \\[2.5mm]\hline&&&&\\[-4mm]2&
2+\Delta_{101}&1&
\left[\arraycolsep=0.2mm\begin{array}{cccc}\c&&\c&\\&\en&&\sn\end{array}\right]&
-5

 \\[2.5mm]\cline{2-4}&&&&\\[-4mm]&2+\Delta_{010}&1&
\left[\arraycolsep=0.2mm\begin{array}{cccc}\sn&&\en&\\&\c&&\c\end{array}\right]&

 \\[2.5mm]\hline&&&&\\[-4mm]3
&\Delta_{101}&1&
\left[\arraycolsep=0.2mm\begin{array}{cccc}\c&&\c&\\&\en&&\db\end{array}\right]&
-9

 \\[2.5mm]\cline{2-4}&&&&\\[-4mm]&\Delta_{010}&1&
\left[\arraycolsep=0.2mm\begin{array}{cccc}\db&&\en&\\&\c&&\c\end{array}\right]&

\\
\hline
\end{array}\
%\ee
%\be
\begin{array}{|c|c|c|c|c|}
\hline
\multicolumn{5}{|c|}{10.\ [2121],\mbox{ the Trefoil knot}}\\
\hline
T\mbox{-grading}&q\mbox{-grading}&|\chi|&\mbox{Diagrams}&
\begin{array}{c}
\mbox{Khovanov} \\\mbox{polynomial,}\\
q\mbox{-gradings}\\\mbox{of the terms}
\end{array}

\\[2.5mm]\hline&&&&\\[-4mm]0
&4+\Delta_{001}&1&
\left[\arraycolsep=0.2mm\begin{array}{cccc}&\en&&\en\\\sn&&\sn&\end{array}\right]&

 \\[2.5mm]\cline{2-4}&&&&\\[-4mm]&4+\Delta_{011}&1&
\left[\arraycolsep=0.2mm\begin{array}{cccc}&\sn&&\sn\\\en&&\en&\end{array}\right]&
-1,

 \\[2.5mm]\cline{2-4}&&&&\\[-4mm]&4+\Delta_{111}&1&
\left[\arraycolsep=0.2mm\begin{array}{cccc}&\en&&\en\\\en&&\en&\end{array}\right]&
-3

 \\[2.5mm]\cline{2-4}&&&&\\[-4mm]&4+\Delta_{000}&1&
\left[\arraycolsep=0.2mm\begin{array}{cccc}&\en&&\en\\\en&&\en&\end{array}\right]&

 \\[2.5mm]\hline&&&&\\[-4mm]1&
2+\Delta_{001}&1&
\left[\arraycolsep=0.2mm\begin{array}{cccc}&\en&&\en\\\db&&\sn&\end{array}\right],
 \left[\arraycolsep=0.2mm\begin{array}{cccc}&\en&&\en\\\sn&&\db&\end{array}\right]&

 \\[2.5mm]\cline{2-4}&&&&\\[-4mm]&2+\Delta_{011}&1&
\left[\arraycolsep=0.2mm\begin{array}{cccc}&\db&&\sn\\\en&&\en&\end{array}\right],
 \left[\arraycolsep=0.2mm\begin{array}{cccc}&\sn&&\db\\\en&&\en&\end{array}\right]&

 \\[2.5mm]\hline&&&&\\[-4mm]2&
2+\Delta_{101}&1&
\left[\arraycolsep=0.2mm\begin{array}{cccc}&\c&&\c\\\sn&&\en&\end{array}\right]&
-5

 \\[2.5mm]\cline{2-4}&&&&\\[-4mm]&2+\Delta_{010}&1&
\left[\arraycolsep=0.2mm\begin{array}{cccc}&\en&&\sn\\\c&&\c&\end{array}\right]&

 \\[2.5mm]\hline&&&&\\[-4mm]3
&\Delta_{010}&1&
\left[\arraycolsep=0.2mm\begin{array}{cccc}&\en&&\db\\\c&&\c&\end{array}\right]&
-9

 \\[2.5mm]\cline{2-4}&&&&\\[-4mm]&\Delta_{101}&1&
\left[\arraycolsep=0.2mm\begin{array}{cccc}&\c&&\c\\\db&&\en&\end{array}\right]&

\\
\hline
\end{array}
\nn\ee

\end{landscape}

\phantom{s}

\vspace{-2.25cm}

\subsection{The ``thick'' knot $8_{19}$ (torus $T[3,4]$)\label{app:T34}}

\be
%\nn\\[-1.68cm]
{\footnotesize
\begin{array}{|c|c|c|p{12cm}|c|c|}
\hline\multicolumn{5}{|c|}{}\\[-3mm]
\multicolumn{5}{|c|}{[1  2  1  2  2  1  2  2],\ \ \ \mbox{knot}\ 8_{19},\ \ \mbox{or torus}\ \ T[3,4]}\\
\hline&&&&\\[-3.5mm]
T\mbox{-gr.}&q\mbox{-gr.}&|\chi|&\mbox{Diagrams}&
\begin{array}{c}
\mbox{Kh.pol.,}\\
q\mbox{-grds.}
\end{array}\\
\hline&&&&\\[-3mm]
 0 &8+\Delta_{0 00}&  1&
 $\arraycolsep=0.2mm\left[  \begin{array}{cccccccc}\en&   &  \en &   &   &  \en &   & \\ &  \en &   &  \en &  \en &   &  \en &  \en\end{array}\right]$  .

 &\\[-1mm]\cline{2-4}  &&&&\\[-3mm]& 8+\Delta_{1 11}&  1&
 $\arraycolsep=0.2mm\left[  \begin{array}{cccccccc}\en  &   &  \en &   &   &  \en &   & \\ &  \en &   &  \en &  \en &   &  \en &  \en\end{array}\right]$  .

 &5,
 \\[-1mm]\cline{2-4}  &&&&\\[-3mm]& 8+\Delta_{0 11}&  1&
 $\arraycolsep=0.2mm\left[  \begin{array}{cccccccc}\sn  &   &  \sn &   &   &  \sn &   & \\ &  \en &   &  \en &  \en &   &  \en &  \en\end{array}\right]$  .

 &7
 \\[-1mm]\cline{2-4}  &&&&\\[-3mm]& 8+\Delta_{0 01}&  1&
 $\arraycolsep=0.2mm\left[  \begin{array}{cccccccc}\en&   &  \en &   &   &  \en &   & \\ &  \sn &   &  \sn &  \sn &   &  \sn &  \sn\end{array}\right]$  .

&\\[-1mm]\hline  &&&&\\[-4.55mm]\hline  &&&&\\ [-3mm]
 1 &6+\Delta_{0 11}&  1 &
 $\arraycolsep=0.2mm\left[  \begin{array}{cccccccc}\db  &   &  \sn &   &   &  \sn &   & \\ &  \en &   &  \en &  \en &   &  \en &  \en\end{array}\right]$  ,
  $\arraycolsep=0.2mm\left[  \begin{array}{cccccccc}\sn  &   &  \db &   &   &  \sn &   & \\ &  \en &   &  \en &  \en &   &  \en &  \en\end{array}\right]$  ,
  $\arraycolsep=0.2mm\left[  \begin{array}{cccccccc}\sn  &   &  \sn &   &   &  \db &   & \\ &  \en &   &  \en &  \en &   &  \en &  \en\end{array}\right]$  .

 &\\[-1mm]\cline{2-4}  &&&&\\[-3mm]& 6+\Delta_{0 01}&  1 &
 $\arraycolsep=0.2mm\left[  \begin{array}{cccccccc}\en&   &  \en &   &   &  \en &   & \\ &  \db &   &  \sn &  \sn &   &  \sn &  \sn\end{array}\right]$  ,
  $\arraycolsep=0.2mm\left[  \begin{array}{cccccccc}\en&   &  \en &   &   &  \en &   & \\ &  \sn &   &  \db &  \sn &   &  \sn &  \sn\end{array}\right]$  ,
  $\arraycolsep=0.2mm\left[  \begin{array}{cccccccc}\en&   &  \en &   &   &  \en &   & \\ &  \sn &   &  \sn &  \db &   &  \sn &  \sn\end{array}\right]$  ,
  $\arraycolsep=0.2mm\left[  \begin{array}{cccccccc}\en&   &  \en &   &   &  \en &   & \\ &  \sn &   &  \sn &  \sn &   &  \db &  \sn\end{array}\right]$  ,
  $\arraycolsep=0.2mm\left[  \begin{array}{cccccccc}\en&   &  \en &   &   &  \en &   & \\ &  \sn &   &  \sn &  \sn &   &  \sn &   \db\end{array}\right]$  .
%\\[-1mm]
&\\[-1mm]\hline  &&&&\\[-4.55mm]\hline  &&&&\\[-3mm]
 2 &6+\Delta_{0 10}&  1&
 $\arraycolsep=0.2mm\left[  \begin{array}{cccccccc}\sn  &   &  \en &   &   &  \en &   & \\ &  \c &   &  \sn &  \sn &   &  \sn &  \c\end{array}\right]$  .

 &9
 \\[-1mm]\cline{2-4}  &&&&\\[-3mm]& 6+\Delta_{1 01}&  1&
 $\arraycolsep=0.2mm\left[  \begin{array}{cccccccc}\c &   &  \sn &   &   &  \c &   & \\ &  \en &   &  \en &  \en &   &  \sn &  \sn\end{array}\right]$  .

&\\[-1mm]\hline  &&&&\\[-4.55mm]\hline  &&&&\\[-3mm]   3 &4+\Delta_{0 10}&  1 &
 $\arraycolsep=0.2mm\left[  \begin{array}{cccccccc}\db  &   &  \en &   &   &  \en &   & \\ &  \c &   &  \sn &  \sn &   &  \sn &  \c\end{array}\right]$  ,
  $\arraycolsep=0.2mm\left[  \begin{array}{cccccccc}\sn  &   &  \en &   &   &  \en &   & \\ &  \c &   &  \db &  \sn &   &  \sn &  \c\end{array}\right]$  ,
  $\arraycolsep=0.2mm\left[  \begin{array}{cccccccc}\sn  &   &  \en &   &   &  \en &   & \\ &  \c &   &  \sn &  \db &   &  \sn &  \c\end{array}\right]$  ,
  $\arraycolsep=0.2mm\left[  \begin{array}{cccccccc}\sn  &   &  \en &   &   &  \en &   & \\ &  \c &   &  \sn &  \sn &   &  \db &  \c\end{array}\right]$  .

 &13
 \\[-1mm]\cline{2-4}  &&&&\\[-3mm]& 4+\Delta_{1 01}&  1 &
 $\arraycolsep=0.2mm\left[  \begin{array}{cccccccc}\c &   &  \db &   &   &  \c &   & \\ &  \en &   &  \en &  \en &   &  \sn &  \sn\end{array}\right]$  ,
  $\arraycolsep=0.2mm\left[  \begin{array}{cccccccc}\c &   &  \sn &   &   &  \c &   & \\ &  \en &   &  \en &  \en &   &  \db &  \sn\end{array}\right]$  ,
  $\arraycolsep=0.2mm\left[  \begin{array}{cccccccc}\c &   &  \sn &   &   &  \c &   & \\ &  \en &   &  \en &  \en &   &  \sn &   \db\end{array}\right]$  .

&\\[-1mm]\hline  &&&&\\[-4.55mm]\hline  &&&&\\[-3mm]
  4 & 4+\Delta_{1 00}&  1&
 $\arraycolsep=0.2mm\left[  \begin{array}{cccccccc}\c &   &  \en &   &   &  \c &   & \\ &  \c &   &  \sn &  \c &   &  \en &  \en\end{array}\right]$  .
&\mathbf{11},\\ %[-1mm]
\cline{2-4}  &&&&\\[-3mm]&2+\Delta_{0 01}&  1&
 $\arraycolsep=0.2mm\left[  \begin{array}{cccccccc}\en&   &  \en &   &   &  \sn &   & \\ &  \db &   &  \db &  \c &   &  \c &  \sn\end{array}\right]$  ,
  $\arraycolsep=0.2mm\left[  \begin{array}{cccccccc}\en&   &  \en &   &   &  \en &   & \\ &  \db &   &  \db &  \sn &   &  \c &  \c\end{array}\right]$  ,
  $\arraycolsep=0.2mm\left[  \begin{array}{cccccccc}\en&   &  \en &   &   &  \en &   & \\ &  \db &   &  \c &  \c &   &  \db &  \sn\end{array}\right]$  ,
  $\arraycolsep=0.2mm\left[  \begin{array}{cccccccc}\en&   &  \en &   &   &  \en &   & \\ &  \db &   &  \c &  \c &   &  \sn &   \db\end{array}\right]$  ,
  $\arraycolsep=0.2mm\left[  \begin{array}{cccccccc}\en&   &  \en &   &   &  \en &   & \\ &  \db &   &  \sn &  \db &   &  \c &  \c\end{array}\right]$  ,
  $\arraycolsep=0.2mm\left[  \begin{array}{cccccccc}\en&   &  \en &   &   &  \db &   & \\ &  \db &   &  \sn &  \c &   &  \c &  \sn\end{array}\right]$  ,
  $\arraycolsep=0.2mm\left[  \begin{array}{cccccccc}\en&   &  \en &   &   &  \sn &   & \\ &  \db &   &  \sn &  \c &   &  \c &   \db\end{array}\right]$  ,
  $\arraycolsep=0.2mm\left[  \begin{array}{cccccccc}\en&   &  \db &   &   &  \en &   & \\ &  \c &   &  \c &  \db &   &  \sn &  \sn\end{array}\right]$  ,
  $\arraycolsep=0.2mm\left[  \begin{array}{cccccccc}\en&   &  \db &   &   &  \en &   & \\ &  \c &   &  \c &  \sn &   &  \db &  \sn\end{array}\right]$  ,
  $\arraycolsep=0.2mm\left[  \begin{array}{cccccccc}\en&   &  \db &   &   &  \en &   & \\ &  \c &   &  \c &  \sn &   &  \sn &   \db\end{array}\right]$  ,
  $\arraycolsep=0.2mm\left[  \begin{array}{cccccccc}\en&   &  \sn &   &   &  \en &   & \\ &  \c &   &  \c &  \db &   &  \db &  \sn\end{array}\right]$  ,
  $\arraycolsep=0.2mm\left[  \begin{array}{cccccccc}\en&   &  \sn &   &   &  \en &   & \\ &  \c &   &  \c &  \db &   &  \sn &   \db\end{array}\right]$  ,
  $\arraycolsep=0.2mm\left[  \begin{array}{cccccccc}\en&   &  \sn &   &   &  \en &   & \\ &  \c &   &  \c &  \sn &   &  \db &   \db\end{array}\right]$  ,
  $\arraycolsep=0.2mm\left[  \begin{array}{cccccccc}\en&   &  \en &   &   &  \en &   & \\ &  \sn &   &  \db &  \db &   &  \c &  \c\end{array}\right]$  ,
  $\arraycolsep=0.2mm\left[  \begin{array}{cccccccc}\en&   &  \en &   &   &  \db &   & \\ &  \sn &   &  \db &  \c &   &  \c &  \sn\end{array}\right]$  ,
  $\arraycolsep=0.2mm\left[  \begin{array}{cccccccc}\en&   &  \en &   &   &  \sn &   & \\ &  \sn &   &  \db &  \c &   &  \c &   \db\end{array}\right]$  ,
  $\arraycolsep=0.2mm\left[  \begin{array}{cccccccc}\en&   &  \en &   &   &  \en &   & \\ &  \sn &   &  \c &  \c &   &  \db &   \db\end{array}\right]$  ,
  $\arraycolsep=0.2mm\left[  \begin{array}{cccccccc}\en&   &  \en &   &   &  \db &   & \\ &  \sn &   &  \sn &  \c &   &  \c &   \db\end{array}\right]$  .

 &13
 \\[-1mm]\cline{2-4}  &&&&\\[-3mm]& 4+\Delta_{1 10}&  1&
 $\arraycolsep=0.2mm\left[  \begin{array}{cccccccc}\en  &   &  \c &   &   &  \c &   & \\ &  \c &   &  \en &  \en &   &  \sn &  \c\end{array}\right]$  .

 &\\[-1mm]\cline{2-4}  &&&&\\[-3mm]& 2+\Delta_{0 11}&  1&
 $\arraycolsep=0.2mm\left[  \begin{array}{cccccccc}\db  &   &  \c &   &   &  \c &   & \\ &  \en &   &  \db &  \sn &   &  \en &  \en\end{array}\right]$  ,
  $\arraycolsep=0.2mm\left[  \begin{array}{cccccccc}\db  &   &  \c &   &   &  \c &   & \\ &  \en &   &  \sn &  \db &   &  \en &  \en\end{array}\right]$  ,
  $\arraycolsep=0.2mm\left[  \begin{array}{cccccccc}\c &   &  \c &   &   &  \db &   & \\ &  \db &   &  \en &  \en &   &  \en &  \en\end{array}\right]$  ,
  $\arraycolsep=0.2mm\left[  \begin{array}{cccccccc}\sn  &   &  \c &   &   &  \c &   & \\ &  \en &   &  \db &  \db &   &  \en &  \en\end{array}\right]$  .

&\\[-1mm]\hline  &&&&\\[-4.55mm]\hline  &&&&\\[-3mm]
  5 & \Delta_{0 01}&  1 &
 $\arraycolsep=0.2mm\left[  \begin{array}{cccccccc}\en&   &  \en &   &   &  \en &   & \\ &  \db &   &  \db &  \db &   &  \c &  \c\end{array}\right]$  ,
  $\arraycolsep=0.2mm\left[  \begin{array}{cccccccc}\en&   &  \en &   &   &  \db &   & \\ &  \db &   &  \db &  \c &   &  \c &  \sn\end{array}\right]$  ,
  $\arraycolsep=0.2mm\left[  \begin{array}{cccccccc}\en&   &  \en &   &   &  \sn &   & \\ &  \db &   &  \db &  \c &   &  \c &   \db\end{array}\right]$  ,
  $\arraycolsep=0.2mm\left[  \begin{array}{cccccccc}\en&   &  \en &   &   &  \en &   & \\ &  \db &   &  \c &  \c &   &  \db &   \db\end{array}\right]$  ,
  $\arraycolsep=0.2mm\left[  \begin{array}{cccccccc}\en&   &  \en &   &   &  \db &   & \\ &  \db &   &  \sn &  \c &   &  \c &   \db\end{array}\right]$  ,
  $\arraycolsep=0.2mm\left[  \begin{array}{cccccccc}\en&   &  \db &   &   &  \en &   & \\ &  \c &   &  \c &  \db &   &  \db &  \sn\end{array}\right]$  ,
  $\arraycolsep=0.2mm\left[  \begin{array}{cccccccc}\en&   &  \db &   &   &  \en &   & \\ &  \c &   &  \c &  \db &   &  \sn &   \db\end{array}\right]$  ,
  $\arraycolsep=0.2mm\left[  \begin{array}{cccccccc}\en&   &  \db &   &   &  \en &   & \\ &  \c &   &  \c &  \sn &   &  \db &   \db\end{array}\right]$  ,
  $\arraycolsep=0.2mm\left[  \begin{array}{cccccccc}\en&   &  \sn &   &   &  \en &   & \\ &  \c &   &  \c &  \db &   &  \db &   \db\end{array}\right]$  ,
  $\arraycolsep=0.2mm\left[  \begin{array}{cccccccc}\en&   &  \en &   &   &  \db &   & \\ &  \sn &   &  \db &  \c &   &  \c &   \db\end{array}\right]$  .

 &\\[-1mm]\cline{2-4}  &&&&\\[-3mm]& 2+\Delta_{1 00}&  1 &
 $\arraycolsep=0.2mm\left[  \begin{array}{cccccccc}\c &   &  \en &   &   &  \c &   & \\ &  \c &   &  \db &  \c &   &  \en &  \en\end{array}\right]$  .

 &15,
 \\[-1mm]\cline{2-4}  &&&&\\[-3mm]& 2+\Delta_{1 10}&  1 &
 $\arraycolsep=0.2mm\left[  \begin{array}{cccccccc}\en  &   &  \c &   &   &  \c &   & \\ &  \c &   &  \en &  \en &   &  \db &  \c\end{array}\right]$  .

 &17
 \\[-1mm]\cline{2-4}  &&&&\\[-3mm]& \Delta_{0 11}&  1 &
 $\arraycolsep=0.2mm\left[  \begin{array}{cccccccc}\db  &   &  \c &   &   &  \c &   & \\ &  \en &   &  \db &  \db &   &  \en &  \en\end{array}\right]$.
&\\
\hline
\end{array}}
\nn\ee

\end{document}